\documentclass{emulateapj}

\usepackage{apjfonts}

\def\cm{\textrm{cm}}

\def\micron{\mu\textrm{m}}

\def\erg{\textrm{ergs}}
\def\kpc{\textrm{kpc}}
\def\pc{\textrm{pc}}
\def\Mpc{\textrm{Mpc}}

\def\Kelv{\textrm{K}}
\def\sr{\textrm{sr}}

\def\ergps{\textrm{ergs}~\textrm{s}^{-1}}

\def\kms{\textrm{km}~\textrm{s}^{-1}}
\def\gcm2{\textrm{g}~\textrm{cm}^{-2}}
\def\ergscm3{\textrm{erg}~\textrm{s}^{-1}~\textrm{cm}^{-3}}
\def\ergcm3{\textrm{erg}~\textrm{cm}^{-3}}
\def\gscm2{\textrm{g}~\textrm{s}^{-1}~\textrm{cm}^{-2}}
\def\ergcmK34{\textrm{erg}~\textrm{cm}^{-3}~\textrm{K}^{-4}}

\def\cms31{\textrm{cm}^{-3}~\textrm{s}^{-1}}
\def\cmg21{\textrm{cm}^{2}~\textrm{g}^{-1}}

\def\phcm2s1{\textrm{photons}~\textrm{cm}^{-2}~\textrm{s}^{-1}}

\def\TeVFUnits{\textrm{TeV}~\textrm{cm}^{-2}~\textrm{s}^{-1}}
\def\eV{\textrm{eV}}
\def\keV{\textrm{keV}}
\def\MeV{\textrm{MeV}}
\def\GeV{\textrm{GeV}}
\def\TeV{\textrm{TeV}}
\def\PeV{\textrm{PeV}}

\def\GHz{\textrm{GHz}}

\def\yr{\textrm{yr}}
\def\Myr{\textrm{Myr}}

\def\muGauss{\mu\textrm{G}}
\def\mGauss{\textrm{mG}}

\def\GeVs1cm3{\textrm{GeV}~\textrm{s}^{-1}~\textrm{cm}^{3}}

\def\2phn{\phn\phn}
\def\Msun{\textrm{M}_{\sun}}
\def\Lsun{\textrm{L}_{\sun}}
\newcommand{\mean}[1]{\ensuremath{\langle #1 \rangle}}

\begin{document}

\title{Diffuse Hard X-ray Emission in Starburst Galaxies as Synchrotron from Very High Energy Electrons}
\author{Brian C. Lacki\altaffilmark{1,2,3,4} \& Todd A. Thompson\altaffilmark{3,4,5}}
\altaffiltext{1}{Janksy Fellow}
\altaffiltext{2}{Institute for Advanced Study, Einstein Drive, Princeton, NJ 08540}
\altaffiltext{3}{Department of Astronomy, The Ohio State University, 140 West 18th Avenue, Columbus, OH 43210.}
\altaffiltext{4}{Center for Cosmology \& Astro-Particle Physics, The Ohio State University, Columbus, Ohio 43210, USA}
\altaffiltext{5}{Alfred P. Sloan fellow}

\begin{abstract}
The origin of the diffuse hard X-ray (2 - 10 keV) emission from starburst galaxies is a long-standing problem.  We suggest that synchrotron emission of 10 - 100 TeV electrons and positrons ($e^{\pm}$) can contribute to this emission, because starbursts have strong magnetic fields.  We consider three sources of $e^{\pm}$ at these energies: (1) primary electrons directly accelerated by supernova remnants; (2) pionic secondary $e^{\pm}$ created by inelastic collisions between CR protons and gas nuclei in the dense ISMs of starbursts; (3) pair $e^{\pm}$ produced between the interactions between 10 - 100 TeV $\gamma$-rays and the intense far-infrared (FIR) radiation fields of starbursts.  We create one-zone steady-state models of the CR population in the Galactic Center ($R \le 112\ \pc$), NGC 253, M82, and Arp 220's nuclei, assuming a power law injection spectrum for electrons and protons.  We consider different injection spectral slopes, magnetic field strengths, CR acceleration efficiencies, and diffusive escape times, and include advective escape, radiative cooling processes, and secondary and pair $e^{\pm}$.  We compare these models to extant radio and GeV and TeV $\gamma$-ray data for these starbursts, and calculate the diffuse synchrotron X-ray and Inverse Compton (IC) luminosities of these starbursts in the models which satisfy multiwavelength constraints.  If the primary electron spectrum extends to $\sim\PeV$ energies and has a proton/electron injection ratio similar to the Galactic value, we find that synchrotron contributes $2 - 20\%$ of their unresolved, diffuse hard X-ray emission.  However, there is great uncertainty in this conclusion because of the limited information on the CR electron spectrum at these high energies.  Inverse Compton emission is likewise a minority of the unresolved X-ray emission in these starbursts, from 0.1\% in the Galactic Center to 10\% in Arp 220's nuclei, with the main uncertainty being the starbursts' magnetic field.  We also model generic starbursts, including submillimeter galaxies, in the context of the FIR--X-ray relation, finding that anywhere between 0 and 16\% of the total hard X-ray emission is synchrotron for different parameters, and up to 2\% in the densest starbursts assuming a $E^{-2.2}$ injection spectrum and a diffusive escape time of $10\ \Myr\ (E / 3\ \GeV)^{-1/2} (h / 100\ \pc)$.  Neutrino observations by IceCube and TeV $\gamma$-ray data from HESS, VERITAS, and CTA can further constrain the synchrotron X-ray emission of starbursts.  Our models do not constrain the possibility of hard, second components of primary $e^{\pm}$ from sources like pulsars in starbursts, which could enhance the synchrotron X-ray emission further.  

\end{abstract}



\section{Introduction}
\label{sec:Introduction}

Starburst galaxies are intense generators of cosmic rays (CRs), which are accelerated by supernova remnants or other star-formation processes.  CR protons in starbursts can produce $\gamma$-rays, neutrinos, and secondary electrons and positrons.  Whatever their origin, CR electrons and positrons can produce emission across the electromagnetic spectrum: bremsstrahlung losses produce $\gamma$-rays; Inverse Compton scattering of ambient photons produces a broadband spectrum extending into $\gamma$-rays; synchrotron emission is responsible for the non-thermal GHz radio emission.

Starburst galaxies are observed to be luminous in hard X-rays (here defined as $\sim 2 - 10\ \keV$) as well.  The total hard X-ray emission from star-formation typically has a luminosity that is $10^{-4}$ times that of the bolometric luminosity of the starburst, and is sometimes used as a star-formation indicator \citep{Franceschini03,Ranalli03,Grimm03,Persic04}.  Most of these X-rays are from point sources, but diffuse emission also is apparent.  The diffuse hard X-ray emission is best studied in the nearby starburst M82 \citep{Strickland07}, but a similar diffuse hard component is observed in NGC 253 \citep{Strickland00-N253,Weaver02}.  The diffuse hard X-ray emission bears some superficial resemblance to the Galactic Ridge emission of the Milky Way \citep{Strickland07}, although that has recently been resolved into stellar sources by \emph{Chandra} \citep{Revnivtsev09}. At the other extreme, unresolved hard X-ray emission is also observed in brighter starbursts including the Luminous Infrared Galaxy (LIRG) NGC 3256 \citep{Moran99,Lira02} and the prototypical Ultraluminous Infrared Galaxy (ULIRG) Arp 220 \citep[e.g.,][]{Clements02,McDowell03}.  Especially in these more distant galaxies, it is unclear whether these components are truly diffuse or simply unresolved sources \citep[e.g.,][]{Lira02}.  

The diffuse hard X-ray emission typically has a power-law continuum spectrum, with the possible addition of softer thermal emission components and spectral lines \citep{Persic02,Strickland07,Lehmer10}.  Thermal emission from hot plasma is one possible source of the hard X-rays: hot gas is predicted by superwind theories \citep[e.g.,][]{Chevalier85,Strickland09}, and thermal emission is clearly detected in soft X-rays \citep[e.g.,][]{Ptak97,Dahlem98,Strickland00}, although \citet{Strickland00} argue that the soft X-rays come from a cooler phase of gas than the bulk of the superwind \citep[see also the discussion in][]{Strickland09}.  The detection of 6.7 keV iron K lines implies the existence of hot plasma that could be a source of the hard X-rays \citep{Persic98,Cappi99,Iwasawa05,Iwasawa09}, although it is not clear that such emission could explain all of the hard continuum \citep{Strickland07}.  The hard X-ray emission is often attributed to unresolved X-ray binaries, particularly high-mass X-ray binaries (HMXBs) \citep[e.g.,][]{Fabbiano82,Fabbiano89,Griffiths90,David92,Persic02,Grimm03,Persic04}.  HMXBs in the Milky Way and Magellanic Clouds have a power law continuum (photon spectra of $dN/dE \propto E^{-\Gamma}$) with a spectral slope of $\Gamma \approx 1.2$ \citep{White83}.  Many starburst galaxies have total hard X-ray emission with $\Gamma \approx 1 - 1.5$ \citep[e.g.,][]{Rephaeli91,Rephaeli95}, which makes HMXBs an attractive candidate for the source of most of the hard X-ray emission.  However, in M82, diffuse hard X-ray emission remains even after subtracting point sources and extrapolating the luminosity function to faint luminosities, and the diffuse emission is softer ($\Gamma_{2-8} \approx 2 - 3$) than expected from HMXBs \citep{Strickland07}, unless the HMXB spectrum cuts off at energies $\la 10\ \keV$.  

Another possible explanation for this X-ray emission is Inverse Compton emission \citep{Hargrave74,Schaaf89,Moran97,Moran99,Persic03}.  CR electrons and positrons ($e^{\pm}$) are known to be present in starbursts from their synchrotron radio emission, and the intense infrared emission of starbursts provides many target photons to be upscattered to higher energies.  The IC spectrum is expected to extend down to the X-rays and even lower energies.  However, recent estimates generally suggest that IC emission is too weak by a factor of $\ga 10$ in M82 and NGC 253 to explain the hard X-ray emission \citep[e.g.,][]{Weaver02,Strickland07}.  As with HMXBs, the spectral slope of the diffuse X-ray emission in some starbursts may be difficult to explain with IC.  The CR $e^{\pm}$ spectrum around $\sim 100\ \MeV$ and the resultant IC spectrum at keV energies is expected to be hard ($\Gamma \approx 1.0 - 1.5$), whereas the diffuse X-ray emission is often softer, as is the case for M82 \citep{Strickland07}.

Here, we consider an alternative source of hard X-rays: \emph{synchrotron emission} from CR $e^{\pm}$ with energies above a TeV.  Synchrotron has previously been considered as a source of Galactic diffuse X-ray emission \citep{Protheroe80,Porter97}, but explaining the diffuse X-ray emission from the Galaxy requires CR electrons of extreme energies ($\sim 100\ \TeV$), and stronger magnetic fields ($B \ga 20\ \muGauss$) than in the diffuse ISM to avoid conflict with constraints on Inverse Compton emission from CASA-MIA \citep{Aharonian00,Bi09}.  Starbursts are expected not only to have large CR populations but also stronger magnetic fields than the Milky Way.  The conclusion of strong magnetic fields in starbursts is supported by a number of lines of evidence: (1) minimum energy estimates applied to radio detections of starbursts imply  $B \approx 50 - 150\ \muGauss$ \citep[e.g.,][]{Voelk89a,Beck05,Thompson06,Persic10,Beck11}; (2) detailed modeling of CR populations in starbursts with fitting of the radio and now $\gamma$-ray spectra imply $B \approx 100\ \muGauss$ in the Galactic Center \citep{Crocker11}, $B \approx 200\ \muGauss$ in the nearby starbursts M82 and NGC 253 \citep{Paglione96,Domingo05,Persic08,deCeaDelPozo09a,Rephaeli09,deCeaDelPozo09b}, and $B \ga 1\ \mGauss$ in the ULIRG Arp 220 \citep{Torres04}; (3) considerations of the linear FIR-radio correlation that applies to quiescent star-forming galaxies and starbursts, and the strong Inverse Compton, bremsstrahlung, and ionization losses expected for CR electrons in these galaxies \citep[e.g.,][]{Voelk89b,Condon91,Thompson06,Murphy09,Lacki10a}; (4) measurements of Zeeman splitting in ULIRGs, although these apply to the denser regions of the starbursts \citep{Robinshaw08}; (5) constraints on leptonic $\gamma$-ray emission from the Galactic Center, limiting the number of CR $e^{\pm}$ \citep{Crocker10}.  

Strong magnetic fields imply not only more synchrotron power per particle, but that lower energy CR electrons can produce synchrotron X-rays.  The synchrotron emission of a CR $e^{\pm}$ peaks near $\nu_C = 3 \gamma^2 e B \sin \alpha / (4 \pi m_e c)$ for an electron with Lorentz factor $\gamma$ with a pitch angle $\alpha$ with respect to a magnetic field of strength $B$ \citep{Rybicki79}.  For an isotropic distribution of electrons, $\mean{\sin \alpha} = \pi / 4$, and this translates to a synchrotron emission energy of $E_c = h \nu_C$:
\begin{equation}
\label{eqn:ECSynch}
E_C \approx 1.0 E_{\rm 10 TeV}^2 B_{200} \keV,
\end{equation}
where $E_{\rm 10 TeV} = E_e / (10\ \TeV)$ and $B_{200} = B / (200\ \muGauss)$ is the magnetic field strength in the starburst.  We see that CR $e^{\pm}$ of energy in the range $10 - 30\ \TeV$ will emit X-ray synchrotron emission in the 1 - 10 keV range.  

In this paper, we show that under the most optimistic assumptions, the synchrotron emission can explain the diffuse hard X-ray emission of starbursts, especially Arp 220.  The synchrotron spectrum ($\nu L_{\nu}$) rises at 1 GHz frequency and peaks near 10 - 100 GHz, \emph{but remains constant (or slowly falling) all the way to X-ray frequencies}.  There are three main reasons for this.  (1)  The synchrotron losses of GHz-emitting $e^{\pm}$ in starbursts are $\sim 1/5 - 1/20$ of the total losses including bremsstrahlung, ionization, and IC \citep{Thompson06,Murphy09,Lacki10a,Lacki11}.  Thus little of the energy in these CR $e^{\pm}$ is transformed into synchrotron emission.  However, while ionization and bremsstrahlung losses increase slowly with CR $e^{\pm}$ energy, synchrotron (and IC) losses become faster increasing with CR $e^{\pm}$ energy.  Therefore, bremsstrahlung and ionization become completely unimportant at $10\ \TeV$ (see the loss time scales in \S~\ref{sec:ElectronLosses}).  (2) Furthermore, IC cooling will be suppressed at energies above 10 TeV.  Most of the starlight generated by starbursts is efficiently absorbed by dust and re-emitted in the IR \citep[c.f.][]{Voelk89b}, and the majority of this IR is blackbody emission from $\sim 40 - 50\ \Kelv$ dust grains.  For FIR photons of energy $\epsilon_{\rm IR}$ and wavelength $\lambda \approx 100\ {\rm \mu m}$, $20\ \TeV$ is near the characteristic energy scale for Klein-Nishina suppression of IC cooling:
\begin{equation}
\label{eqn:EKN}
E_{\rm KN} \approx \frac{m_e^2 c^4}{\epsilon_{\rm IR}} \approx 21\ \TeV (\lambda / {\rm 100\ \mu m}).
\end{equation}
(3) A more subtle but important effect of the $\sim 20\ \TeV$ Klein-Nishina cutoff is that the threshold for pair production for $\gamma$-rays on the FIR emission is near this energy.  Thus the intense IR emission of starbursts converts the $\gamma$-ray emission at 10 - 100 TeV energies into 10 - 100 TeV $e^{\pm}$.  Since the power injected in pionic $\gamma$-rays is greater than the power in pionic secondary $e^{\pm}$, these tertiary pair $e^{\pm}$ can dominate the CR spectrum at 10 - 100 TeV if the pair production optical depth is greater than unity \citep[$\gamma\gamma$ attenuation and pair production at 10 - 100 TeV is expected to be small in galaxies like the Milky Way, see][]{Mastichiadis91,Moskalenko06,Stawarz10}, providing an additional population that can radiate X-ray synchrotron.

We begin in \S~\ref{sec:Motivation} by presenting order of magnitude estimates showing that synchrotron may be important.  Then, in later sections we construct one-zone models to evaluate these effects using standard assumptions for CR modelling and in the context of extant multiwavelength data.  In \S~\ref{sec:Procedure}, we describe our one-zone models of the CR spectra, including the pair produced $e^{\pm}$.  In \S~\ref{sec:Results}, we present the results of these models for the Galactic Center, NGC 253, M82, and the nuclear starbursts of Arp 220.  We discuss some broader implications in \S~\ref{sec:Discussion}, including the synchrotron contribution to X-rays from submillimeter galaxies (\S~\ref{sec:SMGs}) and further constraints on synchrotron X-rays from future neutrino and TeV experiments (\S~\ref{sec:Neutrinos}).  Finally, we conclude by discussing future work that can be done in \S~\ref{sec:Conclusion}.

\section{Motivation}
\label{sec:Motivation}

\subsection{Relevant Cooling Processes}
\subsubsection{$e^{\pm}$ losses}
\label{sec:ElectronLosses}
Starbursts contain dense gas and strong radiation fields, which can cool CR $e^{\pm}$ through bremsstrahlung, ionization, and IC emission.  These cooling processes have associated energy loss rates for individual CR $e^{\pm}$ $b_{\rm brems}$, $b_{\rm ion}$, $b_{\rm IC}$, respectively, which can be comparable to the synchrotron energy loss rate $b_{\rm synch}$.  In addition, CR $e^{\pm}$ can escape by advection or diffusion.  These processes compete with synchrotron emission in the magnetic fields of starbursts for the kinetic energy of a CR $e^{\pm}$.

The synchrotron cooling time of CR $e^{\pm}$ is $E / b_{\rm synch} (E) \approx 6 \pi m^2 c^4 / (c \sigma_T B^2 E)$ \citep{Rybicki79}, or 
\begin{equation}
\label{eqn:tSynch}
t_{\rm synch} (E) \approx 31\ \yr \left(\frac{E}{10\ \TeV}\right)^{-1} \left(\frac{B}{200\ \muGauss}\right)^{-2}.
\end{equation}
For energies greater than 1 TeV, and magnetic fields of at least $200\ \muGauss$, this is about the light crossing time for 100 pc, roughly the scale height for a nuclear starburst.  Therefore, at these energies, $e^{\pm}$ are in the ``calorimeter'' limit, in which the CR $e^{\pm}$ cool before escaping \citep[c.f.,][]{Voelk89b}.

For a gas density $n$, the ionization cooling time of CR $e^{\pm}$ is $E / b_{\rm ion} (E)$, or
\begin{equation}
\label{eqn:tIon}
t_{\rm ion} (E) \approx 1.9 \times 10^9\ \yr \left(\frac{E}{10\ \TeV}\right) \left(\frac{n}{250\ \cm^{-3}}\right)^{-1},
\end{equation}
from e.g., \citet{Schlickeiser02}, where we set the $\ln \gamma$ term to its value at $10\ \TeV$.  The bremsstrahlung cooling time is 
\begin{equation}
\label{eqn:tBrems}
t_{\rm brems} (E) \approx 1.2 \times 10^5\ \yr \left(\frac{n}{250\ \cm^{-3}}\right)^{-1},
\end{equation}
at high energies \citep[assuming $n_{\rm He} \approx 0.1 n_H$; see][]{Strong98}.  Synchrotron losses therefore dominate bremsstrahlung losses as long as $B \ga 10\ \muGauss\ (E / \TeV)^{-1/2} (n / 250\ \cm^{-3})^{1/2}$, and synchrotron losses dominates ionization losses so long as $B \ga 0.2\ \muGauss\ (E / \TeV)^{-1} (n / 250\ \cm^{-3})^{1/2}$.  However, bremsstrahlung and ionization can dominate at the $\sim \GeV$ energies where $e^{\pm}$ are responsible for GHz emission \citep{Thompson06,Murphy09,Lacki10a}. 

The Inverse Compton cooling time is more complex because Klein-Nishina effects appear at the relevant $\sim 10 - 100\ \TeV$ energies in the FIR-dominated radiation fields of starbursts.  \citet{Schlickeiser10} show that the IC loss time in a greybody radiation field with temperature $T$ and radiation energy density $U_{\rm rad}$ can be approximated as 
\begin{equation}
\label{eqn:tIC}
t_{\rm IC} \approx \frac{3 m_e c^2}{4 c \sigma_T U_{\rm rad}} \frac{\gamma_K^2 + \gamma^2}{\gamma \gamma_K^2},
\end{equation} 
where $\gamma_K \approx 0.27 m_e c^2 / (k_B T) = 4.0 \times 10^7 (T / 40\ \Kelv)^{-1}$.  Thus, above $\sim 20\ (T / 40\ \Kelv)^{-1}\ \TeV$, Klein-Nishina effects will cause the IC energy loss time to grow as $\gamma$, while the synchrotron loss time continues to fall as $\gamma^{-1}$.  If $U_B < U_{\rm rad}$, then synchrotron dominates IC when $\gamma \ge \gamma_K \sqrt{U_{\rm rad}/U_{B} - 1}$, or:
\begin{equation}
\label{eqn:ESynGTIC}
E \ga 20\ \TeV\ \left(\frac{T}{40\ \Kelv}\right)^{-1} \left(\frac{U_{\rm rad} - U_B}{U_B}\right)^{1/2}.
\end{equation}

The nearby, prototypical starbursts in M82 and NGC 253 have total infrared luminosities of $\sim 10^{10.5}\ \Lsun$ \citep{Melo02,Sanders03} and radii of $\sim 200\ \pc$ \citep[e.g.,][]{Turner83,Goetz90,Ulvestad97,Williams10}.  Plugging in these specific values for a disk geometry, the radiation field has an energy density of approximately $U \approx L / (2 \pi R^2 c)$, or
\begin{equation}
U_{\rm rad} \approx 1100\ \eV\ \cm^{-3} \left(\frac{L}{10^{10.5} \Lsun}\right) \left(\frac{R}{200\ \pc}\right)^{-2}.
\end{equation}
This gives us an IC loss time of 
\begin{equation}
\label{eqn:tICThomson}
t_{\rm IC}^{\rm Thomson} \approx 300\ \yr \left(\frac{L}{10^{10.5} \Lsun}\right)^{-1} \left(\frac{R}{200\ \pc}\right)^{2} \left(\frac{E}{\TeV}\right)^{-1}
\end{equation}
in the Thomson regime, and
\begin{equation}
t_{\rm IC}^{\rm KN} \approx 71\ \yr\ \left(\frac{L}{10^{10.5} \Lsun}\right)^{-1} \left(\frac{R}{200\ \pc}\right)^{2} \left(\frac{T}{40\ \Kelv}\right)^{2} \left(\frac{E}{100\ \TeV}\right)
\end{equation}
in the extreme Klein-Nishina limit.  For the specific values we have been using -- $L = 10^{10.5} \Lsun$, $R = 200\ \pc$, $T = 40\ \Kelv$ -- equation~\ref{eqn:ESynGTIC} shows that synchrotron dominates over IC above $\sim 5\ \TeV$ when $B = 200\ \muGauss$ and $\sim 35\ \TeV$ when $B = 100\ \muGauss$.

Therefore unless the magnetic field energy density is much lower than the radiation field energy density, synchrotron losses will be the dominant loss process at a few tens of TeV.  

\subsubsection{Proton losses}
\label{sec:ProtonLosses}
CR protons can modify the $e^{\pm}$ population by creating pionic secondary electrons and positrons through collisions with interstellar gas atoms.  The time for CR protons to lose all of their energy through this process is given as
\begin{equation}
\label{eqn:tPion}
t_{\rm pion} = 2 \times 10^5\ \yr \left(\frac{n}{250\ \cm^{-3}}\right)^{-1}
\end{equation}
by \citet{Mannheim94}.  Different sources in the literature give different pionic loss times that can be larger by a  factor $\sim 2 - 3$; however, these decrease slightly with proton energy as the pionic cross section increases (see the discussion in Appendix~\ref{sec:OtherPion}), so that at TeV energies the pionic loss time is closer to equation \ref{eqn:tPion}.

Since protons have much longer cooling times than TeV CR $e^{\pm}$, it is not as obvious whether they can escape.  Advection is clearly present in starbursts in the form of the observed large-scale winds \citep[e.g.,][]{Heckman00,Heckman03,Heesen11}.  The wind crossing time for a starburst of scale height $h$ is $h / v$, or:
\begin{equation}
\label{eqn:tWind}
t_{\rm wind} = 3 \times 10^5\ \yr \left(\frac{h}{100\ \pc}\right) \left(\frac{v}{300\ \kms}\right)^{-1}
\end{equation} 
In starbursts such as M82 and NGC 253, the advective and pionic lifetimes are expected to be roughly equal.  The nuclear starbursts of Arp 220, with $\mean{n} \approx 10^4 \cm^{-3}$ \citep[e.g.,][]{Downes98}, have very short pionic loss times ($\sim 5000\ \yr$), indicating that they are ``proton calorimeters'': most of the power injected into CR protons with energies above the pion-production threshold is lost through pionic interactions.

However, a key uncertainty is whether CRs sample gas of the average density.  The $\gamma$-ray luminosity of the Galactic Center region ($R \le 112\ \pc$) relative to its star-formation rate indicates this is not the case for that region, as does the ratio of synchrotron radio from CR $e^{\pm}$ to infrared emission \citep{Crocker10b,Crocker11}.  \citet{Crocker11} explains these observations as being caused by a powerful wind in the Galactic Center region advecting CRs out of the disk before they can enter the molecular clouds containing most of the gas.  Thus CRs experience gas of much lower density than average.  While the Galactic Center is underluminous in $\gamma$-rays and radio, this is not true for the starbursts M82 and NGC 253, which fall on the FIR-radio correlation and have a larger $\gamma$-ray to star-formation ratio than the Milky Way, consistent with CRs experiencing average gas densities in these starbursts \citep{Lacki11}.

A final uncertainty is whether diffusive escape plays any role.  Diffusive escape in the Milky Way has an energy dependence $t_{\rm diff} \propto E^{-0.3} - E^{-0.6}$ that steepens the CR proton spectrum, since it is the dominant timescale \citep[e.g.,][]{Ginzburg76}. In contrast, the relatively flat GeV-TeV spectra of observed starburst regions (M82, NGC 253, and the Galactic Center) indicate that up to TeV energies, an energy-independent process must determine the lifetimes of CR protons; however, at the still higher energies we are considering, diffusion can dominate.  We consider several values of the diffusive escape time to address this uncertainty.

\subsection{Primary Electrons}
Primary electrons dominate the GHz-emitting electron population in normal galaxies.  However, they are generally expected to be sub-dominant in starbursts with respect to pionic secondary $e^{\pm}$, though still a significant minority \citep{Rengarajan05,Thompson07,Lacki10a}.  Detailed modelling of CR populations in starburst galaxies find that primary electrons are subdominant at GeV energies in M82 \citep{deCeaDelPozo09a}, NGC 253 \citep{Domingo05,Rephaeli09}, and Arp 220 \citep{Torres04}, although models by \citet{Crocker11} find primary electrons dominate in the Galactic Center.  At higher energies, where the protons experience stronger diffusive losses, these models indicate higher primary fractions.  

It is not known where the primary electron injection spectrum ends in starbursts.  As the electrons are accelerated to higher energies, they also experience more severe synchrotron and Inverse Compton cooling.  At some point, the cooling losses balance the rate of acceleration, and there can be no further acceleration of primary electrons.  In the standard supernova acceleration theory, equilibrium between cooling and acceleration occurs when:
\begin{equation}
E_e \approx 27\ \TeV\ v_{5000} B_{{\rm SNR}, 200}^{-1/2},
\end{equation}
where $5000 v_{5000}\ \kms$ is the speed of the supernova shock and $200 B_{{\rm SNR}, 200}\ \muGauss$ is the supernova remnant magnetic field \citep{Gaisser90}.  From equation~\ref{eqn:ECSynch}, the end of the primary synchrotron spectrum will then be at:
\begin{equation}
E_{\rm max}^{\rm synch} \approx 7.3\ \keV\ v_{5000}^2\ \left(\frac{B}{B_{\rm SNR}}\right),
\end{equation}
if the magnetic field strength in the starburst $B$ is similar to the magnetic field strength in the supernova remnant.  X-ray observations of supernova remnants in the Milky Way have revealed synchrotron emission from 10 - 100 TeV electrons, confirming these energies are reached in supernova remnants \citep[e.g.,][]{Koyama95,Allen97,Reynolds99,Reynolds08a,Reynolds08b}.  Thus, SNRs might be able to accelerate primary electrons to the energies where they will produce hard synchrotron X-rays in the diffuse ISM of starburst galaxies.

However, it is not clear that SNRs are responsible for all of the primary CR electrons; other objects such as superbubbles or pulsars may contribute \citep[e.g.,][]{Butt09}.  Pulsar Wind Nebulae (PWNe) powered by pulsar spindown after their birth supernova in particular may inject a very hard component of $e^{\pm}$ dominating at TeV energies, possibly extending to 10 - 100 TeV \citep{Yuksel09,Bamba10}.  Primary CR $e^{\pm}$ from PWNe have been invoked to explain anomalies in the CR electron spectrum observed at Earth.  The birthrate of pulsars is correlated with star-formation, and pulsars should be present in large numbers in starbursts \citep[c.f.][]{Perna04,Mannheim10}.  Their spin-down luminosities could provide enough power to be comparable to the main component of CRs \citep{Lacki11}.  

If a hard primary CR $e^{\pm}$ spectrum extends to 10 - 100 TeV energies or beyond, then the synchrotron X-ray luminosity may be very bright.  Suppose that for every supernova, roughly ${\cal E}_{\rm CR}^e = E_{48} 10^{48}$ erg of CR electrons with energies above 1 GeV are accelerated.  This could occur if 0.1\% of a supernova remnant's kinetic energy ($10^{51}\ \erg$) were converted into primary CR electrons, which is about 1\% of the total energy expected to go into CRs \citep[the exact ratio depends on the poorly constrained low energy CR $e^{\pm}$ spectrum]{Strong10}.  If we restrict our attention to CRs with an energy between 1 GeV and 1 PeV, then for an $E^{-2}$ injection spectrum extending between these energies $E^2 dQ/dE = {\cal E}_{\rm CR}^e / \ln(10^6)$.  At very high energies, a fraction $f_{\rm synch}$ will go into synchrotron, where $f_{\rm synch} = t_{\rm life} / t_{\rm synch} \approx 1 / (1 + t_{\rm synch} / t_{\rm IC})$:
\begin{equation}
\label{eqn:fSynch}
f_{\rm synch} \approx \left[1 + \frac{U_{\rm rad} / U_{\rm B}}{1 + (\gamma/\gamma_K)^2}\right]^{-1}
\end{equation}
(eqs.~\ref{eqn:tSynch} and \ref{eqn:tIC}), which we have argued to be near $1$.  Since the characteristic synchrotron emission energy $E_C$ is proportional to $E_e^2$, a power law electron spectrum spanning a large number of dex in energy will give rise to a synchrotron spectrum spanning roughly twice as many dex in frequency.  We finally have $\nu L_{\nu} (\keV) \approx (1/2) f_{\rm synch} E_e^2 dQ_e/dE_e$ \citep{Loeb06}, or:
\begin{equation}
\label{eqn:LXPrimary}
\nu L_{\nu} (\keV) \approx 1.1 \times 10^{39} E_{48} f_{\rm synch} \left(\frac{\Gamma_{\rm SN}}{\yr^{-1}}\right) \ergps,
\end{equation}
where $\Gamma_{\rm SN}$ is the supernova rate.  For example, M82 is believed to have a supernova rate of roughly $\sim 0.1\ \yr^{-1}$ (although with large uncertainties), suggesting that its synchrotron X-ray luminosity from pionic $e^{\pm}$ may be $1 \times 10^{38} \ergps$, about 3\% of the observed luminosity of the diffuse X-ray emission ($L\ (2 - 8\ \keV) = (4.4 \pm 0.2) \times 10^{39}\ \ergps$, or $\nu L_{\nu} \approx 3.2 \times 10^{39}\ \ergps$ per bin in ln energy; \citealt{Strickland07}).  Softer CR $e^{\pm}$ injection spectra will have still lower synchrotron X-ray luminosities.

If SNRs or other discrete sources do accelerate $\sim 10\ \TeV$ primary electrons, the synchrotron X-ray emission may not be spread continuously throughout the starburst, but concentrated near the CR sources, because the synchrotron cooling time for these electrons is so short (eq.~\ref{eqn:tSynch}).  Even if such electrons free-stream, they will not travel farther than $\sim 10\ \pc$ from their sources.  If there are a small number of CR $e^{\pm}$ accelerators, then the synchrotron X-ray emission should come from a few small diffuse regions, just as patchy ``cells'' of high energy $e^{\pm}$ are predicted for the Milky Way \citep[c.f.,][]{Shen70,Aharonian95,Atoyan95}.  If starbursts, with a high star-formation rate concentrated into a small volume, instead contain many accelerators, the primary $e^{\pm}$ confinement regions around these accelerators will overlap and the X-ray emission will arise throughout the starburst.  Unlike pionic secondary $e^{\pm}$ or pair production $e^{\pm}$ from pionic $\gamma$-rays, both of which depend on the CR proton spectrum, primary $e^{\pm}$ at these energies are unaffected by the escape because cooling is so quick.  

\subsection{Pionic Secondaries}
CR protons can inelastically scatter off protons in the ISM to produce pions, which decay into secondary $e^{\pm}$, $\gamma$-rays, and neutrinos.  From the lifetime calculations given in \S~\ref{sec:ProtonLosses}, starbursts are expected to convert much more of their CR proton energy into pionic products than the Milky Way and approach the ``proton calorimeter'' limit \citep{Loeb06,Thompson07,Lacki10a}.  From the HESS detection, \citet{Acero09} claimed that 5\% of the proton energy was converted into pionic products in NGC 253, although they assumed a hard $\gamma$-ray spectrum.  From the \emph{Fermi}, HESS, and VERITAS detections, \citet{Lacki11} inferred a proton calorimetry fraction of about 1/3 for NGC 253 and M82 from the ratio of the $\ge \GeV$ $\gamma$-ray and IR luminosities.  The secondary $e^{\pm}$ are expected to dominate in starbursts at GeV energies based on physical considerations \citep{Rengarajan05,Loeb06,Thompson07,Lacki10a} and detailed models of M82, NGC 253, and Arp 220 \citep{Torres04,Domingo05,deCeaDelPozo09a,Rephaeli09}.  Primaries probably dominate in the Galactic Center due to a fast wind advecting CR protons from the region before they can interact with the molecular cloud gas \citep{Crocker11}.

In the Milky Way, primary CR protons (and nuclei) are accelerated with a power law spectrum extending to energies of at least $\sim \PeV$, the so-called ``knee'' in the CR spectrum.  Above these energies, the CR spectrum steepens: this may be because Galactic CRs are not accelerated to higher energies and we are seeing the transition to a different component of CRs, or because Galactic CRs propagate differently above the knee.  Essentially nothing is known about the knee in starbursts, but presumably it is also at least at a PeV in energy.

If most of the energy injected into CR protons is lost to pion production for CR protons of up to a PeV in energy, then the large population of secondary pionic 10 - 100 TeV $e^{\pm}$ produced by pion decay can emit bright hard X-ray emission.  Suppose that roughly ${\cal E}_{\rm CR} = 10^{50}$ erg of CR protons are accelerated per supernova (that is, roughly $1/10$ of the SN kinetic power goes into CRs; \citealt{Strong10}).  Once again suppose that we have $E^2 dQ/dE = {\cal E}_{\rm CR} / \ln(10^6)$ for an $E^{-2}$ injection spectrum for CR protons with kinetic energies from 1 GeV to 1 PeV, ignoring CRs of lower energy where the spectrum is uncertain.  A fraction $F_{\rm cal} (E_p)$ of that power is lost to pions for a CR proton energy $E_p$; of that, $\sim 1/6$ will go into secondary $e^{\pm}$ \citep[e.g.,][]{Steigman71,Loeb06}\footnote{This is because the charged pions receive 2/3 of the energy, and each charged pion ultimately decays into a positron or electron and three neutrinos of roughly equal energy.  Therefore, $e^{\pm}$ recieve $1/4 \times 2/3 = 1/6$ of the pion energy.  Similarly, neutrinos receive 1/2 of the pion energy, and $\gamma$-rays recieve the remaining 1/3, from the neutral pions.} and $f_{\rm synch}$ of the secondary $e^{\pm}$ power in turn goes into synchrotron emission.  From a similar argument as the primaries, we finally have $\nu L_{\nu} (\keV) = (1/2) f_{\rm synch} E_e^2 dQ_e/dE_e = (F_{\rm cal} / 12) f_{\rm synch} E_p^2 dQ_p/dE_p$ \citep{Loeb06}, or:
\begin{equation}
\label{eqn:LXSecondary}
\nu L_{\nu} (\keV) \approx 1.9 \times 10^{40} F_{\rm cal}(100\ \TeV) f_{\rm synch} \left(\frac{\Gamma_{\rm SN}}{\yr^{-1}}\right) \ergps,
\end{equation}
where $\Gamma_{\rm SN}$ is the supernova rate.  For a supernova rate in M82 of $0.1\ \yr^{-1}$, we get $2 \times 10^{39} \ergps$ for $f_{\rm synch} = F_{\rm cal} = 1$, more than half the observed luminosity of the diffuse X-ray emission \citep{Strickland07}.  In practice, energy-dependent diffusive escape of the primary protons, advective escape in the starburst superwind, and softer CR proton injection spectra will reduce the secondary $e^{\pm}$ luminosity.

In contrast with primary electrons, pionic $e^{\pm}$ will be generated throughout the starburst, instead of being concentrated near the source of CR protons, because CR protons can travel a longer distance during their lifetime.  Therefore the synchrotron X-ray emission secondary $e^{\pm}$ will not have as patchy a structure as the primaries, but they will be generated wherever there is gas being sampled by CRs.  

\subsection{Pair Production Tertiaries}
Starburst galaxies are generally predicted to become opaque to $\gamma$-rays above a few TeV, because $\gamma$-rays will pair produce $e^{\pm}$ with the IR light in the starbursts \citep{Torres04,Domingo05,Inoue11}.  Thus, the pionic $\gamma$-ray emission at 10 - 100 TeV will be efficiently converted into 10 - 100 TeV $e^{\pm}$.  What is less appreciated is that these $e^{\pm}$ will lose much of their energy to synchrotron cooling, emitting mostly X-rays.  Thus, if multi-TeV CR protons in starbursts have efficient pionic losses, then up to half of the CR proton energy at these energies will go into synchrotron X-rays, with a similar fraction escaping as neutrinos.

The optical depth of a starburst to multi-TeV photons is $\tau_{\gamma\gamma} \approx h n_{\rm IR} \sigma_{\gamma\gamma}$, where $h$ is the height of the starburst disk, $\sigma_{\gamma\gamma}$ is the cross section of pair-production, and $n_{\rm IR}$ is the number density of target IR photons.  Near threshold (where $E_{\gamma} \approx E_{\rm KN}$ from eq.~\ref{eqn:EKN}), $\sigma_{\gamma\gamma} \approx \sigma_T / 4$.  The average energy of IR photons in a greybody field is $\pi^4 \zeta(3) kT / 30 \approx 2.70 kT$, where $\zeta$ is the Riemann zeta function.  The number density of IR photons for a starburst disk of radius $R$ (and emitting area $2 \pi R^2$) is approximately $n_{\rm IR} \approx L_{\rm IR} / (2 \pi R^2 c \times 2.70 k_B T)$, where $T$ is the typical temperature of the IR photon.  We have
\begin{equation}
\tau_{\gamma\gamma} \approx 5.8 \left(\frac{L_{\rm FIR}}{10^{10.5}\ \Lsun}\right) \left(\frac{h}{100\ \pc}\right) \left(\frac{R}{200\ \pc}\right)^{-2} \left(\frac{T}{40\ \Kelv}\right)^{-1},
\end{equation}
demonstrating that luminous starbursts are opaque to $\sim 30\ \TeV$ photons, turning them into pair $e^{\pm}$.  In practice, since about half of the total infrared radiation is FIR \citep{Calzetti00}, the expected $\tau_{\gamma\gamma}$ will be lower by a factor of a few, but still of order unity.  This is in contrast to the Milky Way, which is essentially transparent to $\gamma$-rays \citep{Mastichiadis91,Moskalenko06,Stawarz10}.

Pionic $\gamma$-rays are expected to dominate the VHE $\gamma$-ray luminosity of starbursts, so the calculation of the synchrotron power from tertiary pair $e^{\pm}$ is similar to that for secondary pionic $e^{\pm}$.  CR protons roughly inject 2 times more energy in pionic $\gamma$-rays than in pionic secondary $e^{\pm}$, so the synchrotron power should likewise be twice as great for pair $e^{\pm}$ than direct secondaries (eq.~\ref{eqn:LXSecondary}):
\begin{equation}
\nu L_{\nu} (\keV) \approx 3.8 \times 10^{40} F_{\rm cal}(100\ \TeV) f_{\rm synch} \left(\frac{\Gamma_{\rm SN}}{\yr^{-1}}\right) \ergps,
\end{equation}
for an $E^{-2}$ injection spectrum between kinetic energies of 1 GeV and 1 PeV with $10^{50}\ \erg$ per supernova in CR protons.  Again comparing to M82 with a supernova rate of $\Gamma_{\rm SN} \approx 0.1\ \yr^{-1}$, we find that $\nu L_{\nu}$ may be as high as $4 \times 10^{39}\ \ergps$ (with $F_{\rm cal} = f_{\rm synch} = 1$), equal to the observed diffuse X-ray emission.  As with secondary pionic $e^{\pm}$, diffusive and advective losses (lower $F_{\rm cal}$) and softer injection spectra reduce the predicted synchrotron X-ray luminosity.  On the other hand, non-pionic $\gamma$-rays, such as from discrete sources like pulsars, will also produce pairs in the starburst radiation field, and this can enhance the electron population and synchrotron X-ray luminosity further.

Like secondary $e^{\pm}$, tertiary $e^{\pm}$ will not be concentrated near CR accelerators, since the pionic $\gamma$-rays that generate them are emitted everywhere in the starburst region.  However, the pair $e^{\pm}$ production may be enhanced near luminous IR sources within the starbursts.  A treatment of this effect requires a radiative transfer calculation, which is beyond the scope of this work.  

\section{Modelling Assumptions}
\label{sec:Procedure}
To understand the synchrotron and Inverse Compton X-ray emission of starburst galaxies, we create models of the steady-state CR populations of the Galactic Center, NGC 253, M82, and the nuclei of Arp 220.  Our goal here is to sketch out the parameter space allowed by multiwavelength data using a few standard assumptions, and investigate the synchrotron and IC emission that arises under these assumptions.  We model the starbursts as one-zone disks of radius $R$ and midplane-to-edge scale heights $h$.  The evolution of CR population is governed by the diffusion-loss equation \citep[e.g.,][]{Ginzburg76,Strong07}.  In steady-state one-zone models with no spatial or temporal dependence, the diffusion-loss equation reduces to the leaky box equation:
\begin{equation}
\frac{N(E)}{t_{\rm life} (E)} - \frac{d}{dE}[b(E)N(E)] - Q(E) = 0.
\end{equation}
Here, $Q(E)$ is the injection spectrum of CRs, $b(E)$ is the cooling rate of CRs (including ionization, bremsstrahlung, IC, and synchrotron), and $t_{\rm life} (E)$ is the lifetime of CRs from escape (both diffusive and advective) and pionic losses for protons.  The CR lifetime $t_{\rm life} (E)$ is calculated as $[t_{\rm life} (E)]^{-1} = [t_{\rm diff}(E)]^{-1} + t_{\rm adv}^{-1} + [t_{\pi} (E)]^{-1}$, where $t_{\rm diff} (E)$ is the diffusive escape time, $t_{\rm adv}$ is the advective escape time, and $t_{\pi} (E)$ is the pionic loss time for protons.  We use the numerical code described in \citet{Lacki10a} to find the steady-state CR spectra, employing a Green's function given in \citet{Torres04}. See Table~\ref{table:ModelParameters} for the parameters we used for each starburst.  

\begin{deluxetable*}{llccccc}
\tabletypesize{\scriptsize}
\tablecaption{Model Parameters}
\tablehead{\colhead{Parameter} & \colhead{Units} & \colhead{Galactic Center} & \colhead{NGC 253 Core} & \colhead{M82} & \colhead{Arp 220 West} & \colhead{Arp 220 East}}
\cutinhead{Assumed parameters}
$B$                           & $\muGauss$ & 50 - 100             & 50 - 400             & 50 - 400             & 250 - 16000        & 250 - 16000\\
$\Sigma_g$                    & $\gcm2$    & 0.003 - 0.1          & 0.10                 & 0.17                 & 10                 & 10\\
$h$                           & $\pc$      & 42                   & 50                   & 100                  & 50                 & 50\\
$R$                           & $\pc$      & 112                  & 150                  & 250                  & 50                 & 50\\
$\ell_{\oplus}$\tablenotemark{a} & $\pc$   & 112                  & 150                  & 250                  & 50                 & 50\\
$D$                           & $\Mpc$     & 0.008                & 3.5                  & 3.6                  & 79.9               & 79.9\\
$v_{\rm wind}$                & $\kms$     & 600                  & 300                  & 300                  & 300                & 300\\
$L_{\rm TIR}$                 & $\Lsun$    & $4 \times 10^8$      & $2 \times 10^{10}$   & $5.9 \times 10^{10}$ & $3 \times 10^{11}$ & $3 \times 10^{11}$\\
$L_X$ (diffuse)\tablenotemark{b} & $\ergps$   & $7.4 \times 10^{36}$\tablenotemark{c} & $8.5 \times 10^{38}$\tablenotemark{d} & $4.4 \times 10^{39}$\tablenotemark{e} & $4 \times 10^{40}$\tablenotemark{f} & $1.5 \times 10^{40}$\tablenotemark{f}\\
\cutinhead{Fiducial Parameters}
$B$                           & $\muGauss$ & 100    & 100    & 150    & 4000    & 4000\\
$\Sigma_g$                    & $\gcm2$    & 0.003  & 0.10   & 0.17   & 10      & 10\\
$p$                           & \nodata    & 2.2    & 2.2    & 2.2    & 2.2     & 2.2\\
$t_{\rm diff} (3\ \GeV)$      & $\Myr$     & 1      & 1      & 10     & 10      & 10\\
$\gamma_{\rm max}^{\rm prim}$ & \nodata    & $10^9$ & $10^9$ & $10^9$ & $10^9$  & $10^9$\\
$\eta$                        & \nodata    & 0.14   & 0.40   & 0.10   & 0.10    & 0.10\\
$\xi$                         & \nodata    & 0.0066 & 0.027  & 0.0077 & 0.021   & 0.0083\\
$S_{\rm therm} (1.0\ \GHz)$   & Jy         & 370    & 0.63   & 1.3    & \nodata & \nodata\\
\cutinhead{Fiducial Results}
$L_X$ (synch)\tablenotemark{g}    & $\ergps$ & $4.3 \times 10^{35}$ & $7.2 \times 10^{37}$ & $9.6 \times 10^{37}$ & $6.0 \times 10^{39}$ & $5.2 \times 10^{39}$\\
$\Gamma$ (synch)\tablenotemark{g} & \nodata  & 2.12   & 2.08 & 2.15 & 2.12 & 2.12\\
$L_X$ (IC)\tablenotemark{g}       & $\ergps$ & $9.0 \times 10^{33}$ & $2.7 \times 10^{37}$ & $4.1 \times 10^{37}$ & $2.6 \times 10^{39}$ & $1.9 \times 10^{39}$\\
$\Gamma$ (IC)\tablenotemark{g}    & \nodata  & 1.58   & 1.35 & 1.31 & 1.44 & 1.44\\
$\tilde\delta$                    & \nodata  & 84     & 59   & 51   & 18   & 48\\
$F_{\rm cal}$                     & \nodata  & 0.0037 & 0.15 & 0.33 & 0.97 & 0.97\\
$F_{\rm cal} (\ge 10\ \TeV)$      & \nodata  & $6 \times 10^{-4}$ & 0.016 & 0.11 & 0.93 & 0.93
\enddata
\label{table:ModelParameters}
\tablenotetext{a}{Adopted sightline distance through starburst to Earth, for computing observed TeV $\gamma$-ray spectrum.}
\tablenotetext{b}{Adopted diffuse hard X-ray emission, with which synchrotron and IC are compared.}
\tablenotetext{c}{Galactic Center luminosity in 2 - 10 keV X-ray band from \citet{Koyama96}, as extrapolated to $|\ell| \le 0.8^{\circ}$ and $|b| \le 0.3^{\circ}$ assuming a constant surface brightness.}
\tablenotetext{d}{NGC 253 disk diffuse 2 - 10 keV X-ray luminosity from \citet{Bauer08}.  Note that this includes the outlying regions of the galaxy and not just the starburst core.}
\tablenotetext{e}{Luminosity of diffuse hard X-ray excess in 2 - 8 keV band from \citet{Strickland07}.  The uncertainty is $0.2 \times 10^{39}\ \ergps$.}
\tablenotetext{f}{Absorption-corrected Arp 220 X-ray luminosities from \citet{Clements02}.  We assume that Arp 220 X-1 is the western nucleus and Arp 220 X-4 is the eastern nucleus.}
\tablenotetext{g}{Synchrotron and IC luminosities and photon indexes, for the hard X-ray energy bands given for $L_X$ (diffuse) (2 - 8 keV for M82, 2 - 10 keV for the other starbursts).}
\end{deluxetable*}

\subsection{Injection}
We calculate the star-formation rate by directly converting the TIR (total infrared; $8 - 1000\ \micron$) luminosity of the starburst disk:
\begin{equation}
\label{eqn:TIRtoSFR}
{\rm SFR} = L_{\rm TIR} / (\varepsilon c^2),
\end{equation}
where $\varepsilon = 3.8 \times 10^{-4}$ is a dimensionless factor relating the luminosity to the instantaneous star formation rate, and is IMF dependent \citep{Kennicutt98}.  The factor $\varepsilon$ is derived assuming a starburst that is continuous over 10 - 100 Myr and a Salpeter IMF \citep{Kennicutt98}.  The CR energy injection rate per unit volume is assumed to be proportional to star-formation rate:
\begin{eqnarray}
\epsilon_{\rm CR,e} & = & 9.2 \times 10^{-6} E_{51} \psi_{17} L_{\rm TIR} (\xi / 0.001) / V\\
\epsilon_{\rm CR,p} & = & 9.2 \times 10^{-4} E_{51} \psi_{17} L_{\rm TIR} (\eta / 0.1) / V\\
                    & = & \epsilon_{\rm CR,e} \delta\\                    
\end{eqnarray}
for electrons and protons respectively, where $\xi$ is the electron acceleration efficiency, $\delta$ is the ratio of proton accleration efficiency $\eta$ to $\xi$, $\psi_{17} = (\beta_{\rm SN} / \varepsilon)/(17\ \Msun^{-1})$, $V = 2 \pi R^2 h$ is the starburst volume, and $\beta_{\rm SN}$ is the SN rate per unit star formation.  The total SN rate in the starbursts is $\Gamma_{\rm SN} = 17 L_{\rm TIR} / (\Msun c^2)$:
\begin{equation}
\Gamma_{\rm SN} = 0.036\ \yr^{-1} \psi_{17} \left(\frac{L_{\rm TIR}}{10^{10.5}\ \Lsun}\right).
\end{equation}
We assume $\psi_{17} = 1$ throughout this work.  The efficiency of primary CR electron acceleration is described by the $\xi$ parameter, and the ratio of energy going into CR protons and CR electrons is $\delta$.  

We assume that CR protons and electrons are respectively injected with a momentum power law spectrum
\begin{eqnarray}
dQ_p/dq & = & C_p q^{-p}\\
dQ_e/dq & = & C_e q^{-p}
\end{eqnarray}
per unit volume, where $q$ is the CR proton or electron momentum.  A minimum kinetic energy cutoff of $K_{\rm min} = 1\ \MeV$ was used with these spectra.  The CR proton spectrum is assumed to extend to $\gamma_{\rm max}^p = 10^6$, corresponding to an energy of 938 TeV ($\gamma_{\rm max}^p = 10^5$ and $10^7$ are considered in Appendix~\ref{sec:OtherGammaP}).  We try different cutoffs in the Lorentz factor $\gamma_{\rm max}^{\rm prim}$ of the primary electron spectrum, both $10^6$ and $10^9$.  We note that we are assuming the ``test particle'' approach to CR acceleration here.  Nonlinear effects can result in more complicated spectra, with breaks in the power law at low energy and spectral hardening at high energy \citep[e.g.,][]{Berezhko99,Ellison00,Malkov01,Blasi05}.  We also neglect additional, harder components of $e^{\pm}$ from different CR accelerators.  In general, a hardening of the injected electron spectrum at high energy will result in more synchrotron X-ray emission and a softening of the injected electron spectrum will result in less synchrotron X-ray emission.

The normalizations of the CR injection spectra are then set by calculating the integral of the kinetic energy injected and equating with $\epsilon_{\rm CR}$:
\begin{eqnarray}
\epsilon_{\rm CR, p} & = & C_p \int_{q_{\rm min, p}}^{q_{\rm max, p}} q^{-p} \left(\sqrt{q^2 c^2 + m_p^2 c^4} - m_p c^2\right) dq\\
\epsilon_{\rm CR, e} & = & C_e \int_{q_{\rm min, e}}^{q_{\rm max, e}} q^{-p} \left(\sqrt{q^2 c^2 + m_e^2 c^4} - m_e c^2\right) dq
\end{eqnarray}
where $q_{\rm min} = (1 / c) \sqrt{K_{\rm min}^2 + 2 K_{\rm min} m c^2}$ and $q_{\rm max} = m c \sqrt{(\gamma_{\rm max}^{\rm prim})^2 - 1}$.  The ratio of energy in CR electrons to CR protons injected at high energies can be approximated as:
\begin{equation}
\label{eqn:deltaTilde}
\tilde\delta = \frac{C_p}{C_e} \approx \left(\frac{m_p c^2}{K_{\rm min}}\right)^{p - 2} \delta.
\end{equation}
In the Milky Way, $\tilde\delta \approx 50 - 100$, which is expected from charge conservation in the acceleration region for a $p \approx 2.2$ momentum power law injection spectrum \citep{Schlickeiser02}.  This value of proton/electron injection ratio is also inferred from propagation studies that compare with observations of CRs in the Milky Way \citep[e.g.,][]{Strong10}, and from the CR pressure derived from shock structure in Tycho's supernova remnant \citep{Warren05}.  It is also the approximate ratio of the CR proton and electron energy densities at Earth \citep[e.g.,][]{Ginzburg76}.

\subsection{Propagation}
In solving the leaky box equation for CR $e^{\pm}$, we consider energy losses from ionization, bremsstrahlung, synchrotron, and IC.  CR protons experience continuous ionization losses, which are important at low energies where pionic losses are negligible.  All CRs can also escape, either through energy-dependent diffusion, or energy-independent advection (winds).  We assume an $E^{-1/2}$ energy dependence to the diffusive escape times; the exact energy dependence is not known, although our assumption is conservative in that CR protons will tend to easily escape at higher energy without producing pionic $e^{\pm}$ and $\gamma$-rays.  The normalization of the diffusive escape time in starbursts is also not known, so we consider a variety of normalizations:
\begin{equation}
\label{eqn:tDiff}
t_{\rm diff} = t_{\rm diff} (3\ \GeV) \left(\frac{E}{3\ \GeV}\right)^{-1/2},
\end{equation}
with $t_{\rm diff} (3\ \GeV)$ ranging from $1\ \Myr$ to $\infty$ (in the Milky Way, $t_{\rm diff} (3\ \GeV) \approx 30\ \Myr$; \citealt{Connell98,Webber03}).  

The power law dependence of the diffusion escape time is not precisely known, even for the Milky Way.  By interpreting local CR nuclei data with CR propagation models, \citet{DiBernardo10} showed that the diffusion escape time in the Milky Way has a $E^{-0.3}$ to $E^{-0.6}$ dependence.  By comparing the diffusion lengths of CR electrons and protons of different energies, as traced by radio and $\gamma$-rays, \citet{Murphy12} found that the diffusion constant has a $\sim E^{-0.75}$ form in the super star cluster 30 Doradus, but with a smaller value at GeV energies than in the Milky Way disk.  Meanwhile, \citet{Abramowski12} could only place small upper limits on the diffusion constant for 30 TeV protons in the starburst NGC 253 from the lack of an observed spectal break.  Different energy dependences would alter the CR proton population at TeV energies, and we naturally expect more secondary and pair $e^{\pm}$ (and their accompanying synchrotron X-rays) if there is a weaker energy dependence in $t_{\rm diff}$ than in equation~\ref{eqn:tDiff}.  However, the TeV $\gamma$-ray luminosity will also be different, so in starbursts with a TeV $\gamma$-ray luminosity constraint, the allowed parameter space would change.  This would weaken the effect of different energy dependence in $t_{\rm diff}$, since there is a shorter ``lever arm'' between the population traced by TeV $\gamma$-rays ($\sim 10 - 100$ TeV protons for 1 - 10 TeV $\gamma$-rays) and synchrotron X-rays ($\sim 100 - 1000$ TeV protons).  For $E^{-0.3}$ energy dependence (the weakest usually expected), the diffusion time is greater than 300 kyr at PeV energies as long as $t_{\rm diff} (3\ \GeV) \ga 10\ \Myr$, meaning that advection (and any pionic losses) will be more important than diffusive escape (see equation~\ref{eqn:tWind}).  Thus, relatively slow diffusive escape with a $E^{-0.3}$ energy dependence is equivalent to no diffusive escape at all ($t_{\rm diff} (3\ \GeV) \to \infty$).  In addition, in models where primary electrons dominate the synchrotron X-ray flux (the Galactic Center and low $B$ models of other starbursts), diffusive escape will have little effect, since the synchrotron cooling time is so short at these energies (equation~\ref{eqn:tSynch}). 

Advective escape times are more well known; a wind with a speed of a few hundred kilometers per second will carry a CR out of the starburst in a few hundred kyr (eq~\ref{eqn:tWind}).  Our models of NGC 253, M82, and Arp 220 assume that $v_{\rm wind} = 300\ \kms$.  We use a wind speed of $600\ \kms$ for the Galactic Center models, as \citet{Crocker11} infers for the Galactic Center region.  

To calculate the effects of pionic losses on the CR proton spectrum, we directly integrate all of the energy going into pionic secondary products:
\begin{equation}
\label{eqn:tPiCS}
t_{\pi} = K_p \left[n_H \beta_{\rm CR} c \sum_{e^{\pm}, \gamma,\nu} \int_0^{E_p} E_{\rm sec} \frac{d \sigma (E_p, E_{\rm sec})}{dE_{\rm sec}} dE_{\rm sec}\right]^{-1}
\end{equation}
The spectra of the pionic secondary $e^{\pm}$, $\gamma$-rays, and neutrinos are calculated using the \citet{Kamae06} cross sections for proton energies below 500 TeV and the \citet{Kelner06} cross sections for proton energies above 500 TeV.  This method has the advantage of being consistent between the differential cross sections and the pionic lifetime.   We also consider other cross section parametrizations and pionic lifetimes in Appendix~\ref{sec:OtherPion}; we generally found that these produced similar results.  

Calculating the injection rate of pair-produced $e^{\pm}$ requires both the low-energy target photon and $\gamma$-ray spectra within the starburst.  Assuming a planar geometry, with $\gamma$-rays traversing vertically out of the starburst disk\footnote{Note that $\gamma$-rays do not always travel straight out of the disk plane, but also at horizontal angles through the disk.  Thus we underestimate the mean $\tau$ by a geometrical factor of order unity.}, the $\gamma$-ray number density is
\begin{equation}
\label{eqn:NGammaOneZone}
N_{\gamma}(E_{\gamma}) = \frac{Q_{\gamma}(E_{\gamma}) h}{c \tau_{\gamma\gamma} (E_{\gamma})} [1 - \exp(-\tau_{\gamma\gamma}(E_{\gamma}))],
\end{equation}
where $Q_{\gamma}$ is the injection rate of $\gamma$-rays per unit volume in photons per unit energy per unit volume.  For a photon number density spectrum $n (\epsilon)$, we calculate the pair production opacity $\tau_{\gamma\gamma}$ using
\begin{equation}
\tau_{\gamma\gamma} (E_{\gamma}) =  \int h n (\epsilon) \sigma_{\gamma\gamma}(\epsilon, E_{\gamma}) d\epsilon
\end{equation}
where we use the approximation in \citet{Aharonian04b} (equation 3.23) for the differential pair-production cross section $\sigma_{\gamma\gamma}(\epsilon, E_{\gamma})$ given in full in \citet{Gould67}, and where we have also assumed that the radiation field is isotropic.  When calculating the $\gamma$-ray spectrum observed at Earth, we replace $h$ with $\ell_{\oplus}$, the sightline to the center of the starburst disk (equal to $R$ for perfectly edge-on disks).

\begin{figure*}
\centerline{\includegraphics[width=8cm]{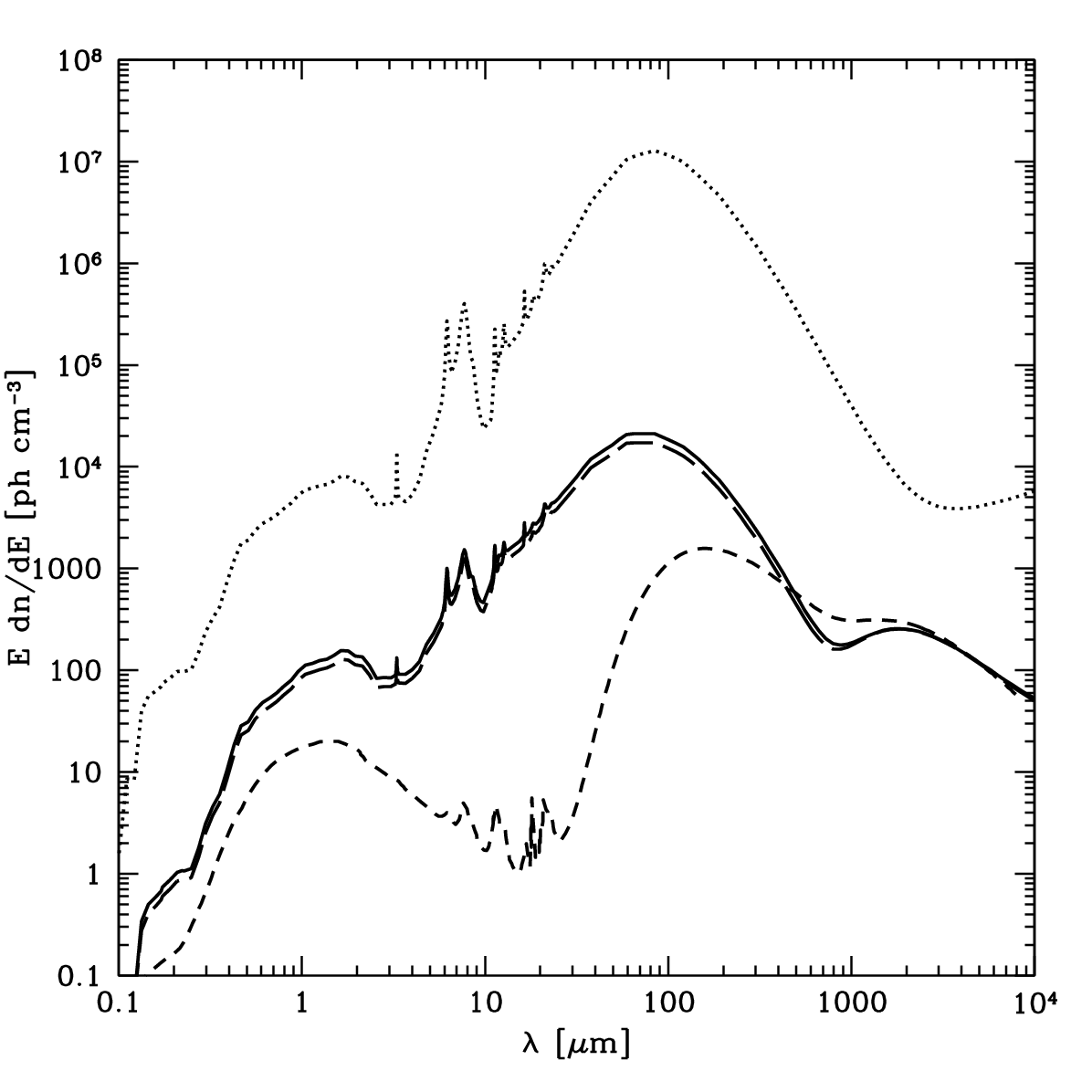}\includegraphics[width=8cm]{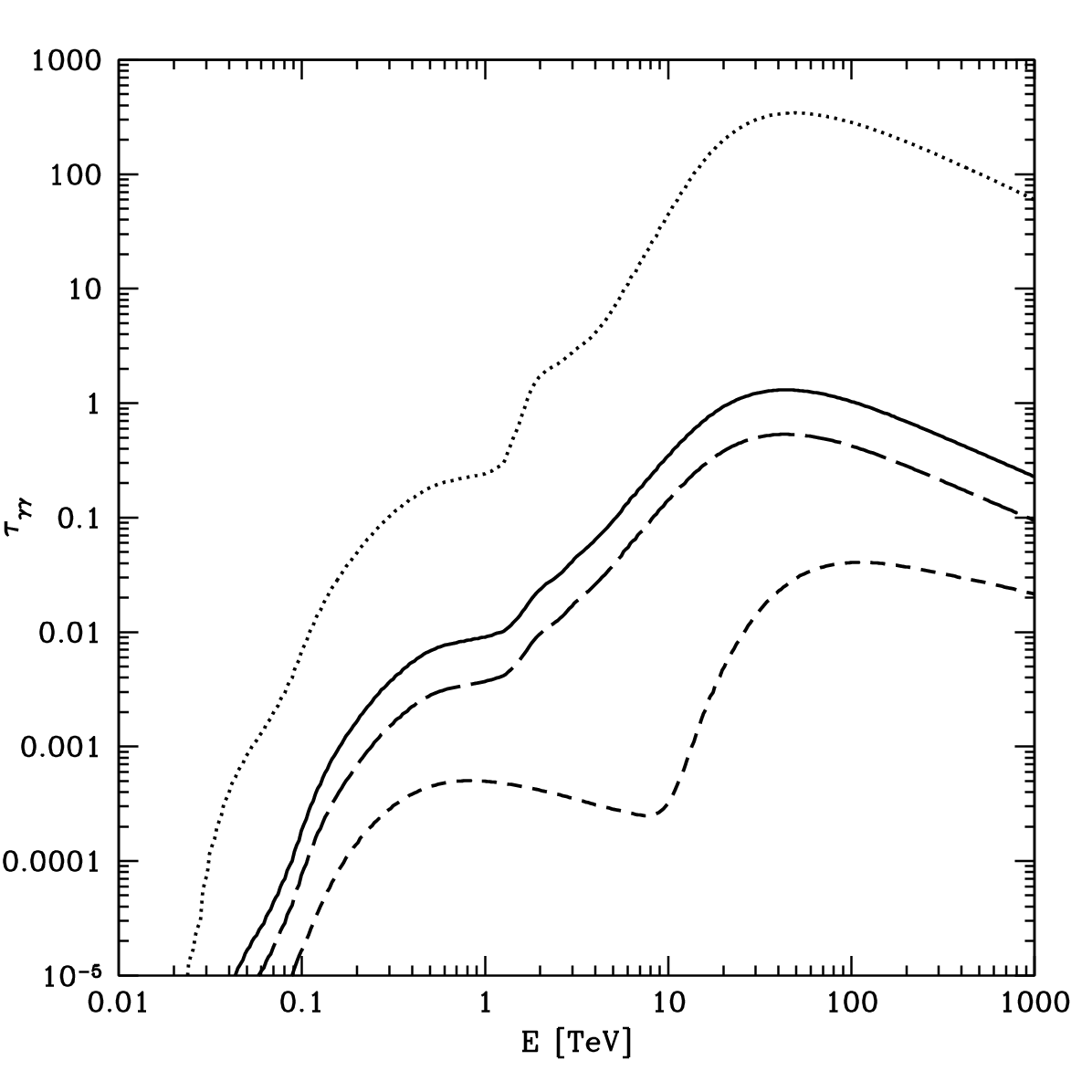}}
\figcaption[simple]{On the left, we show the photon densities of the radiation fields of the Galactic Center (short-dashed), NGC 253 (long-dashed), M82 (solid), and a starburst nucleus of Arp 220 (dotted).  The FIR emission from dust dominates the photon population, except in the Galactic Center where the number of CMB photons is comparable.  On the right we show the pair-production optical depths of these starbursts, along a vertical sightline out of the starburst disk.  Note that M82 and Arp 220 are opaque at tens of TeV.\label{fig:GammaTau}}
\end{figure*}

We use the GRASIL SEDs for the radiation fields of M82 and Arp 220, scaled to the correct luminosities \citep{Silva98}.\footnote{The GRASIL SEDs are available at http://adlibitum.oat.ts.astro.it/silva/grasil/modlib/fits/fits.html.}  For NGC 253, we use the M82 GRASIL SED scaled to the NGC 253 starburst's luminosity.  We use the SED of \citet{Porter08} for the Galactic Center, adding an infrared greybody dust component from star-formation of temperature $20\ \Kelv$ \citep{Launhardt02}.  The GRASIL luminosities are converted to energy densities as $U = L / (2 \pi R^2 c)$.  We then add the CMB to the SEDs of NGC 253, M82, and Arp 220's nuclei (the CMB is already present in the \citealt{Porter08} radiation field for the Galactic Center).  We show the resultant $\gamma$-ray optical depths for these SEDs in Figure~\ref{fig:GammaTau}.  M82 and Arp 220 are both optically thick in the 10 - 100 TeV range.  NGC 253's optical depth peaks at $\sim 35\ \TeV$ with $\tau \approx 0.5$.  However, the Galactic Center is transparent at all considered energies.  Opacity from three radiation components are visible in Figure~\ref{fig:GammaTau}: near infrared radiation from old stars, far infrared radiation from dust grains, and the CMB.  In NGC 253, M82, and Arp 220, the far infrared radiation completely dominates the $\gamma\gamma$ opacity.

The source function $Q_{\rm pair} (E_e)$ of pair production $e^{\pm}$ is then calculated from the $\gamma$-ray and IR spectra \citep[e.g.,][]{Aharonian83,Boettcher97}.  We have implemented pair production in our code.  We use the \citet{Aharonian83} approximations for the source function, as given in equation 32 of \citet{Boettcher97}, which are known to be accurate at high energies.\footnote{Note that both equations 26 and 32 in \citet{Boettcher97} are for electrons only (Aharonian \& Khangulyan 2011, private communication; B\"ottcher 2011, private communication).  Our attempts at calculating the pair spectrum using the \citet{Boettcher97} formula gives results that are a factor of 2 too small, both by comparing with the \citet{Aharonian83} pair spectrum and checking energy conservation.  A previous version of this paper assumed that \citet{Boettcher97} equation 32 was for both $e^{\pm}$; thus the pair injection rate was underestimated a factor of 2 in that version.}

After calculating the $\gamma\gamma$ pair $e^{\pm}$ spectrum, we then calculate the $\gamma$-ray luminosity from these pairs.  We then in turn calculate the pair $e^{\pm}$ from these $\gamma$-rays, using the same procedure as we did for the $\gamma$-rays from protons and primary and secondary $e^{\pm}$.  We found that these higher-order pair $e^{\pm}$ were typically only a small fraction of the $\gamma\gamma$ pair $e^{\pm}$ population, reaching a maximum of 50\% in the Arp 220 nuclei models even with $B = 250\ \muGauss$ and under 8\% for M82, NGC 253, and the Galactic Center.  

\subsection{Constraints}
\label{sec:Constraints}
We run grids in $\eta$, $B$, $p$, and the normalization of the diffusive escape time $t_{\rm diff}$ (see eq.~\ref{eqn:tDiff}).  Hadronic models (with CR protons, secondary $e^{\pm}$, and the pair $e^{\pm}$ associated with the $\gamma$-rays from these) are run independently of leptonic models (with primary CR electrons and the pair $e^{\pm}$ from the $\gamma$-rays they generate), giving us a hadronic and leptonic template for each parameter set.  The two are then added together by scaling the hadronic template with $\eta$ and the leptonic template with $\xi$.  For starbursts where there are error bars in the radio data (Galactic Center, NGC 253, and M82), we then select the free-free emission flux $S_{\rm therm} (\GHz)$ at 1 GHz based on chi-square fitting of the radio data, scaling $\xi$ for each $S_{\rm therm} (\GHz)$ so that the 1.4 GHz synchrotron radio emission of the starbursts equals our model predictions.  We use the interferometric measurements in \citet{Williams10} of M82 and the starburst core of NGC 253 and the $\ge \GHz$ radio measurements in \citet{Crocker11} for the Galactic Center (``HESS region'') radio flux.  We use the radio data compiled in \citet{Torres04} for the east and west nuclei of Arp 220; since these data do not have error bars, we simply normalize $\xi$ to match the observed 5 GHz flux.  Models with negative $\xi$ (if the secondaries alone overproduced the radio flux) were not allowed.  

We then require the predicted 0.3 - 10 GeV \citep{Abdo10b} and TeV \citep{Acero09,Acciari09} emission of M82 and NGC 253 to match the observed values to within a factor 2.  For the Galactic Center region, we require the predicted TeV emission to match the observed emission \citep{Aharonian06} to within a factor 2, and the predicted GeV emission to be lower than the observed emission within $|\ell| \le 1^{\circ}$ and $|b| \le 1^{\circ}$ from EGRET \citep{Hunter97}.  For Arp 220, we require both the predicted GeV and TeV to be lower than the upper limits from the Fermi-LAT one year catalog \citep{Abdo10b} and MAGIC \citep{Albert07}, respectively.

We also run a purely calorimetric hadronic model for each $p$, $B$, and density, with no advective or diffusive losses.  By integrating up the volumetric power generated in secondary $e^{\pm}$, $\gamma$-rays, and neutrinos in a model, and comparing the yield with the calorimetric model, we can quantify the efficiency of pionic losses as the calorimetric fraction $F_{\rm cal}$: 
\begin{equation}
F_{\rm cal} = \frac{\int_0^{\infty} (E \frac{dQ_{\gamma}}{dE} + E \frac{dQ_{e}}{dE} + E \frac{dQ_{\nu}}{dE}) dE}{\int_0^{\infty} (E \frac{dQ_{\gamma}^{\rm cal}}{dE} + E \frac{dQ_{e}^{\rm cal}}{dE} + E \frac{dQ_{\nu}^{\rm cal}}{dE}) dE}.
\end{equation}
We also quantify the efficiency of pionic losses for generating VHE products by calculating $F_{\rm cal} (\ge 10\ \TeV)$, in which we change the lower bound of integration to 10 TeV.

Usually, a range of $\eta$ will be compatible with these constraints.  For the sake of brevity, for a given parameter set ($\Sigma_g$, $B$, $p$, $t_{\rm diff} (3\ \GeV)$), we consider only those models that either have (a) the minimum allowed $\eta$, (b) the maximum allowed $\eta$, (c) the $\eta$ which predicts the closest match to the $\sim\GeV$ $\gamma$-ray emission (in M82 and NGC 253) or the TeV $\gamma$-ray emission (in the Galactic Center), or (d) $\eta$ fulfills (a), (b), or (c) for some other $\gamma_{\rm max}^{\rm prim}$ and the model otherwise satisfies our constraints.  For Arp 220's nuclear starbursts, we simply assume that $\eta = 0.1$ since there are no $\gamma$-ray detections yet.

\section{Results and Comparison With Observations}
\label{sec:Results}
At GeV energies, synchrotron losses must compete with extremely strong bremsstrahlung, ionization, and IC losses, which cool $e^{\pm}$ before they radiate much synchrotron.  In order to account for the observed radio emission, there must be more CR $e^{\pm}$ than one would naively expect for primary electrons with Milky Way-like acceleration efficiencies.  In general, we find that two different kinds of models work, given the radio constraints and the $\gamma$-ray observations:

1.) In models with high $B$, the GeV CR $e^{\pm}$ spectrum is dominated by pionic secondaries, which enhance the radio emission.  In this limit, models are not affected much by variations in sufficiently small $\xi$, since any small $\xi$ will result in sub-dominant primaries.  The amount of synchrotron emission from secondaries (and tertiary pair $e^{\pm}$ from pionic $\gamma$-rays) is set by the efficiency of CR proton acceleration ($\eta \approx 0.1$, in turn set by the $\gamma$-ray observations), the efficiency of pionic losses $F_{\rm cal}$ and pair production (set by the gas density and radiation fields, not $B$), and the power of synchrotron with respect to other losses $f_{\rm synch}$ (which is highly dependent on $B$ for GHz emission, but is $\sim 1$ for X-ray emitting energies in high $B$ models).  Thus the hadronic component asymptotes to a constant synchrotron X-ray luminosity as $B$ increases.  However, the GHz radio synchrotron emission does increase with $B$, because of the dominant non-synchrotron losses; in models with too high $B$, the secondaries overproduce the radio emission at fixed $\eta$.  This sets an upper bound on $B$.

2.) At low $B$, the synchrotron emission of secondary $e^{\pm}$ is insufficient to explain the observed GHz radio.  The only way for there to be enough CR electrons is to increase the primary CR electron acceleration efficiency greatly.  Thus, low $B$ models favor primary electrons dominating the CR $e^{\pm}$ spectrum at GeV energies.\footnote{\citet{Lacki10a} postulated a ``high-$\Sigma_g$ conspiracy'' that sets the radio synchrotron luminosity of starbursts: the suppression of starbursts' radio emission by non-synchrotron losses is compensated by the appearance of secondary $e^{\pm}$ and the dependence of the critical synchrotron frequency on $B$.  In the low $B$ models, the non-synchrotron losses are still present and are even more influential because $B$ is smaller, but they are instead compensated by a much larger primary CR electron acceleration efficiency than in the Milky Way.  Thus, there still is a conspiracy, but a different one involving high $\xi$ instead of secondaries.}  If the value of $\gamma_{\rm max}^{\rm prim}$ is relatively small, the 10 - 100 TeV CR $e^{\pm}$ spectrum only consists of secondaries and tertiaries.  If the primary CR electron spectrum extends to higher energies, though, the primary electrons in these models may overwhelm the secondaries and tertiaries and greatly enhance the synchrotron X-ray emission.  Although we consider them, we feel that these models are contrived: in those starbursts with $\gamma$-ray constraints, these models require the CR proton acceleration efficiency to be the same or less than the in the Milky Way ($\eta \la 0.1$), so that the the total hadronic and leptonic $\gamma$-ray emission not exceed observations.  However, they also require that primary CR electrons be accelerated much more efficiently ($\xi \approx 0.05 - 0.2$) in starbursts compared to $\xi \la 0.01 - 0.02$ in the Milky Way (\citealt{Lacki10a}; the numerical value depends on the shape of the CR electron spectrum at energies $\le \GeV$).  They can be distinguished observationally from the high $B$ secondary-dominant models by $\gamma$-ray emission below $\sim 100\ \MeV$: models with low $B$ have intense bremsstrahlung and IC emission which flattens out the pionic ``bump'' at lower energies.  Indeed, the leptonic $\gamma$-ray emission sets a lower limit on $B$ for the starbursts we model.

This basic idea that different $B$ require different electron acceleration efficiency to match the GHz radio constraints has been described before in \citet{Persic08}, \citet{deCeaDelPozo09a}, and \citet{Rephaeli09}.  

In Tables~\ref{table:GalCenXRayLuminosities}, \ref{table:NGC253XRayLuminosities}, \ref{table:M82XRayLuminosities}, and \ref{table:A220XRayLuminosities}, we list the synchrotron X-ray emission in models for the Galactic Center, NGC 253, M82, and the nuclei of Arp 220, respectively.  We find that the synchrotron X-ray emission is very weak in the Galactic Center and generally weak in NGC 253 and M82, especially in models where the primary $e^-$ spectrum cuts off below a TeV ($\gamma_{\rm max}^{\rm prim} = 10^6$).  In Arp 220, however, the synchrotron contribution to the diffuse X-ray emission is significant, even with a low energy primary cutoff.  The IC emission is also a small minority of the X-ray emission.  We discuss individual starbursts in \S~\ref{sec:GalCen} - \ref{sec:Arp220}.  

Some general trends are apparent in the tables.  First, the synchrotron X-ray fraction is higher if the injection spectra are harder, simply because there are more $e^{\pm}$ and protons at higher energy.  Second, the amount of X-ray emission depends on the primary electron maximum energy.  The cutoff dependence is especially strong in low $B$ models where the primaries dominate the CR $e^{\pm}$ spectrum.  Even in high $B$ models, where secondaries dominate the CR $e^{\pm}$ population at $\sim \GeV$ energies, primaries can still dominate the $\sim \TeV$ $e^{\pm}$ population, because diffusive escape becomes quicker at high energies and removes CR protons before they can interact with starburst gas.  Third, a shorter diffusive escape time reduces the synchrotron X-ray contribution, particularly in high B models for M82 and NGC 253 where the secondary $e^{\pm}$ at multi-TeV energies is dominant in some models and sub-dominant in others.

To simplify our presentation, we also subjectively choose a ``fiducial'' model for each considered starburst.  To qualify as a fiducial model, $\tilde\delta$ must be near its approximate Milky Way value, within the range 50 - 100, whenever this is possible among the allowed models.  This criterion selects the higher $B$ models.  Because of the coarseness of our grids in $B$, however, we were not able to get $\tilde\delta$ to match in all of our fiducial models: $\tilde\delta$ is 18 in the fiducial model for Arp 220's western nucleus, and 48 - 84 in the other starbursts.  We also consider only $p = 2.2$ models (which also means $13 \le \delta \le 25$, from eqn.~\ref{eqn:deltaTilde}).  Finally, we try to choose models that are close fits to any existing $\gamma$-ray observations.  The parameters for these fiducial models are listed in Table~\ref{table:ModelParameters}.

\subsection{The Galactic Center}
\label{sec:GalCen}
\emph{Introduction} -- The Galactic Center region in many ways resembles a mini-starburst; we refer the reader to \citet{Crocker11} for a complete discussion of its properties.  Diffuse hard ($\Gamma \approx 2.3$) TeV $\gamma$-rays have been observed with HESS in the region with $|\ell| < 0.8^{\circ}$ and $|b| < 0.3^{\circ}$ ($R = 112\ \pc$; $h = 42\ \pc$; \citealt{Aharonian06,Crocker11}), which we take as our modelled region.  The source of the CRs responsible for this emission has been conjectured to be Sgr A$^{\star}$ \citep[e.g.,][]{Ballantyne07}, supernovae in the region \citep[e.g.,][]{Buesching07,Crocker11}, and diffusive reacceleration of CRs in the intercloud medium \citep{Wommer08,Melia11}.  The hard CR spectrum could either be explained by either a recent non-steady-state injection of CRs \citep{Aharonian06,Buesching07}, or steady-state energy-independent transport such as advection \citep{Crocker11}.  For the purposes of this paper, we assume like \citet{Crocker11} that the diffuse TeV emission is from a steady-state population of CRs accelerated by star-formation processes, and then briefly discuss the consistency of this scenario with our results.  

GeV emission from the nuclear starburst specifically, as opposed to the inner Galaxy, is not yet well constrained, although it probably is present \citep{Crocker10}.  Radio emission is observed both from this region and a surrounding halo-like structure that extends out to $|\ell| < 3^{\circ}$ and $|b| < 1^{\circ}$ ($R = 420\ \pc$, $h = 140\ \pc$) which contains most of the observed radio emission.  The luminosity of the dust in the region is $4 \times 10^8\ \Lsun$ \citep{Launhardt02}; taking that as $L_{\rm TIR}$, we derive a star-formation rate of 0.07 $\Msun\ \yr^{-1}$ (eq.~\ref{eqn:TIRtoSFR}), which is compatible with the star-formation rate over the past few million years found by \citet{YusefZadeh09} (see also the extensive discussion in \citealt{Crocker10b,Crocker11}).\footnote{\citet{Launhardt02} find a \emph{total} stellar luminosity of $2.5 \times 10^{9}\ \Lsun$, mostly from the nuclear star cluster, but this luminosity with eq.~\ref{eqn:TIRtoSFR} implies a much higher star-formation rate of $0.3\ \Msun\ \yr^{-1}$ if it mostly came from young stars.  Since the total radio emission is mostly from primaries in our models, this would require $\xi$ to be $\sim 6$ times smaller than our derived values, or $0.001$.  The TeV $\gamma$-ray emission, if it is hadronic, would imply $\sim 6$ times smaller $\eta$ than in our models, $\la 0.02$.  \citet{Crocker11} however conclude that the supernova rate is consistent with the lower rate we use.}  This IR luminosity implies that the inner starburst region observed with HESS falls off the FIR-radio correlation observed in starbursts like M82 and NGC 253, but the surrounding halo does lie on the correlation \citep{Crocker10b}.  This combined with the $\gamma$-ray faintness of the starburst region given its supernova rate implies a strong wind, with a speed we take to be $600\ \kms$ following \citet{Crocker11}.  The radiation energy density from the dust luminosity is $42\ \eV\ \cm^{-3}$.  For the background radiation field we add this star-formation TIR radiation field to the inner Milky Way radiation field in \citet{Porter08}, which includes the contributions from stars outside the Galactic Center starburst proper.  This radiation field has an energy density $15\ \eV\ \cm^{-3}$, for a total radiation energy density $57\ \eV\ \cm^{-3}$ ($B_{\rm rad} = \sqrt{8 \pi U_{\rm rad}} = 48\ \muGauss$).  

The gas surface density that CRs traverse in the Galactic Center region is not very well constrained.  \citet{PiercePrice00} find a gas mass of $5 \times 10^7 \Msun$ in the inner 200 pc, corresponding to $\Sigma_g = 0.09~\gcm2$.  \citet{Ferriere07} find that $\Sigma_g \approx 0.03~\gcm2$ in the inner few hundred pc, although steeply falling with distance from the Galactic Center.  \citet{Crocker11} find that CRs have to sample gas of less than average density to fit the multiwavelength data.  We run a grid in $\Sigma_g$ ranging from $0.003~\gcm2$ to $0.1~\gcm2$ to account for these uncertainties.  We assume a magnetic field strength of $50$ or $100~\muGauss$ \citep{Crocker10}.

The Galactic Center region also has a hard X-ray and soft $\gamma$-ray excess along the Galactic ridge \citep[e.g.,][]{Worrall82}.  The central square degree has a 2 - 10 keV surface brightness of $10^{-9}\ \ergps\ \cm^{-2}\ \sr^{-1}$ \citep{Koyama96}; we scale this to the modeled region, assuming constant surface brightness, to get a luminosity of $7.4 \times 10^{36}\ \ergps$.  Although the X-ray ridge emission was once attributed to very hot gas, most of the emission has been resolved into faint X-ray sources, mainly associated with stars \citep{Revnivtsev06,Revnivtsev09}.  At higher energies $\ga 50\ \keV$ (particularly soft $\gamma$-rays), Inverse Compton may dominate the emission \citep{Porter08}.  Synchrotron has previously been considered as a source of the Galactic ridge X-ray emission \citep{Protheroe80,Porter97}, but requires stronger magnetic fields than are present in much of the Milky Way to avoid overproducing TeV Inverse Compton emission \citep{Aharonian00}.  Given the strong magnetic fields and relatively high star-formation rate of the Galactic Center starburst, we reconsider whether synchrotron emission may contribute to the X-rays from this region.

\begin{figure}
\centerline{\includegraphics[width=8cm]{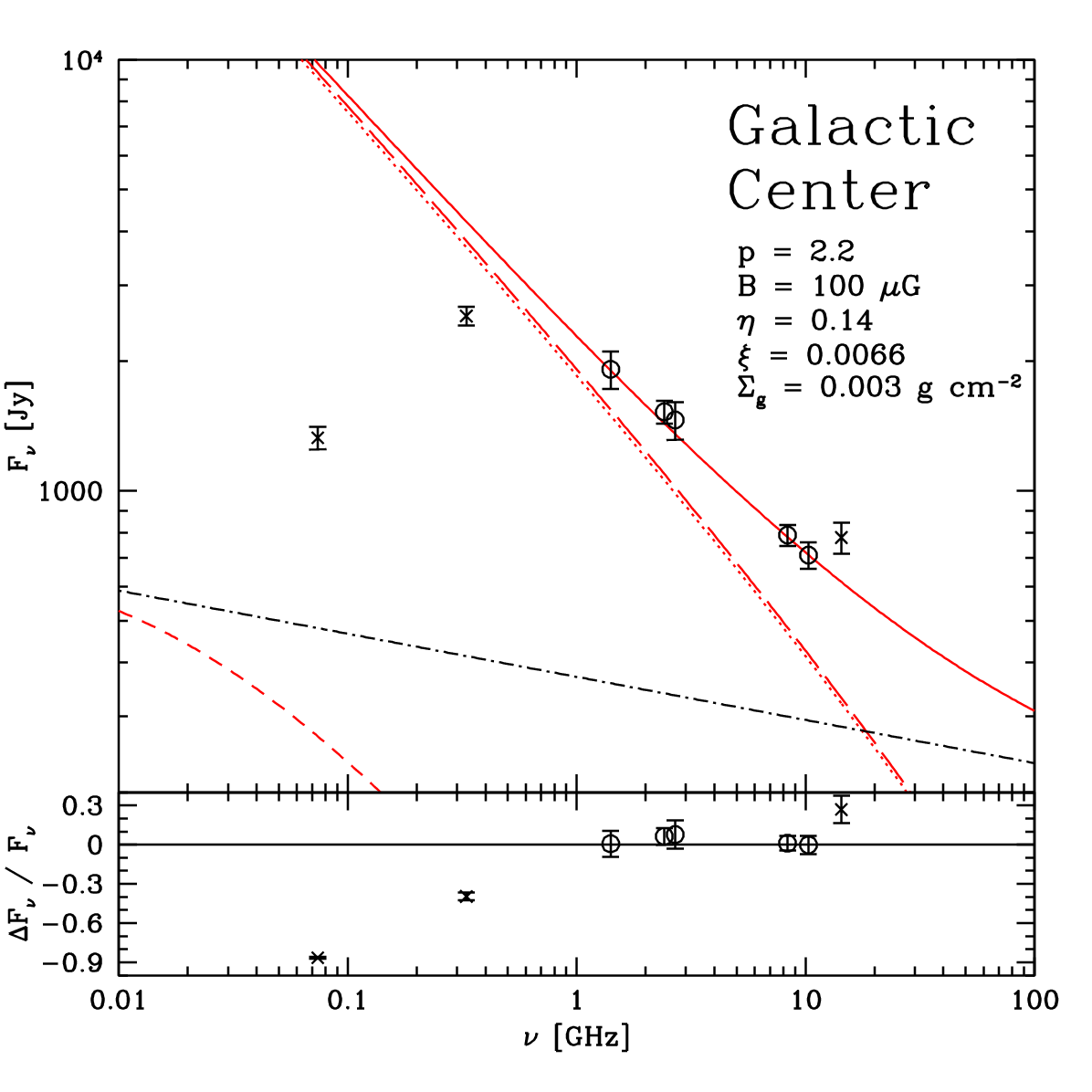}}
\figcaption[figure]{Our predicted total (synchrotron + free-free) radio spectrum (red, solid) compared to the observations compiled in \citet{Crocker11}.  Open circles are fitted while Xs are not fitted.  The synchrotron spectrum itself is the red long-dashed line, with components from primary and secondary $e^{\pm}$ are shown as red dotted and red short-dashed lines. The black dash-dotted line is our fit to the thermal free-free emission with these parameters.\label{fig:GalCenterRadio}}
\end{figure}

\emph{Modelled Radio and $\gamma$-rays} -- The Galactic Center region has few radio data points, so it is not surprising that we are able to fit the $\ge \GHz$ radio data.  In the allowed models, the radio emission is dominated by primaries \citep[c.f.][]{Crocker11}, which have $\xi = 0.005 - 0.01$ for $B = 100\ \muGauss$ and $\xi = 0.03 - 0.06$ for $B = 50\ \muGauss$.  The 74 and 330 MHz data points fall well below our fits, which is probably due to the strong free-free absorption known to be present in the region \citep[e.g.,][]{Brogan03}.

\begin{figure*}
\centerline{\includegraphics[width=9cm]{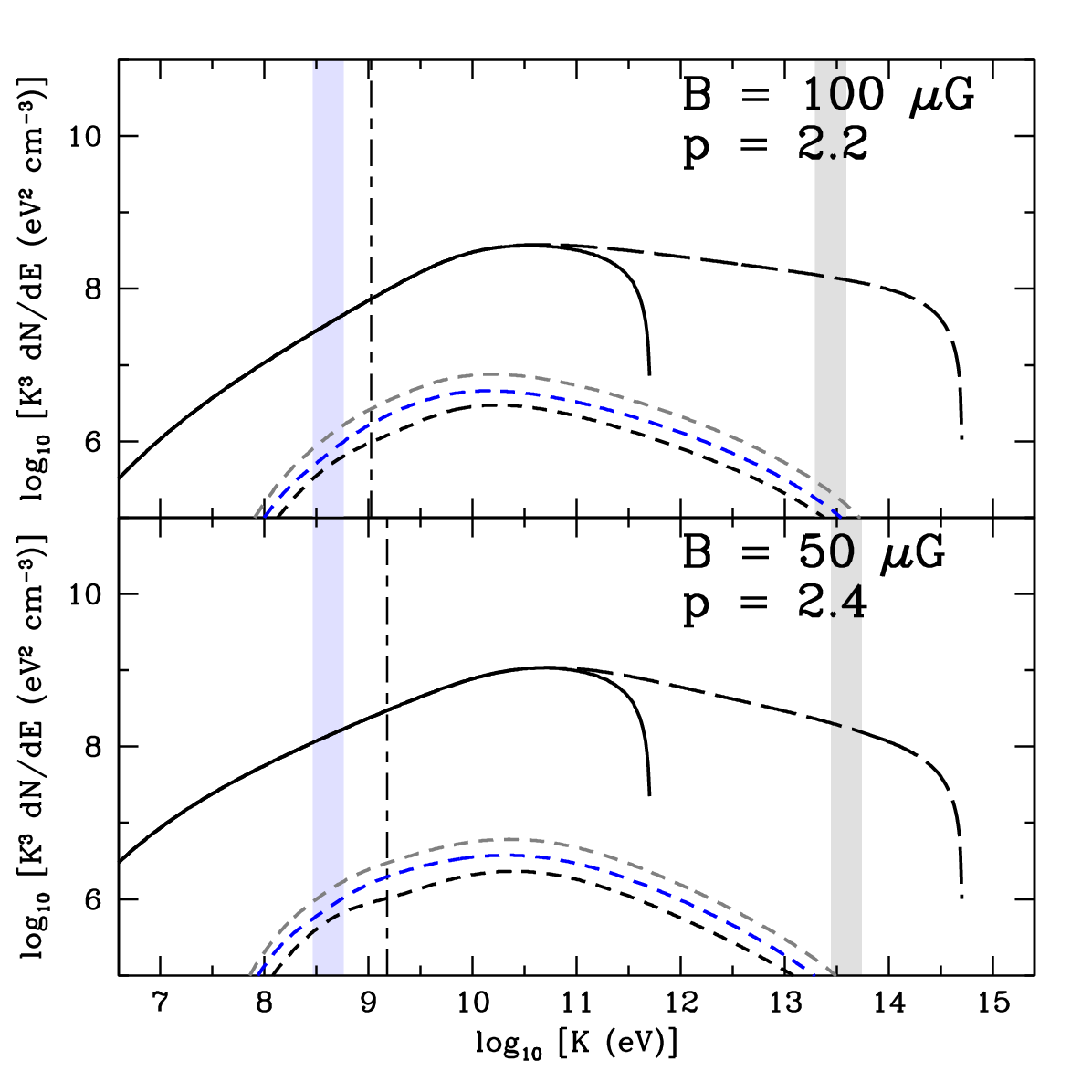}\includegraphics[width=9cm]{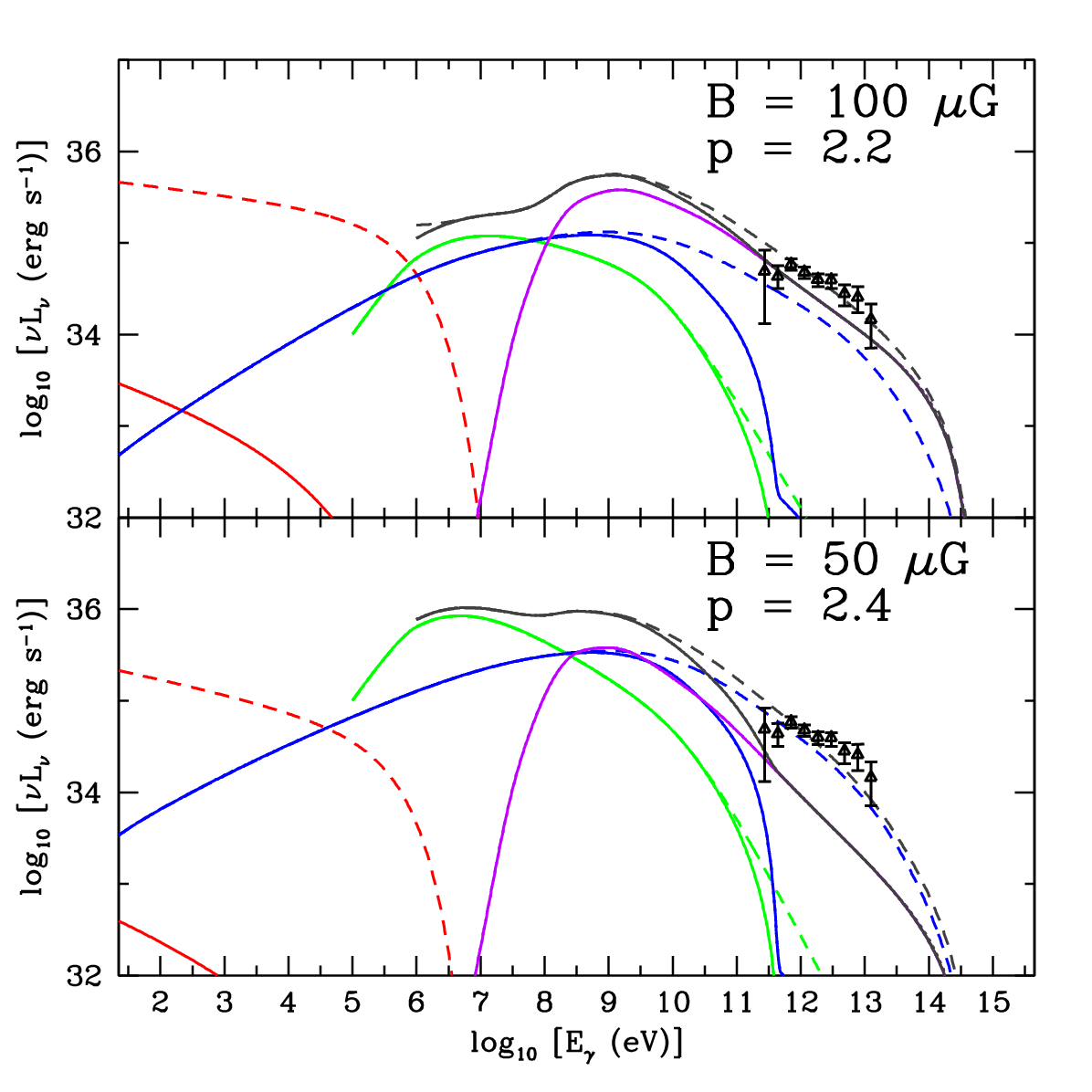}}
\figcaption[simple]{The CR electron (\emph{left}) and photon (\emph{right}) spectra of the Galactic Center region.  On the left, primary electrons are solid ($\gamma_{\rm max}^{\rm prim} = 10^6$) and long-dashed ($\gamma_{\rm max}^{\rm prim} = 10^9$), pionic secondaries are short-dashed (blue is $e^+$, black is $e^-$, and grey is total).  Pair and knock-off electrons are insignificant.  Blue and black shading denote the approximate $e^{\pm}$ energies that radiate in the 2 - 10 keV band through IC and synchrotron, respectively.  The vertical line indicates the approximate energy that radiates synchrotron at 1.4 GHz.  On the right, synchrotron is red, bremsstrahlung is green, IC is blue, pionic is violet, and total is black.  Solid lines are the Earth-observed absorbed luminosities in the $\gamma_{\rm max}^{\rm prim} = 10^6$ models.  Dashed lines show $\gamma\gamma$ absorbed luminosities in $\gamma_{\rm max}^{\rm prim} = 10^9$ models. The TeV data points are from \citet{Aharonian06}.\label{fig:GalCenterSpectra}}
\end{figure*}

Models with $\Sigma_g = 0.003\ \gcm2$ are strongly preferred by the $\gamma$-ray constraints (see Table~\ref{table:GalCenXRayLuminosities} for the allowed models).  This supports recent modeling by \citet{Crocker11} implying that CRs do not sample the average density in the Galactic Center region.  Low gas surface densities means that $F_{\rm cal}$ is also small, $\sim 0.3 - 0.8\%$ for $\Sigma_g = 0.003\ \gcm2$ (contrast this with $F_{\rm cal} \approx 8 - 21\%$ for $\Sigma_g = 0.1\ \gcm2$).  Thus, in many of our models, the Galactic Center is actually less of a proton calorimeter than the Milky Way as a whole.  Above 10 TeV, diffusive escape can reduce this even further, to $\sim 0.4\%$ when $t_{\rm diff} (3\ \GeV) = 10\ \Myr$ and $\sim 0.06\%$ when $t_{\rm diff} (3\ \GeV) = 1\ \Myr$ for $\Sigma_g = 0.003\ \gcm2$.  Secondaries are therefore a minority compared to the primaries (dashed lines in Figure~\ref{fig:GalCenterSpectra}, left).  Tertiary pair $e^{\pm}$ are negligible at all energies, because the Galactic Center is transparent to 10 - 100 TeV $\gamma$-rays (Figure~\ref{fig:GammaTau}, short-dashed line).

\emph{Synchrotron X-rays} -- Except in a few of the most optimistic models, only a small fraction of the hard X-ray emission from the Galactic Center is synchrotron (see Table~\ref{table:GalCenXRayLuminosities}).  This is perhaps not surprising, given that most of the X-ray emission has already been resolved.  With CR protons and VHE $\gamma$-rays escaping so readily, there are few secondary or tertiary $e^{\pm}$ at 10 - 100 TeV energies to emit synchrotron X-rays (Figure~\ref{fig:GalCenterSpectra}).  Therefore the X-ray emission strength is determined by primary electrons.  Furthermore, the relatively low $B$ of the Galactic Center compared to the other starbursts means that a given synchrotron frequency corresponds to higher $e^{\pm}$ energies, where the cutoff in CR protons causes a steep falloff in the secondary $e^{\pm}$ spectrum.  We find synchrotron fractions of $0.01\%$ to $6\%$, with a fiducial value of $5.9\%$ ($L_{2-10}^{\rm synch} = 4.3 \times 10^{35}\ \ergps \approx 3 \times 10^{-7}\ L_{\rm TIR}$).  This is consistent with the resolution of $88 \pm 12\%$ of the X-ray ridge emission at energies of $\sim 7\ \keV$ \citep{Revnivtsev09}.  For $p = 2.2$, we find $\gamma_{\rm max}^{\rm prim} = 10^9$ models have synchrotron fractions of $\sim 6\%$; almost all allowed models with $\gamma_{\rm max}^{\rm prim} = 10^6$ have synchrotron fractions of $0.1\%$ or less.

The synchrotron emission has a very soft spectrum ($\Gamma_{2-10} =  2.35 - 2.50$) in models with $\gamma_{\rm max}^{\rm prim} = 10^6$.  This is because in the weaker $50 - 100\ \muGauss$ magnetic fields of the Galactic Center, the 2 - 10 keV X-ray emission traces $e^{\pm}$ of very high energy (grey shading in left panel of Figure~\ref{fig:GalCenterSpectra}).  The CR secondary $e^{\pm}$ spectrum starts to cutoff at these energies, because the CR proton spectrum ends at a PeV in our model (see Appendix~\ref{sec:OtherGammaP} for the effects of altering the maximum proton energy).  In models with $\gamma_{\rm max}^{\rm prim} = 10^9$, the synchrotron spectrum has $\Gamma_{2-10} = 2.12 - 2.27$, since the primary electron spectrum remains hard at these energies (left panel of Figure~\ref{fig:GalCenterSpectra}).  

\emph{Inverse Compton X-rays} -- The IC contribution to the observed 2 - 10 keV emission from the TeV-emitting region is also small.  The main factor affecting the IC contribution is $B$: in models with $B = 50\ \muGauss$, IC makes up 0.5\% of the 2 - 10 keV emission; in models with $B = 100\ \muGauss$, IC makes up 0.09 - 0.16\% of the 2 - 10 keV emission.  The IC emission has a spectral index of $1.45 - 1.67$, due to the strong advective losses flattening the electron spectrum.  

The ratio of synchrotron and IC emission in the 2 - 10 keV band has a large range (0.029 - 55; fiducial: 48) in the allowed models.  However, in all of the $\gamma_{\rm max}^{\rm prim} = 10^9$ models, synchrotron is brighter than IC, with typical ratios of $\sim 3 - 5$ for $p = 2.4$ models and $\sim 50$ for $p = 2.2$ models.  The large values of these ratios even when $p = 2.4$ arises because $U_B \ga U_{\rm rad}$ and winds remove electrons below 10 GeV before they can radiate much IC (eqns.~\ref{eqn:tICThomson} and~\ref{eqn:tWind}).  By contrast, in most of the $\gamma_{\rm max}^{\rm prim} = 10^6$ models, the synchrotron to IC ratio is almost always less than unity.  

\emph{Conclusion} -- The Galactic Center in some respects is a poor target for synchrotron X-ray emission.  This is because it has a large X-ray background probably from stellar X-ray sources, contains a fast wind that removes protons before they can interact with the gas and produce secondary $e^{\pm}$ and $\gamma$-rays that could cascade, and a weak radiation field compared to other starbursts that means few pairs are produced from the $\gamma$-rays.  The synchrotron emission is a few percent of the observed emission only if the primary electron spectrum extends to high enough energies.  IC emission in the 2 - 10 keV X-ray band is also very weak ($< 1\%$), because the Galactic Center wind removes GeV electrons before they can IC upscatter ambient radiation.  However, the proximity of the Galactic Center means that many point sources can be resolved out, leaving the truly diffuse X-ray emission \citep[e.g.,][]{Revnivtsev09}.

A potential concern about our results is whether time-independence is a good approximation.  Our assumed TIR luminosity of $4 \times 10^8\ \Lsun$ translates to a supernova rate of $(2200\ \yr)^{-1}$, while our fiducial diffusive escape time is $t_{\rm diff} = 10^6\ \yr\ (E / 3\ \GeV)^{-1/2} = 1700\ \yr\ (E / \PeV)^{-1/2}$.  Thus the steady-state approximation roughly holds for protons in our models until a PeV.  The advective lifetime in our models, $63000\ \yr$ (c.f. eqn.~\ref{eqn:tWind}) is long enough for $\sim 30$ supernovae to go off in the HESS region, which would smooth out stochastic effects at energies where advection dominates.  Advective escape also naturally produces the hard spectrum \citep{Crocker11}.  We cannot rule out that diffusive escape is much faster than we suppose, though.  For primary electrons, the total $\ge 10\ \TeV$ population of the entire region is not affected by diffusive escape since the synchrotron lifetime alone at these energies is $t_{\rm synch} = 120\ \yr\ (E / 10\ \TeV)^{-1} (B / 100\ \muGauss)^{-2}$ (eqn.~\ref{eqn:tSynch}), faster than any possible escape process.  However, the level of primary CR electrons may vary over kyr timescales if the time in which CR accelerators accelerate 10 - 100 TeV electrons is shorter than $\sim 2200$ years.  Thus, there could be a stochastic element in the synchrotron X-ray emission from the Galactic Center region.  

Another issue with these is that the Galactic Center is embedded in a 420 pc wide radio halo, in turn embedded in the Galaxy as a whole.  It is therefore likely that our one-zone approximations cannot capture the complex geometry of the region.  We generally expect the background sea of CRs from the Galaxy to be only a minor part of the CR population in the $R \le 112\ \pc$ region.  As \citet{Aharonian06} noted, the TeV detection indicates a CR energy density several times greater than that of the Galaxy at large ($\sim 1\ \eV\ \cm^3$).  Using their own one-zone models of the same region, \citet{Crocker11} also find that CR energy densities of $2.5 - 50\ \eV\ \cm^{-3}$ are consistent with the multiwavelength data, with the best-fit model having $U_{\rm CR} = 20\ \eV\ \cm^{-3}$.  

In the models presented here that are consistent with constraints, we find CR energy densities of $1.8 - 38\ \eV\ \cm^{-3}$.  In all of these models, then, the Center-accelerated CR population is at least comparable, and often overwhelms any background Galactic population.  The models at the low end of the range have small $\eta$ ($\sim 0.01$ compared to the canonical $0.1$) and TeV luminosities that are relatively small.  If we restrict our attention to only those models that also have $0.05 \le \eta \le 0.2$ (within a factor of 2 of the typically assumed $0.1$), we find CR energy densities of $4.9 - 37\ \eV\ \cm^{-3}$.  Our fiducial model has a CR energy density of $20\ \eV\ \cm^{-3}$, just as \citet{Crocker11} find.  

Since we are interested in the high energy electron population, we also compare the energy density in CRs with $\ge 10\ \GeV$.  In the local Milky Way, this energy density is just $6 \times 10^{-4}\ \eV\ \cm^{-3}$ \citep[see][and references therein]{Adriani11}.  In the models consistent with constraints, however, the $\ge 10\ \GeV$ $e^{\pm}$ energy density is $0.016 - 0.10\ \eV\ \cm^{-3}$, with a value of $0.035\ \eV\ \cm^{-3}$ in our fiducial model.  These high energy densities occur even in the face of the strong synchrotron and IC losses that CRs experience, requiring a large luminosity density of CR $e^{\pm}$ to sustain.  We therefore conclude that the Galactic Center CR population is distinct from that of the large-scale Galaxy.

The presence of the 420 pc radius radio halo which actually has a larger GHz synchrotron luminosity than the central molecular zone may also pose a challenge for our use of one-zone models.  \citet{Crocker11} interpret this halo as being the emission from CR $e^{\pm}$ advected out of the smaller TeV-emitting region within.  In addition, \citet{Launhardt02} find this halo region emits an additional $2 \times 10^8\ \Lsun$ of bolometric luminosity, which translates to a star-formation rate half that of the inner 112 pc.  Multi-TeV $e^{\pm}$ generated in the inner region do not live long enough to contribute to any synchrotron X-ray luminosity from the radio halo.  However, GeV $e^{\pm}$ that can emit IC X-rays are likely present.  To estimate how big of a foreground IC X-rays from the radio halo is, we find the ratio of 1.4 GHz radio surface brightnesses of the halo region ($|\ell| \le 3^{\circ}$, $|b| \le 1^{\circ}$) and the central molecular zone ($|\ell| < 0.8^{\circ}$, $|b| < 0.3^{\circ}$) from \citet{Crocker11}.  The central molecular zone has a 1.4 GHz surface brightness that is $\sim 3$ times that of the outer radio zone.  Thus, while the central molecular zone CR population dominates the observed radio emission in that area of the sky, the radio halo also contributes significantly.  A multidimensional model of the region is necessary to fully capture the CR transport in the region, although for order-of-magnitude purposes our one-zone approach is sufficient.  

\subsection{NGC 253 Starburst Core}
\label{sec:N253}
\emph{Introduction} -- NGC 253 contains one of the nearest starbursts (we adopt $D = 3.5\ \Mpc$; 1\farcs = 17 pc) and one of the two detected in both GeV (by Fermi-LAT; \citealt{Abdo10}) and TeV $\gamma$-rays (by HESS; \citealt{Acero09}).  Previous modelling of the nonthermal emission from NGC 253 has been done by \citet{Paglione96}, \citet{Domingo05}, and \citet{Rephaeli09}, and the latter predictions are in good agreement with the \emph{Fermi} and HESS results.  Unlike the Galactic Center, the starburst core of NGC 253 lies on the FIR-radio correlation and is $\gamma$-ray bright compared to the Milky Way.  NGC 253 has a large reservoir of gas in its center; the estimated mass is $(2 - 5) \times 10^7\ \Msun$ from molecular lines \citep[e.g.,][]{Mauersberger96,Harrison99,Bradford03,Sakamoto11}, corresponding to central gas surface densities of $0.07 - 0.15\ \gcm2$.  The starburst appears as a disk with a diameter of 20\farcs - 30 \farcs\ (340 - 510 pc) in radio continuum \citep{Turner83,Ulvestad97} and 15\farcs - 30\farcs\ (250 - 510 pc) in infrared and molecular lines \citep{Peng96,Ulvestad00,Sakamoto11}; we adopt a radius of 150 pc.  The vertical extent of the radio disk is 8\farcs\ (136 pc) \citep{Ulvestad97}, implying a maximum vertical scale height of 68 pc.  On the other hand, a cylindrical geometry for NGC 253's radio disk, given an inclination of $78^{\circ}$ and a radio diameter of 340 pc, implies that $\sim 71\ \pc$ of the vertical extent is the projection of the diameter; the remainder implies a scale height of only 33 pc.  Due to the uncertainties in the geometry, we adopt a scale height of 50 pc (giving a gas density of 190 $\cm^{-3}$).  Finally, the total IR luminosity of NGC 253 is $4 \times 10^{10}\ \Lsun$ \citep{Sanders03}, of which half comes from the nuclear starburst \citep{Melo02}.  We therefore adopt a starburst bolometric luminosity of $2 \times 10^{10}\ \Lsun$, which corresponds to a low supernova rate of $0.025\ \yr^{-1}$, and a radiation energy density $1200\ \eV\ \cm^{-3}$, equal to that of a magnetic field $B_{\rm rad} = 220\ \muGauss$.  

NGC 253 has also been studied with \emph{Chandra} and XMM-Newton \citep[e.g.,][]{Strickland00-N253,Weaver02,Strickland02,Bauer08}, revealing the effects of a starburst wind in soft X-rays and hard emission from the galactic disk.  XMM-Newton measured the diffuse (point source subtracted) 2 - 10 keV luminosity from the optical disk as $8.5 \times 10^{38}\ \ergps$ \citep{Bauer08}.  We adopt this as the X-ray luminosity to compare against, although it includes the regions outside the starburst; these outer regions host 30\% of the total star-formation rate \citep{Melo02}.  

\emph{Modelled Radio and $\gamma$-rays} -- The combination of radio, GeV, and TeV constraints still leaves a large and complex parameter space of models that work.  Selected models are shown in Table~\ref{table:NGC253XRayLuminosities}.  

\begin{figure}
\centerline{\includegraphics[width=8cm]{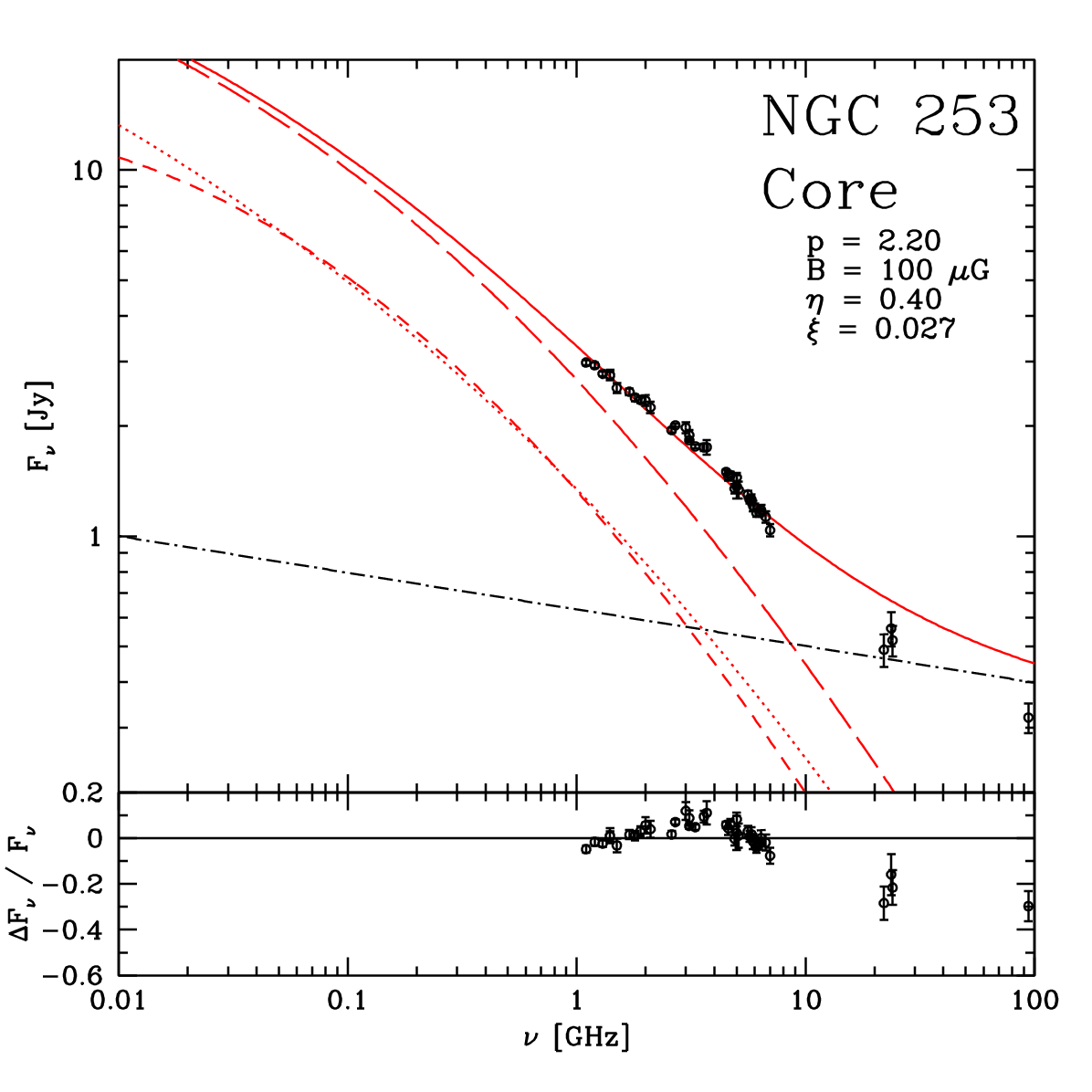}}
\figcaption[figure]{Our predicted total radio spectrum for the starburst core of NGC 253 (red, solid) compared to observations compiled in \citet{Williams10} (open circles).  The line styles are the same as in Figure~\ref{fig:GalCenterRadio}.  The radio emission at high frequencies is overproduced by the large thermal component, suggesting that the synchrotron spectrum is intrinsically flatter than shown here and that escape, bremsstrahlung, or ionization are stronger and $\xi$ is higher in reality.\label{fig:N253Radio}}
\end{figure}

We show radio emission in a typical model in Figure~\ref{fig:N253Radio}.  The flat radio spectrum favors high GHz thermal fractions in our models, with a span of $\sim 4 - 28\%$ (fiducial: 21\%) selected by our criteria.  As seen in Figure~\ref{fig:N253Radio}, there is evidence for unaccounted spectral curvature in the residuals, and the high thermal fraction overproduces the observed emission at 20 - 100 GHz.  This implies that the synchrotron spectrum is actually flatter than predicted by our models at 1 GHz, and then grows steeper at higher frequency, with a lower thermal fraction.  The flatter spectrum can arise from loss mechanisms with a flatter energy dependence than synchrotron or IC, such as bremsstrahlung, ionization, or advection losses.  Enhancing advection losses decreases the yield of secondary electrons, but increasing the density (and bremsstrahlung and ionization losses) increases the yield. Furthermore, more primaries would be needed to account for the smaller fraction of electron power going into GHz synchrotron emission if there are additional losses, so the results of such changes are complex.  Secondaries dominate the GHz synchrotron emission in models with high $B$, but in low to intermediate $B$, primaries dominate.

\begin{figure*}
\centerline{\includegraphics[width=9cm]{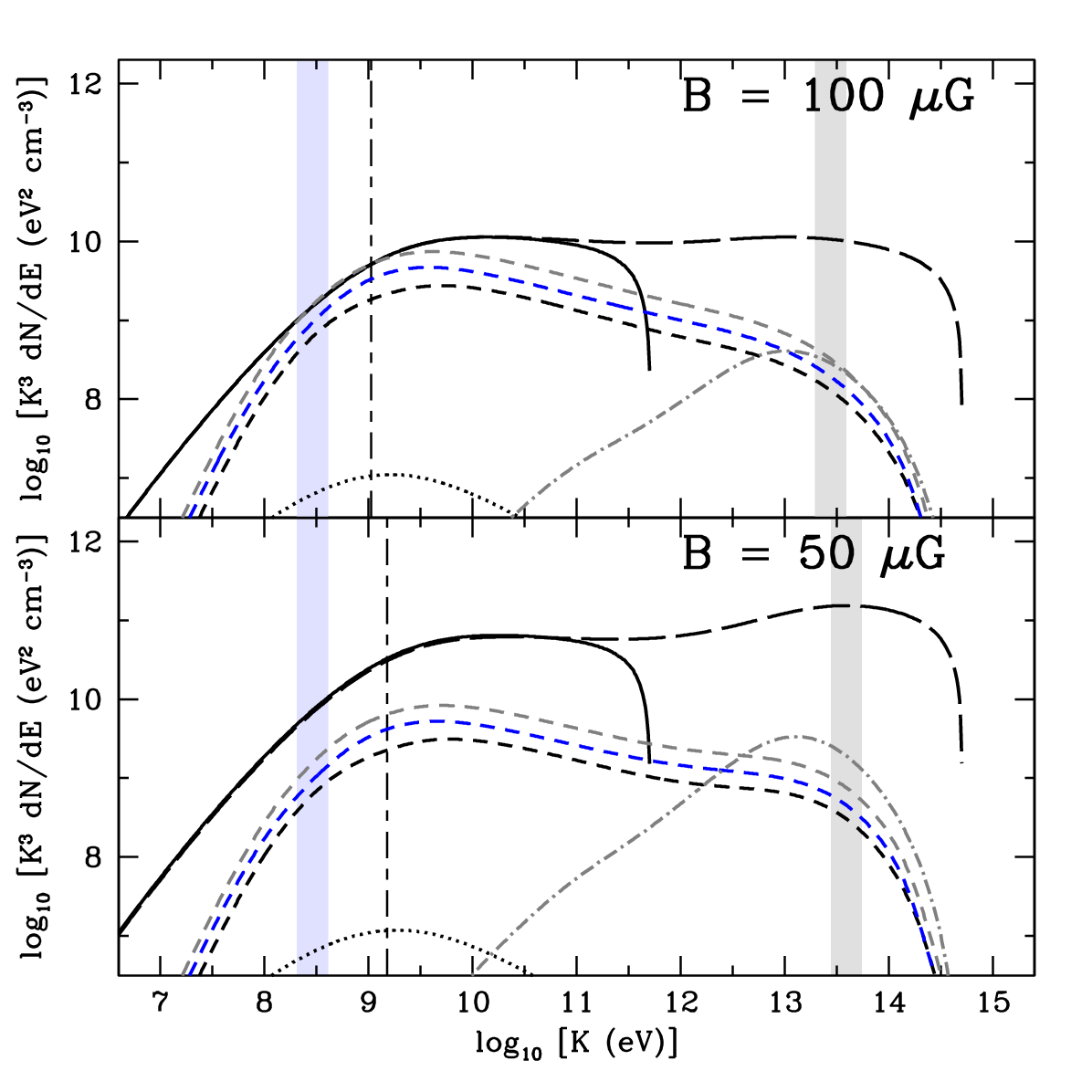}\includegraphics[width=9cm]{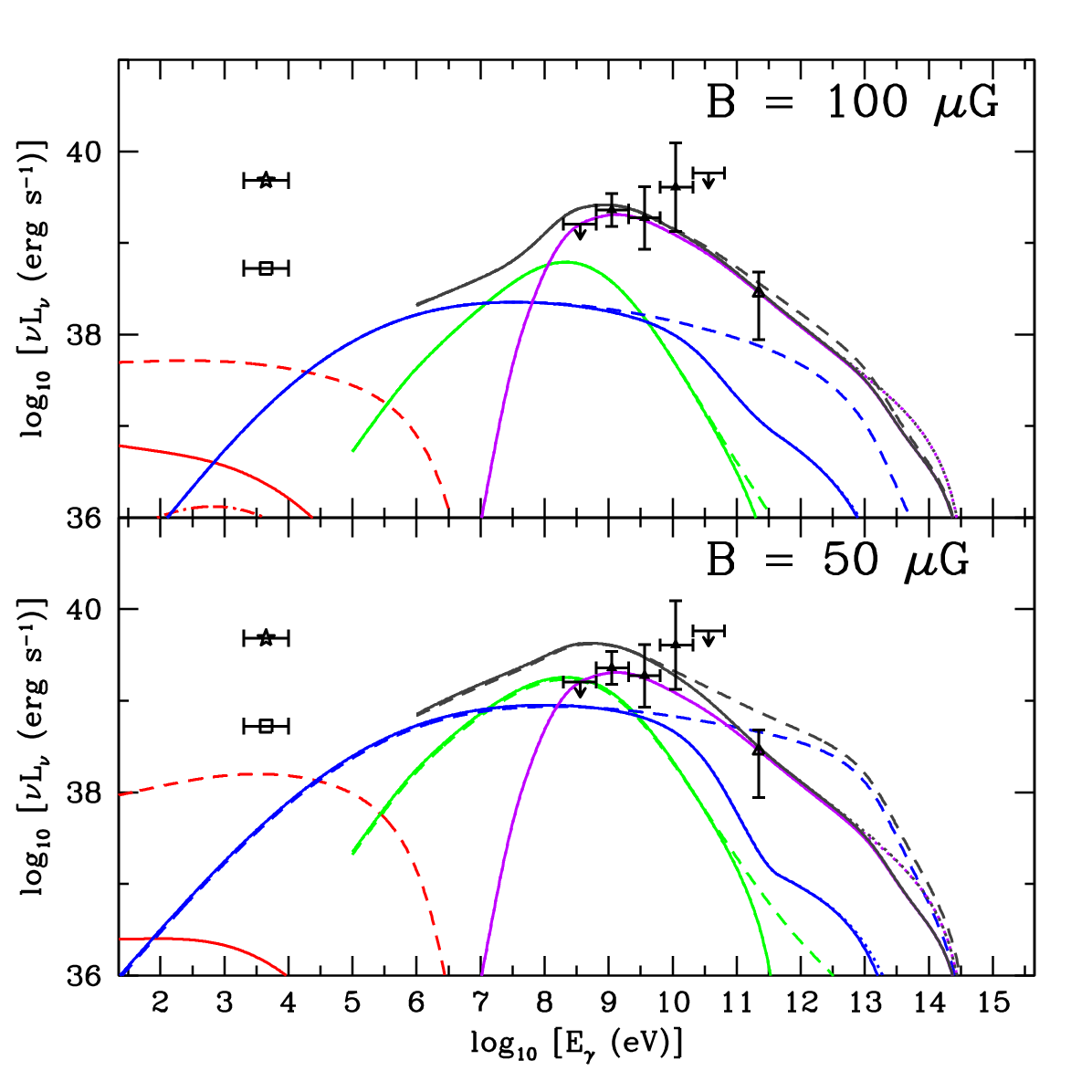}}
\figcaption[simple]{The CR electron (\emph{left}) and photon (\emph{right}) spectra of NGC 253's core for models with $p = 2.2$, $t_{\rm diff} (3\ \GeV) = 10^6\ \yr$, and $\eta = 0.40$.  On the left, primary electrons are solid ($\gamma_{\rm max}^{\rm prim} = 10^6$) and long-dashed ($\gamma_{\rm max}^{\rm prim} = 10^9$), pionic secondaries are short-dashed (blue is $e^+$, black is $e^-$, and grey is total), knock-off electrons are dotted, and pair-production tertiaries (using the $\gamma$-ray spectrum for $\gamma_{\rm max}^{\rm prim} = 10^9$) are dash-dotted.  Note that pair production $e^{\pm}$ are comparable to pionic secondary $e^{\pm}$ at $10\ \TeV$.  Blue and black shading denote the approximate $e^{\pm}$ energies that radiate in the 2 - 10 keV band through IC and synchrotron, respectively.  The vertical line indicates the approximate energy that radiates synchrotron at 1.4 GHz.  On the right, synchrotron is red, bremsstrahlung is green, IC is blue, pionic is violet, and black is total.  The dotted lines are without $\gamma\gamma$ absorption ($\gamma_{\rm max}^{\rm prim} = 10^6$), while solid are $\gamma\gamma$ absorbed according to the sightline observed from Earth ($\gamma_{\rm max}^{\rm prim} = 10^6$).  Dashed lines show $\gamma\gamma$ absorbed luminosities in $\gamma_{\rm max}^{\rm prim} = 10^9$ models.  Note that the $B = 50\ \muGauss$ model with $\gamma_{\rm max}^{\rm prim} = 10^9$ is not allowed by our constraints.  Observed luminosities from from Fermi and HESS (triangles), as well as the diffuse emission from XMM-Newton (square) and total emission from BeppoSAX \citep[star;][]{Cappi99}, are plotted.  X-ray luminosities are scaled to one ln bin in energy assuming $\Gamma = 2.0$.  The effects of neutral hydrogen absorption are not shown on right.  \label{fig:N253Spectra}}
\end{figure*}

The high energy emission from example models is shown in Figure~\ref{fig:N253Spectra}.  The predicted GeV $\gamma$-ray emission changes from being mainly leptonic at $B = 50\ \muGauss$ to nearly all hadronic at $B = 150\ \muGauss$.  This implies that NGC 253's starburst cannot have a magnetic field much lower than $50\ \muGauss$, or else the leptonic $\gamma$-ray emission would exceed the observed values.  Similarly, the magnetic field cannot be much higher than $150\ \muGauss$, given our assumed density and radiation field, or else the radio emission from the secondaries would be overpredicted, and in fact no models with $B \ge 200\ \muGauss$ fit our criteria.  Similar values have been derived from previous modeling of NGC 253's starburst \citep{Domingo05,Rephaeli09}.  

In $B \ge 100\ \muGauss$ models, high $\eta$ models are favored, and our fiducial model has $\eta = 0.40$, three to four times higher than the other starbursts.  This is because the proton calorimetry fraction is less than one half of the value for M82 in our fiducial model, even though the $\gamma$-ray to bolometric luminosity ratio is the same.  The actual CR acceleration efficiency could be $\sim 0.1$ and still consistent with the $\gamma$-ray data if: (1) the supernova rate in NGC 253's core is higher than we assume (and $\delta$ is the same as in our fiducial model), since only the injection rate of CRs matter, (2) the gas density is higher than we assume, increasing $F_{\rm cal}$, or (3) advective losses are weaker, again increasing $F_{\rm cal}$ \citep[c.f.][]{Lacki11}.  

The HESS detection also constrains the proton and electron populations at TeV energies.  In low $B$ leptonic models, the primary electron spectrum cannot extend to $\gamma_{\rm max}^{\rm prim} = 10^9$ with p = 2.0, or else the IC emission is overproduced.  Even with p = 2.2, the leptonic models are severely constrained and just barely fit the TeV constraint (see Table~\ref{table:NGC253XRayLuminosities}).  The high B scenarios allow a few p = 2.0 models, but only when diffusive escape time is quick and the TeV/GeV ratio is still $\sim 3$ times higher than observed.

In our models, NGC 253 is not a true proton calorimeter because of its strong wind, but it is much more calorimetric than the Milky Way.  The calorimetry fraction obtained by integrating the pionic products over all energies is $F_{\rm cal} \approx 10 - 30\%$ \citep[c.f.,][]{Lacki11}.  However, energy-dependent diffusive escape plays an increasing role at higher energies.  Above 10 TeV, $F_{\rm cal}$ drops to $\sim 12\%$ when $t_{\rm diff} (3\ \GeV) = 10\ \Myr$ and $\sim 2\%$ when $t_{\rm diff} (3\ \GeV) = 1\ \Myr$.

\emph{Synchrotron X-rays} -- The wide span of the parameter space allowed by the data (Table~\ref{table:NGC253XRayLuminosities}) means the 10 - 100 TeV electron population is not very well constrained, even with the HESS detection.  Thus, the synchrotron X-ray fraction of NGC 253's diffuse X-ray emission ranges from $0.19\%$ to $120\%$, with a fiducial value of 8\% ($L_{2-10}^{\rm synch} = 7.2 \times 10^{37}\ \ergps \approx 9 \times 10^{-7} L_{\rm TIR}$).  For $p = 2.2$ models, the synchrotron fraction is typically tens of percent in low $B$ models with high $\gamma_{\rm max}^{\rm prim}$, a few percent in high $B$ models, and a few percent to less than one percent in low $B$ models with low $\gamma_{\rm max}^{\rm prim}$.  Models with large synchrotron fractions ($\ga 1/3$) overproduce the observed TeV emission by a factor $\sim 2$.

The spectral index of synchrotron emission is $2.28 \le \Gamma_{2-10} \le 2.42$ for $\gamma_{\rm max}^{\rm prim} = 10^6$ and $2.00 \le \Gamma_{2-10} \le 2.29$ for $\gamma_{\rm max}^{\rm prim} = 10^9$.

\emph{Inverse Compton X-rays} -- The calculated IC X-ray emission in the 2 - 10 keV band varies because the magnetic field in NGC 253's starburst is unknown.  The models with $B = 50\ \muGauss$, $100\ \muGauss$, and $150\ \muGauss$ have IC fractions of $7 - 19\%$, $2 - 4\%$, and $1.3 - 1.9\%$, respectively.  The synchrotron-to-IC ratio varies widely from 0.013 to 49 (fiducial: 2.8).  Models with high $\gamma_{\rm max}^{\rm prim}$ have higher synchrotron/IC ratios (see Table~\ref{table:NGC253XRayLuminosities}).  The IC emission is consistently much harder than the synchrotron emission with $1.26 \le \Gamma_{2-10} \le 1.40$. 

\emph{Conclusion} -- The GeV and TeV detections of NGC 253's starburst inform our knowledge of the synchrotron X-ray emitting $e^{\pm}$ population, but leave room for large variations.  While the synchrotron fraction of NGC 253 can be as low as 0.2\% or as high as 120\%, in the fiducial model it is 8\%.  A key uncertainty is the primary electron energy cutoff, because the $e^{\pm}$ population is dominated by primary electrons at these energies.  The IC emission is also a minority of the X-ray emission, with the largest uncertainty being the unknown magnetic field in NGC 253.  

The higher resolution measurements of NGC 253 with \emph{Chandra} can help reduce the competing emission from other sources.  First, \emph{Chandra} can resolve out more point sources with its higher angular resolution.  Second, the emission from the outlying disk can be subtracted, leaving only the emission from the nuclear starburst proper.  \emph{Chandra} has revealed that there is extended hard X-ray emission in the nuclear starburst of NGC 253, but little is known about it \citep{Strickland00-N253,Weaver02}.

\subsection{M82}
\label{sec:M82}
\emph{Introduction} -- M82 (D = 3.6 Mpc; 1\farcs = 17\ \pc) is also detected in GeV and TeV $\gamma$-rays, with Fermi-LAT and VERITAS respectively \citep{Abdo10,Acciari09}.  \citet{deCeaDelPozo09a} and \citet{Persic08} previously modelled the nonthermal emission from M82, and there is again good agreement with the observed fluxes \citet{deCeaDelPozo09b}.  M82 lies on the FIR-radio correlation and is $\gamma$-ray bright with respect to the Milky Way.  Most of the star-formation is in the starburst in the center of M82.  \citet{Weiss01} find a gas mass of $2.0 \times 10^8\ \Msun$ in the inner 280 pc (for D = 3.6 Mpc), corresponding to a gas surface density of $0.17\ \gcm2$, which we use here.  The infrared emission of M82 is concentrated in a ring with radius 225 pc, and the gas peaks at a radius of 250 pc (see the compilation in Table 4 of \citealt{Goetz90}; see also \citealt{Kennicutt98}).  \citet{Williams10} find a radio extent for M82 of 35\farcs\ by 10\farcs, corresponding to a radius of 300 pc and a scale height of 90 pc.  We therefore adopt 250 pc as the radius and 100 pc as the scale height of the modeled starburst region.  Other studies find radio scale heights of 50 - 200 pc \citep{Klein88,Seaquist85}.  Values for the wind speed vary; however, \citet{Greve04} find that the wind accelerates over a 200 pc scale height to $400\ \kms$ and \citet{Westmoquette09} find H$\alpha$ FWHM of $100 - 300\ \kms$ in the inner few hundred parsecs.  The wind speed may asymptote to $\ga 1000\ \kms$, but in the standard theory this happens outside most of the starburst; inside the starburst we expect wind speeds near the sound speed \citep{Chevalier85}, which is $\sim 200 - 300\ \kms$ \citep{Greve04}.  We therefore again adopt an advection speed of $300\ \kms$.  The total IR luminosity of M82 is $5.9 \times 10^{10}\ \Lsun$ from \citet{Sanders03}, corresponding to a supernova rate of $0.06\ \yr^{-1}$.  This also implies a radiation energy density $1300\ \eV\ \cm^{-3}$, equal to that of a magnetic field $B_{\rm rad} = 230\ \muGauss$.  

The diffuse X-ray emission of M82 is well-studied.  The diffuse hard emission has a luminosity of $4.4 \times 10^{39}\ \ergps$ and is spatially extended over a few hundred pc, with a scale height estimated at $h = 175\ \pc$ \citep{Strickland07}.  The spectral slope of the diffuse hard X-ray emission is $\Gamma \approx 2 - 3$ and it is difficult to explain as thermal bremsstrahlung or IC emission \citep{Strickland07}.

\begin{figure}
\centerline{\includegraphics[width=8cm]{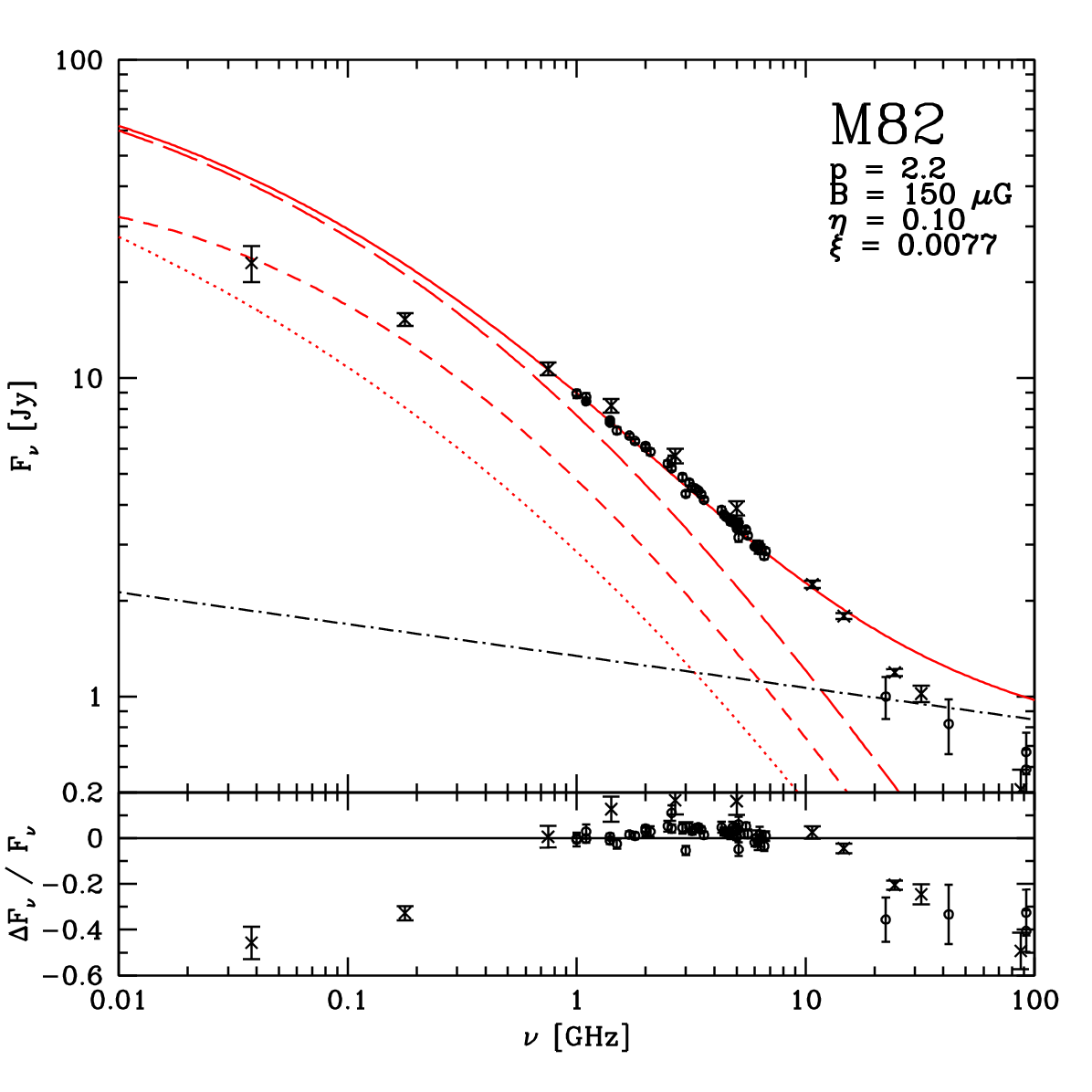}}
\figcaption[figure]{Our predicted synchrotron radio spectrum (red, solid) compared to observations compiled in \citet{Williams10} (open circles) and the unfit data in \citet{Klein88} (Xs).  The residuals indicate the synchrotron spectrum is too steep at high frequencies, suggesting that escape, bremsstrahlung, or ionization are stronger and $\xi$ is higher in reality.  The line styles are the same as in Figure~\ref{fig:GalCenterRadio}.  \label{fig:M82Radio}}
\end{figure}

\begin{figure*}
\centerline{\includegraphics[width=9cm]{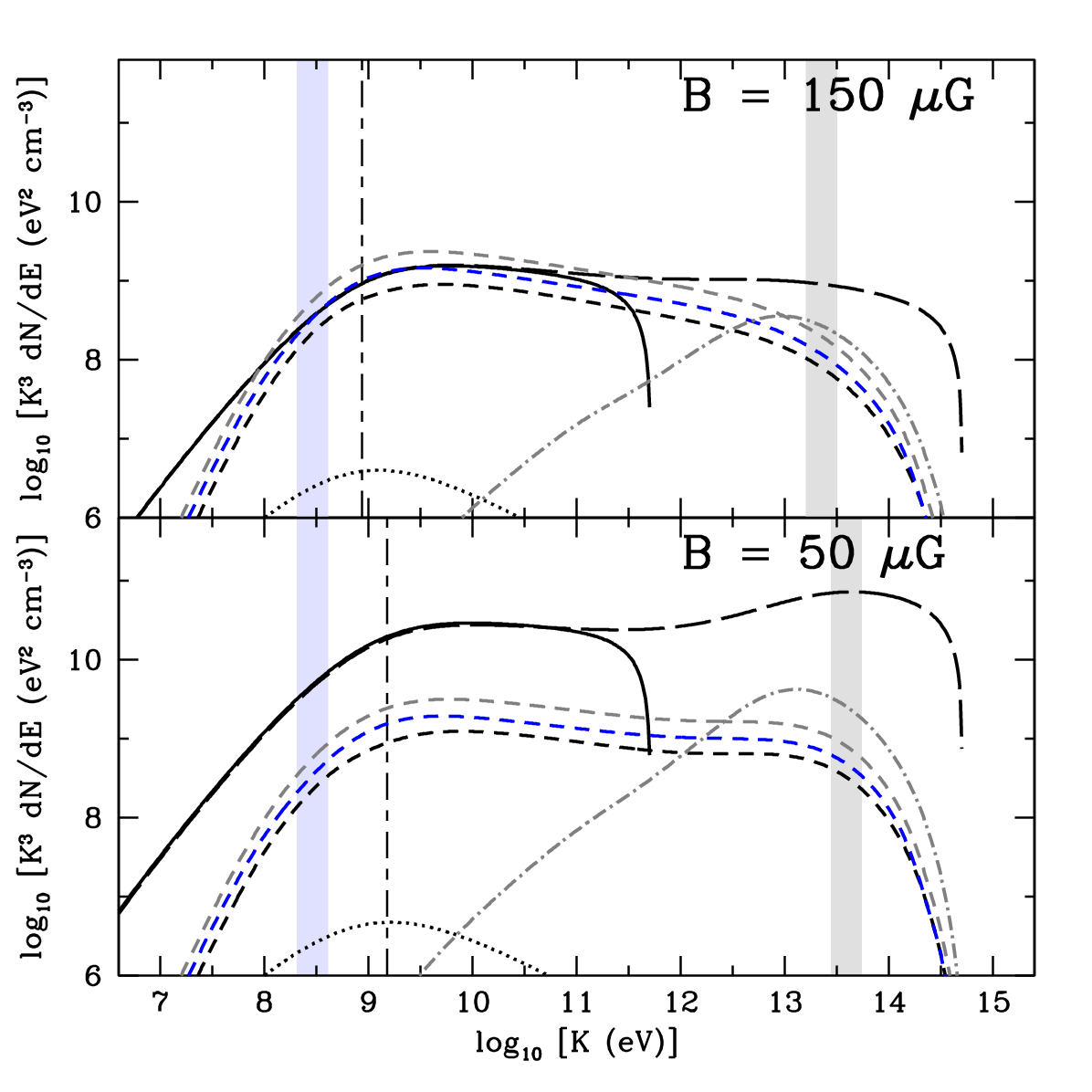}\includegraphics[width=9cm]{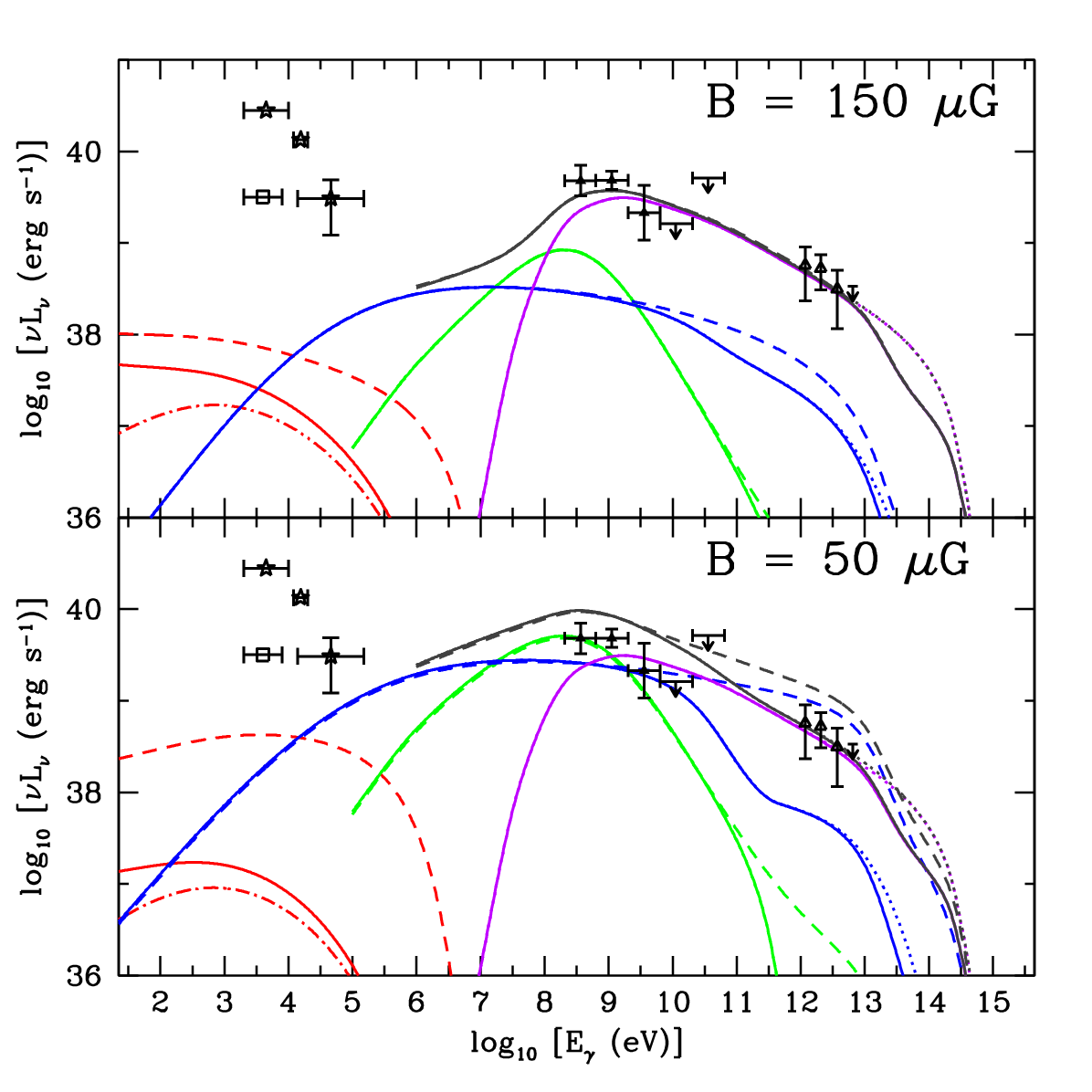}}
\figcaption[simple]{The CR electron (\emph{left}) and photon (\emph{right}) spectra of M82 for models with $p = 2.2$, $t_{\rm diff} (\GeV) = 10^7\ \yr$, and $\eta = 0.1$.  The line and shading styles are the same as Fig.~\ref{fig:N253Spectra}.  Synchrotron emission can dominate IC emission at energies less than a few keV in high $B$ models.  Data plotted on right are Fermi and VERITAS (triangles), as well as the diffuse emission from Chandra (square) and total emission from BeppoSAX \citep{Cappi99}, Suzaku \citep{Miyawaki09}, and Swift \citep{Cusumano10} (stars). X-ray luminosities are scaled to one ln bin in energy assuming $\Gamma = 2.0$.  \label{fig:M82Spectra}}
\end{figure*}

\emph{Modelled Radio and $\gamma$-rays} -- The GHz synchrotron radio spectrum combined with the GeV and TeV $\gamma$-ray detections leaves us with the parameter space in Table~\ref{table:M82XRayLuminosities}.  The predicted synchrotron radio emission tends to be too large below a GHz (as shown in Figure~\ref{fig:M82Radio}); this may be due to free-free absorption \citep{Klein88}.  Conversely, the synchrotron radio spectrum falls off steeply at high frequencies, with $\alpha \approx 0.8 - 0.9$ instead of the observed $0.7$ \citep{Klein88,Williams10}.  In order to accommodate the observed spectrum, our fitting requires higher 1 GHz thermal fractions (8 - 21\%; with a fiducial value of 15\%) than derived in \citet{Williams10} ($6 \pm 2\%$).  These high thermal fractions overproduce the observed total radio emission at 20 - 100 GHz.  The underlying synchrotron spectrum is therefore probably flatter, which could occur with higher bremsstrahlung, ionization, or escape losses.  When $B \ge 150\ \muGauss$, the GeV $e^{\pm}$ spectrum is dominated by pionic secondaries (short-dashed lines in left panel of Figure~\ref{fig:M82Spectra}), and the $\ge 100\ \MeV$ emission is almost entirely pionic $\gamma$-rays (purple line in right panel of Figure~\ref{fig:M82Spectra}).  These models have low $\xi$, and thus relatively little power going into primary electrons.  When $B \le 100\ \muGauss$, however, the GeV $e^{\pm}$ spectrum is mostly primary electrons (long-dashed line in Figure~\ref{fig:M82Spectra}); these models usually have much higher $\xi$ ($\tilde\delta \la 10$ for $p = 2.2$) than the Milky Way.

As with NGC 253, the GeV $\gamma$-ray emission is primarily leptonic when $B = 50\ \muGauss$ and hadronic when $B = 150 - 200\ \muGauss$.  The predicted GeV $\gamma$-ray flux is already at least 40\% higher than observed in the $B = 50\ \muGauss$ models because of the strong bremsstrahlung and IC emission; therefore, M82's magnetic field is probably greater than $50\ \muGauss$.  Indeed, most of these models only work by either cutting off the primary $e^{\pm}$ spectrum so that the IC emission doesn't greatly overproduce the observed TeV emission, or by having very low $\eta$ so there's virtually no pionic emission and all of the TeV emission is leptonic.  Thus, the TeV detection of M82 place constraints on the high energy electron spectrum when $B$ is low.  Conversely, $B$ is unlikely to be much higher than $200\ \muGauss$ (if our assumptions about the gas density and radiation field are correct), or else the secondary $e^{\pm}$ that accompany the pionic $\gamma$-rays produce too much radio emission.  These values for $B$ are in accord with previous modeling for M82's starburst \citep{Persic08,deCeaDelPozo09a}.  The efficiency of proton acceleration can be small in the low $B$ leptonic models, but when $B \ge 100\ \muGauss$ and $p = 2.2$, $\eta$ is of order 0.1, the usual value assumed in models of star-forming galaxies.  

In our models, M82 is slightly more proton calorimetric than NGC 253, with $F_{\rm cal} \approx 10 - 50\ \%$ integrated over all energies.  At the energies above 10 TeV relevant for synchrotron X-ray production, $F_{\rm cal}$ is only $\sim 11\%$ when $t_{\rm diff} (3\ \GeV) = 10\ \Myr$ and $\sim 1\%$ when $t_{\rm diff} (3\ \GeV) = 1\ \Myr$.

\emph{Synchrotron X-rays} -- In Table~\ref{table:M82XRayLuminosities}, we list the predicted X-ray synchrotron emission for a variety of model parameters.  Overall, the results are similar to those for NGC 253: the diffuse synchrotron fraction varies from 0.4\% to 54\% but has a fiducial value of $\sim 2\%$ ($L_{2-8}^{\rm synch} = 9.7 \times 10^{37}\ \ergps$; assuming $\Gamma \approx 2.0$, $L_{2-10}^{\rm synch} \approx 5 \times 10^{-7} L_{\rm TIR}$).  For $p = 2.2$ models, the synchrotron fraction is typically $\sim 15\%$ for low $B$ models with a high energy primary $e^{\pm}$ cutoff, $0.8 - 5\%$ for high $B$ models, and a fraction of $1\%$ for low $B$ models with a low energy primary $e^{\pm}$ cutoff.  Like NGC 253, most of the models at the high end of the range of synchrotron fraction have TeV luminosities that are $>1.5$ times higher than observed.  

Interestingly enough, the spectral slope of synchrotron emission in the 2-8 keV energy range ($\Gamma \approx 1.98 - 2.34$) also matches the observed diffuse hard X-ray emission ($\Gamma \approx 2 - 3$).  

\emph{Inverse Compton X-rays} -- The IC fraction of the diffuse X-ray emission is largest (4 - 10\%) in the $B = 50\ \muGauss$ models and becomes smaller as $B$ increases: 1.4 - 2.4\% when $B = 100\ \muGauss$, 0.8 - 1.1\% when $B = 150\ \muGauss$, 0.6\% when $B = 200\ \muGauss$.  This confirms the expectation in \citet{Strickland07} that IC does not make up the observed diffuse hard X-ray emission in M82.  Because of the poorly constrained synchrotron X-ray fraction, the synchrotron-to-IC ratio is anywhere between 0.05 and 36, with a fiducial value of 2.3.  We find that the IC emission has $1.23 \le \Gamma_{2 - 8} \le 1.36$, significantly harder than the observed spectral slope of the diffuse hard emission in M82.  

\emph{Conclusion} -- The synchrotron fraction of M82's diffuse hard X-ray emission is largely unknown, even with the GeV and TeV detections, but is of order $2\%$ in our fiducial model.  The unknown primary electron energy cutoff is one of the largest contributors to the uncertainty.  IC is a minority ($0.5 - 10\%$) of the diffuse hard X-ray emission, with its contribution depending mainly on magnetic field strength.  The synchrotron-to-IC ratio in our models is $\sim 2$ in our fiducial model.

Unlike NGC 253, foreground hard X-ray emission from the host galaxy has already been removed, and the extant studies with \emph{Chandra} already take into account the bright point sources.  However, the diffuse hard X-ray emission has a different morphology than the radio and IR disks (and therefore, the expected sources of CRs) with $h \approx 175\ \pc$.  Restricting attention to the radio disk would help studies of the nonthermal contribution.  Furthermore, the IC contribution might be identified spectrally since it has a hard spectrum.  Finally, the models with the largest synchrotron X-ray contributions are dominated by primaries; so large synchrotron fractions could be tested by looking for concentrations of X-ray emission near CR accelerators.

\subsection{Arp 220}
\label{sec:Arp220}
\emph{Introduction} -- Arp 220 consists of a starburst disk containing most of the gas and possibly most of the star-formation, and two intense but smaller starburst nuclei \citep{Downes98}.  While there is radio data for the individual starbursts, and the galaxy as a whole, there are no flux values in the literature for the starburst disk specifically.  We therefore consider the nuclear starbursts separately and ignore the surrounding disk.  We adopt $D = 79.9\ \Mpc$ from \citet{Sanders03}.  The bolometric luminosity of Arp 220 is $1.6 \times 10^{12}\ \Lsun$ \citep{Sanders03}; if it is entirely due to star-formation, this would imply a supernova rate of $2\ \yr^{-1}$.  Estimates of the bolometric luminosities of the nuclear starbursts vary, but we adopt $3 \times 10^{11}\ \Lsun$ for each, based off \citet{Downes98} and consistent with \citet{Sakamoto08}.  These imply supernova rates of $0.34\ \yr^{-1}$ in each nucleus.  We note that some authors argue that the bolometric luminosity of Arp 220's western starburst is dominated by an AGN, based on its compactness, which would reduce the IR luminosity attributable to star-formation \citep{Downes07}.  On the other hand, the actual supernova rates may in fact be up to near $3\ \yr^{-1}$ in the western starburst and $1\ \yr^{-1}$ in the eastern starburst, based on the appearance of radio-bright supernovae (\citealt{Lonsdale06}; earlier, \citealt{Rovilos05} found $0.7\ \yr^{-1}$).  According to \citet{Lonsdale06}, the radio sources in Arp 220's nuclear starbursts are concentrated into regions $\sim 100\ \pc$ wide, and the radio images of the nuclear starbursts indicate diameters of 80 - 120 pc.  Following \citet{Sakamoto99}, we assume that the nuclear starbursts are small disks, each with radius and scale height 50 pc.  The radiation energy density in each nuclear starburst is then $1.6 \times 10^5\ \eV\ \cm^{-3}$ ($B_{\rm rad} = 2.5\ \mGauss$).  The gas mass in each disk is $\sim 10^9\ \Msun$ within a 100 pc radius of each \citep{Downes98,Sakamoto99}, for gas surface densities of $7\ \gcm2$.  Given the uncertainties in the gas mass and distribution, we use $\Sigma_g = 10\ \gcm2$.  The magnetic field strength in Arp 220 is not known, but it is very large.  Estimates range from the minimum energy value of $0.3\ \mGauss$ to $30\ \mGauss$, the maximum allowed by hydrostatic balance \citep{Thompson06}.  \citet{Torres04} found that $B \approx \mGauss$ in the main disk, but $B \approx 5\ \mGauss$ in the starburst nuclei.  We run a grid of models in $B$ for values of $0.25\ \mGauss$ to $16\ \mGauss$ for each nucleus.

Unlike the Galactic Center, NGC 253, and M82, Arp 220 has no $\gamma$-ray detection in either GeV or TeV bands.  For a $\sim\GeV$ upper limit, we use the $\ge 100\ \MeV$ flux limit for $5\sigma$ sources in the \emph{Fermi}-LAT one year catalog \citep{Abdo10b}.  \citet{Albert07} provide upper limits in the VHE range for Arp 220 (specifically, an integrated photon flux from 0.36 to 1.8 TeV).  To keep the parameter space manageable, the CR proton acceleration efficiency $\eta$ is assumed to be a standard value of $0.1$.

The X-ray emission of Arp 220 has been studied by \emph{Chandra} and XMM-Newton \citep{Iwasawa01,Clements02,McDowell03,Iwasawa05}.  XMM-Newton has detected a Fe K line, suggesting at least some thermal emission \citep{Iwasawa05}; but there is also a hard X-ray continuum of unknown origin.  There are several diffuse components, including a nuclear source roughly cospatial with the western nuclear starburst (Arp 220 X-1), a ``hard halo'' with a radius of a few hundred parsecs, a circumnuclear halo extended on kiloparsec scales, and vast ``plumes'' stretching out 10 kpc from the galactic center \citep{Clements02,McDowell03}.  The central unresolved source Arp 220 X-1 is approximately coincident with the western nucleus, with an unabsorbed 2-10 keV luminosity estimated at $4 \times 10^{40}\ \ergps$ \citep{Clements02}.  A second X-ray source, Arp 220 X-4 ($L_{2 - 10} \approx 1.5 \times 10^{40}\ \ergps$, after correcting for absorption), is tentatively identified with the eastern nucleus, although the geometry does not exactly match the radio morphology.  For this paper, we assume that Arp 220 X-1 is the western starburst and Arp 220 X-4 is the eastern starburst; in Table~\ref{table:A220XRayLuminosities}, we list the (unabsorbed) 2 - 10 keV synchrotron luminosities so they can be compared with other values for the luminosity.

\begin{figure}
\centerline{\includegraphics[width=8cm]{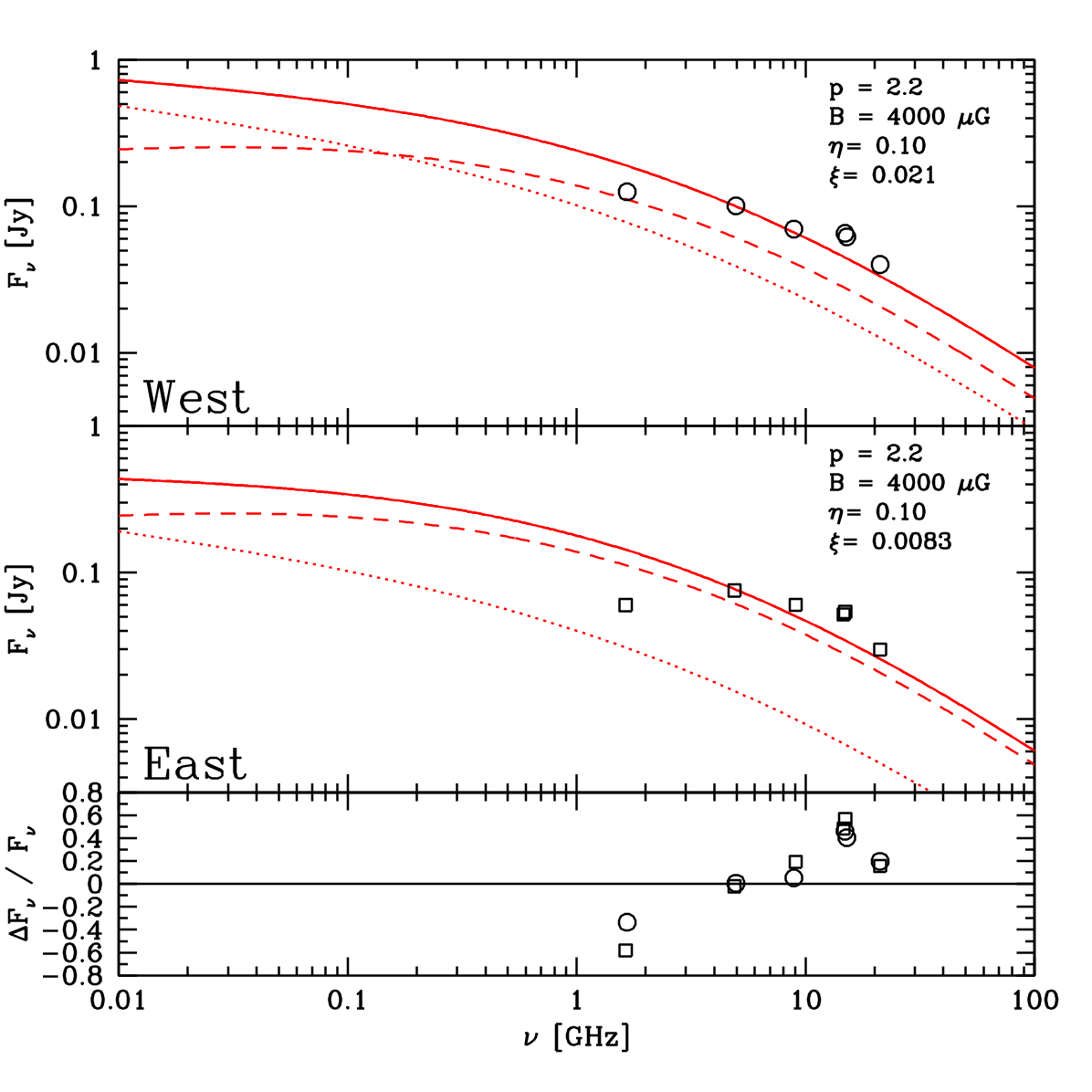}}
\figcaption[figure]{Our predicted synchrotron radio spectrum (red, solid) compared to observations compiled in \citet{Torres04}.  The open circles are for the western nucleus and the open squares are for the eastern nucleus.  The line and shading styles are the same as in Figure~\ref{fig:GalCenterRadio}.\label{fig:A220Radio}}
\end{figure}

\emph{Modelled Radio and $\gamma$-rays} -- We show an example radio spectrum for the western and eastern nuclei in Figure~\ref{fig:A220Radio}.  Secondaries dominate the GHz synchrotron emission for $B \ge 4\ \mGauss$, while primaries dominate for the lower magnetic field strengths.   As with M82 and NGC 253, the radio data for each nucleus appears to be flatter than predicted in our models.  The discrepancy at $\sim 10 - 20\ \GHz$ becomes larger in low $B$ models, since we are then probing the higher energy (more IC and synchrotron cooled) parts of the $e^{\pm}$ spectrum.  However, it is possible that a thermal emission component flattens the radio spectra.  The thermal dust contribution is expected to be important only at 100 GHz and above \citep{Downes98,Torres04}.  Furthermore, the 15 GHz data points appear systematically high for both the west and east nuclei, and are high even in the models of \citet{Torres04}.  As \citet{Torres04} note, the radio data use different beam sizes at different frequencies.  Finally, it is possible that bremsstrahlung and ionization are even stronger than expected.  In order for advective escape to be strong enough to flatten the spectrum, wind escape times would have to be $\la 1000\ \yr$ (compare eqns.~\ref{eqn:tBrems} and \ref{eqn:tWind}), requiring speeds of $\gg 10^4\ \kms$, which is unlikely.  

The non-detection of either nuclear starburst by \emph{Fermi} already implies that $B \ge 250\ \muGauss$ in each of them, otherwise the leptonic emission alone would be observable.  Note that we applied the GeV constraint to each nucleus independently; considering them together would push the lower limit on $B$ towards $\sim 500\ \muGauss$.  With more data, \emph{Fermi} will be able to constrain the leptonic $\gamma$-ray emission further and increase the lower limit on $B$. Furthermore, note that in the $B = 500\ \muGauss$ and $B = 1\ \mGauss$ models, we require that more power goes into CR electrons than protons.  This is unlikely, considering that the reverse situation holds for the Milky Way.

In all of our models, the nuclear starbursts of Arp 220 are proton calorimeters, with $F_{\rm cal} = 0.84 - 0.98$.  Even at energies above 10 TeV, where diffusion is strongest, $F_{\rm cal} \approx 0.6$ when $t_{\rm diff} (3\ \GeV) = 1\ \Myr$.  

\emph{Synchrotron X-rays} -- In the extremely dense gas of Arp 220, the CR protons at 10 - 100 TeV are efficiently converted into pionic products, including secondary $e^{\pm}$ and $\gamma$-rays.  In turn, the extreme FIR radiation field converts 10 - 100 TeV $\gamma$-rays into pair $e^{\pm}$.  Finally, the secondary and tertiary $e^{\pm}$ emit prodigious synchrotron X-rays.  Models with $p \approx 2.0$ and slow diffusive escape can explain most of the observed X-ray emission from the nuclei as synchrotron.   Without any significant TeV $\gamma$-ray constraints on $p$, we cannot rule out these models: it is possible that \emph{all} of the hard X-ray emission from Arp 220 is synchrotron (though, see the later caveats about X-ray absorption).  

For $p = 2.2$, the allowed synchrotron fractions are $0.9 - 42\%$ (fiducial: 15\%) for the western starburst and $2 - 84\%$ (fiducial: 34\%) for the eastern starburst.  In high $B$ models ($\sim 4\ \mGauss$) with $p = 2.2$, the synchrotron fraction is $\sim 10\%$ for the western starburst and $\sim 30\%$ for the eastern starburst, consistent with previous expectations that HMXBs dominate the hard X-ray emission of starbursts.  However, synchrotron is not so minor that it can be ignored, especially considering the uncertainties.  Lower $B$ models have a larger synchrotron fraction if $\gamma_{\rm max}^{\rm prim} = 10^9$ and a smaller fraction if $\gamma_{\rm max}^{\rm prim} = 10^6$.  At still higher $p = 2.4$, synchrotron X-ray emission is reduced to a few percent of the observed diffuse X-ray flux in each starburst.  
 
\begin{figure*}
\centerline{\includegraphics[width=9cm]{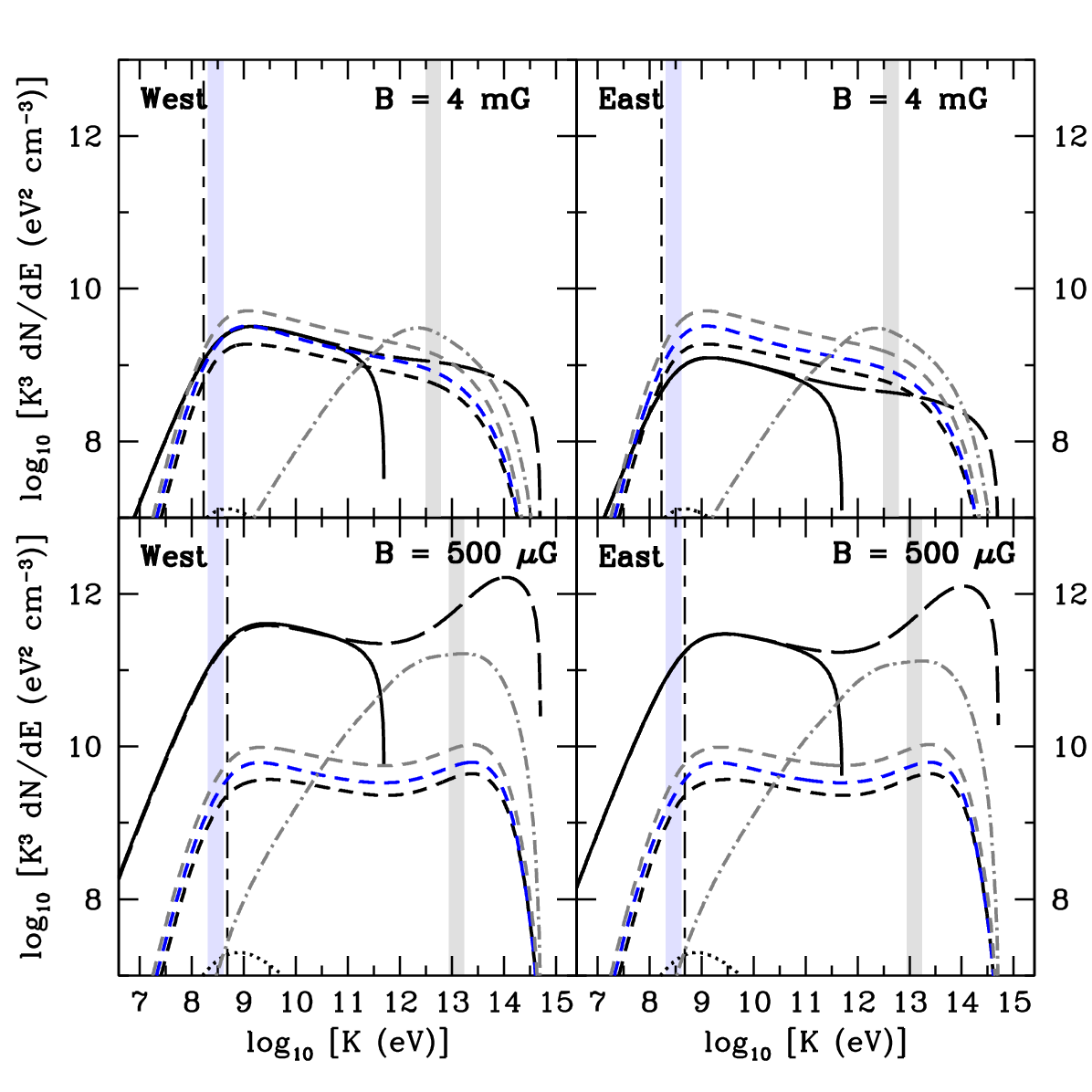}\includegraphics[width=9cm]{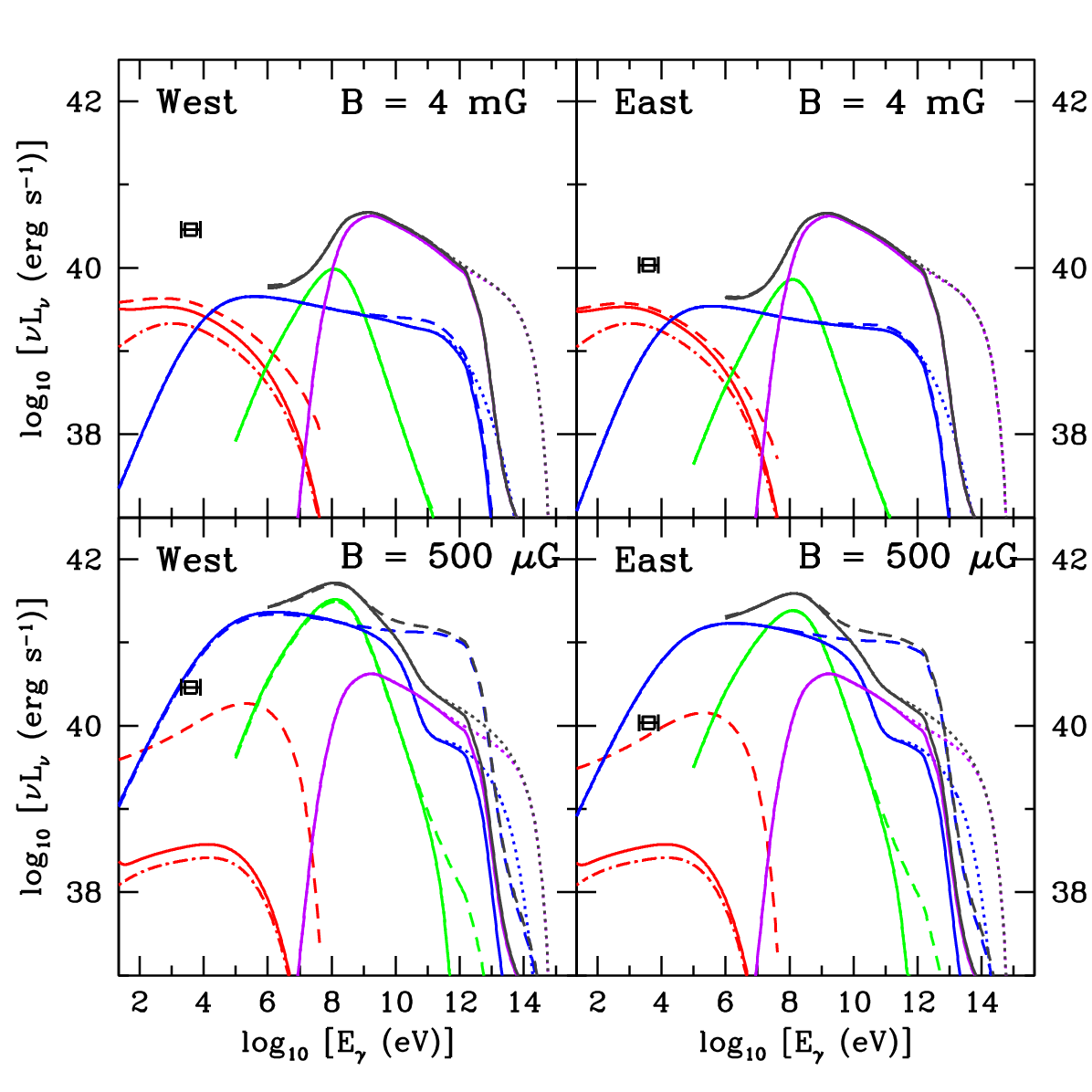}}
\figcaption[simple]{The CR electron (\emph{left}) and photon (\emph{right}) spectra of Arp 220 for models with $p = 2.2$ and $t_{\rm diff} (3\ \GeV) = 10^7\ \yr$.  The line styles and colors are the same as in Figure~\ref{fig:N253Spectra}.  The effects of neutral hydrogen absorption are not shown on right.  Observed absorption-corrected luminosities from Chandra are plotted, scaled to one ln bin in energy assuming $\Gamma = 2.0$.  Synchrotron emission dominates IC emission at energies less than $\sim 10\ \keV$ in high B models.\label{fig:A220Spectra}}
\end{figure*}

We find that X-ray synchrotron emission in Arp 220's nuclear starburst is harder than in the other starbursts, with $\Gamma_{2-10} \approx 1.84 - 2.23$.  The hardest synchrotron spectra arise in low $B$ models, where the Klein-Nishina bump is very prominent.  With $B$ greater than in M82 and NGC 253, lower electron energies are being probed (grey shading in left panel of Figure~\ref{fig:A220Spectra}), where the cutoff in the proton spectra does not matter as much.  To compare, the spectral slope of the diffuse hard X-ray emission in Arp 220 is not well constrained observationally, but is probably harder than synchrotron emission: \citet{Clements02} measured $\Gamma = 1.4 \pm 1$, while \citet{Iwasawa01} measured $\Gamma = 1.7$ for the total X-ray emission of Arp 220, and \citet{Iwasawa05} measured $\Gamma = 1.2^{+0.4}_{-0.7}$, though for a Galactic absorption column.

\emph{Inverse Compton X-rays} -- The contribution of Inverse Compton emission depends on the magnetic field strength.  The contributions are highest in the low $B$ models: for $B = 500\ \muGauss$, IC is 1.6 - 2.8 times the unabsorbed 2 - 10 keV luminosities in the western starburst and 3.1 - 5.7 times the eastern starburst's luminosity.  However, the actual X-ray absorption is uncertain, so these cannot be said to constrain the magnetic field with any certainty.  At the other extreme of $B = 4\ \mGauss$, IC makes up $\sim 6 - 7\%$ (western) and $12 - 14\%$ (eastern) of the unabsorbed 2 - 10 keV luminosity.  Thus, IC emission in our models can be much more inefficient than usually expected \citep[c.f.][]{Iwasawa01}.  This is because $B$ is allowed to be much higher than the usual minimum-energy estimate; with the higher magnetic field strengths, IC is much weaker when scaling from the synchrotron radio emission.

The synchrotron luminosity can be anywhere from 0.05\% to 57 times the IC emission in Arp 220's western nucleus and 0.06\% to 35 times the IC emission in Arp 220's eastern nucleus, an even greater range than in the other starbursts because $p$ is not yet constrained by TeV $\gamma$-ray observations.  Restricting ourselves to $p = 2.2$ models narrows the range ($0.4 - 240\%$ for the western nucleus; $0.6 - 280\%$ for the eastern nucleus), with fiducial values for the synchrotron-to-IC ratio of $2.3$ for the western nucleus and $2.7$ for the eastern nucleus. 

Like the other starbursts, IC emission is predicted to be harder than synchrotron emission, with $\Gamma_{2-10} \approx 1.30 - 1.69$, which is a closer match to the observed X-ray emission.  It is possible, then, that Arp 220 has low magnetic fields of less than a milliGauss and its X-ray emission is IC emission.  However, both HMXBs and thermal emission are also expected to have the same spectral shape and dominate the X-ray spectrum \citep{Persic02,Iwasawa05}.  

\emph{Conclusion} -- Arp 220 is extremely efficient at converting VHE CR proton energy into synchrotron X-rays.  Without the TeV $\gamma$-ray detections that exist for the Galactic Center, NGC 253, and M82, we cannot rule out hard spectra extending to very high energies.  Therefore, it is entirely possible to generate or even exceed the observed absorption-corrected X-ray luminosities of Arp 220 X-1 and Arp 220 X-4 (assumed to be the western and eastern nucleus, respectively).  Therefore, \emph{if} the unabsorbed luminosities are correct, we can constrain $p = 2.0$ electron spectra extending to multi-TeV energies based on the X-rays alone.  Even in $p = 2.2$ models, synchrotron can easily be $\sim 10\%$ or more of the X-ray luminosities of each starburst.  

A large uncertainty in our models of Arp 220 is the amount of hydrogen absorption.  A surface density of $10~\gcm2$ implies $N_H \approx 6 \times 10^{24}\ \cm^{-2}$, which is Compton thick.  If this is correct, then the true X-ray luminosity of Arp 220 may be much higher.  By contrast, the X-ray emission indicates hydrogen column depths of $N_H \approx 3 \times 10^{22} \cm^{-2}$ \citep[$\Sigma_g = 0.05\ \gcm2$;][]{Clements02}.  It is possible, however, that CR $e^{\pm}$ emitting synchrotron X-rays are not cospatial with the nuclear starbursts themselves, either instead diffusing out of the nuclei (though this is unlikely given the extremely rapid losses in the extreme environments; \citealt{Voelk89b,Condon91}) or are pair-produced from $\gamma$-rays outside the nuclei, in which case hydrogen absorption may be relatively small.  It is also possible that the covering fraction of the hydrogen is less than unity, in which case, some X-ray emission will be able to escape.  Hydrogen absorption will also affect other explanations of the diffuse X-ray flux in Arp 220, such as IC emission and high-mass X-ray binaries.  Indeed, it might explain why Arp 220 seems underluminous in hard X-rays for its star-formation rate, as noted by \citet{Iwasawa05} and supported by \citet{Lehmer10}.

\section{Discussion}
\label{sec:Discussion}

\subsection{Synchrotron, the FIR-X-ray Correlation, and Submillimeter Galaxies}
\label{sec:SMGs}
Star-forming galaxies show a correlation between their FIR luminosities and their 2 - 10 keV X-ray luminosities, with $L_{2-10} \approx 10^{-4} L_{\rm FIR}$ when AGNs do not contribute to the X-ray emission \citep{David92,Grimm03,Persic04,Persic07,Lehmer10}.  The correlation holds better for the most intense starbursts like ULIRGs, where there is less infrared and X-ray emission from older stellar populations \citep{Persic07}.  X-ray luminosity has often been used as a star-formation rate indicator \citep{Ranalli03,Grimm03,Persic04}.  The X-ray luminosity is often attributed to HMXBs, compact objects accreting matter from a massive companion.  HMXBs are expected to trace the recent stellar population.  Is it possible that synchrotron accounts for much of the X-ray luminosity in some galaxies?

Our models show a trend, with increasing synchrotron X-ray emission with increasing gas density: synchrotron is a tiny fraction of the Galactic Center's X-ray luminosity if primaries cut off at low energies, a small portion of M82's diffuse hard X-ray luminosity, and possibly a substantial fraction of Arp 220's hard X-ray luminosity.  Three basic quantities affect the luminosity from secondary and pair $e^{\pm}$: (1) the amount of gas, which CR protons collide with to create pionic secondaries; (2) the amount of star-formation, which creates IR photons that pair produce off VHE $\gamma$-rays, and which is generally correlated with the amount of gas through the Schmidt Law; and (3) the physical size of the starburst.

At the extreme end of synchrotron X-ray luminosity from secondary and pair $e^{\pm}$, it is worth considering the diffuse synchrotron emission from submillimeter galaxies (SMGs), well-studied extreme starbursts at high redshift ($z \ga 2$).  SMGs contain large amounts of gas, so they should be relatively efficient at converting energy in CR protons into $\gamma$-rays and secondary $e^{\pm}$.  Furthermore, they are much more extended spatially than the more compact starbursts observed at low redshift \citep{Chapman04,Tacconi06,Genzel08,Law09,Younger10}.  This means that both diffusive and advective escape will be less effective at transporting CR protons out of SMGs, simply because they have to travel farther, further supporting the case for proton calorimetry at multi-TeV energies \citep{Lacki10b}.  Finally, the radiation energy density in SMGs is very high, and because the SMGs are ``puffy'' with long sightlines through them, the optical depth to VHE $\gamma$-rays will be high.  Therefore, we may expect that SMGs, like Arp 220, will be very efficient at converting energy in VHE CR protons into synchrotron X-ray emission with $L^{\rm synch}_{\rm X} > 10^{-6} L_{\rm SF}$.  We can evaluate how good SMGs are at creating secondaries with the estimates in \S~\ref{sec:ProtonLosses}.  \citet{Tacconi06} find that SMGs have typical gas surface densities of $\Sigma_g \approx 0.4~\gcm2$.  Using a typical midplane-to-edge scale height $h$ of 1 kpc, we find densities of $\mean{n} \approx 40\ \cm^{-3}$, comparable to those found by \citet{Tacconi06}.  From eqn.~\ref{eqn:tPion}, the typical pionic loss time in SMGs is $t_{\rm pion} \approx 1.3\ \Myr$.  Eqn.~\ref{eqn:tWind} gives us advective loss times of $t_{\rm wind} \approx 3.0\ \Myr\ (h / \kpc) (v_{\rm wind} / 300\ \kms)^{-1}$.  The ratio of these times suggests that SMGs are proton calorimeters, depending on the unknown diffusive escape time and to a lesser extent on wind speeds (though they would have to be $\ga 700\ \kms$ to reduce $F_{\rm cal}$ significantly).

To study this effect, we modeled generic galaxies of several gas surface densities from $\Sigma_g = 0.01~\gcm2$ to $30~\gcm2$.  The magnetic fields were chosen to scale as $B = 6\ \muGauss (\Sigma_g / 0.0025~\gcm2)^{0.7}$, in line with our previous work on the radio emission of star-forming galaxies \citep{Lacki10a}.  The star-formation rate was calculated using the Schmidt law in \citet{Kennicutt98}.  The interstellar radiation field was taken to consist of the CMB, a $T = 10000\ \Kelv$ greybody field corresponding to the unobscured starlight from young stars, and a greybody corresponding to dust-obscured starlight (40 K in the compact starbursts and 30 K in the puffy starbursts).  Scale heights of $h = 100\ \pc$ and $1000\ \pc$ were modeled.  We chose $\eta = 0.1$ and varied $\xi$ from 0.003 to 0.03; several values of $t_{\rm diff} (3\ \GeV)$ and $\gamma_{\rm max}^{\rm prim}$ were tried.

\begin{figure}
\centerline{\includegraphics[width=8cm]{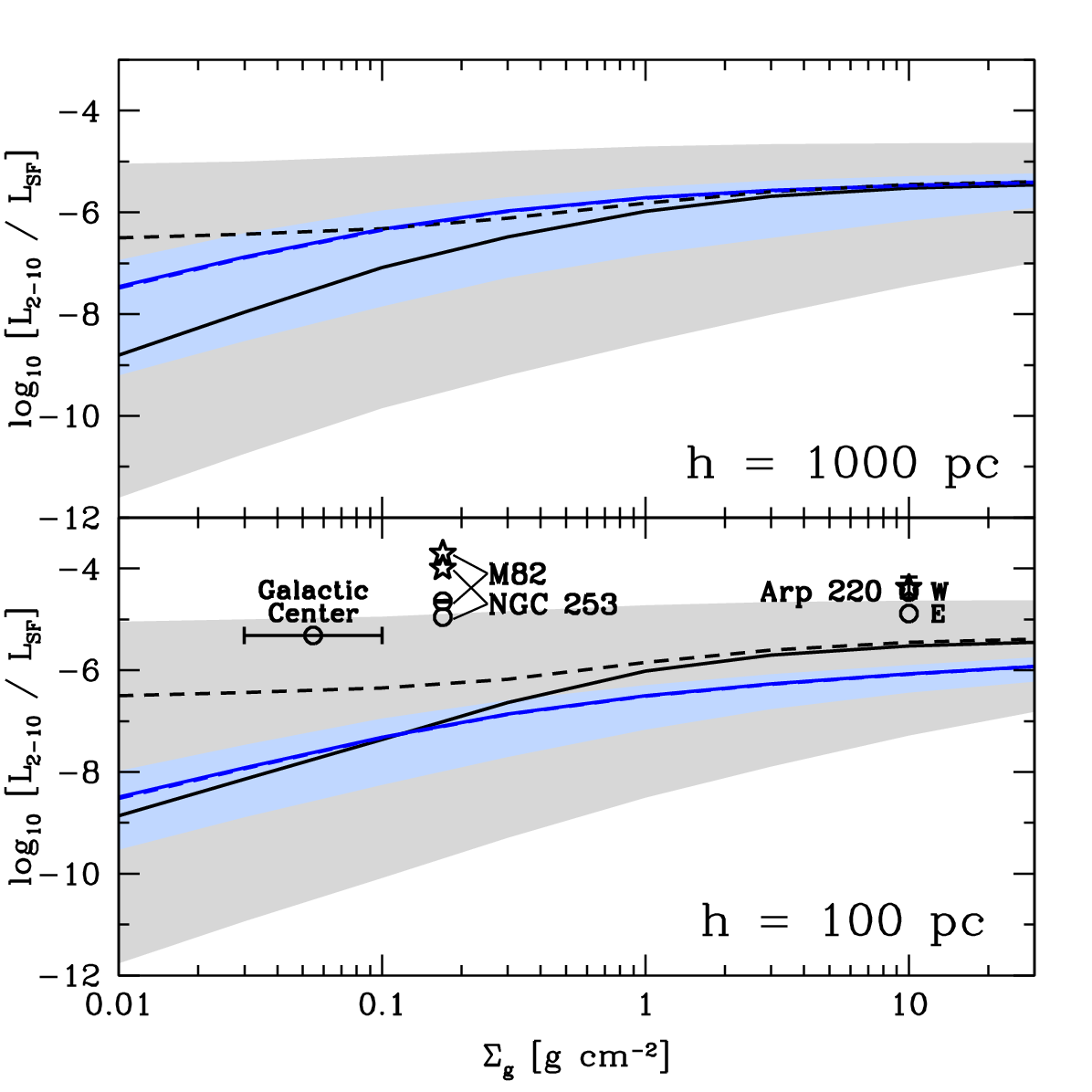}}
\figcaption[simple]{Fraction of the bolometric flux from star-formation $L_{\rm SF}$ in 2 - 10 keV synchrotron (black/grey) and IC (blue) emission.  On the bottom, we show $h = 100\ \pc$; the top shows $h = 1000\ \pc$, as is the case for SMGs. The lines indicate $p = 2.2$, $\xi = 0.01$, and $t_{\rm diff} (3\ \GeV) = 10\ \Myr$ for compact starbursts and $t_{\rm diff} (3\ \GeV) = 100\ \Myr$ for puffy starbursts.  Solid is $\gamma_{\rm max}^{\rm prim} = 10^6$, while dashed is $\gamma_{\rm max}^{\rm prim} = 10^9$.  Shading indicates possible values for other values of those parameters.  For a typical star-formation hard X-ray luminosity $L_{2 - 10} = 10^{-4} L_{\rm SF}$, synchrotron and IC are subdominant; for optimistic choices of parameters, synchrotron can be $10^{-5} L_{\rm SF}$.  The circles represent the diffuse X-ray emission for the studied starbursts, while the stars represent the total (point sources included) X-ray emission that would be seen at large distances (from the \citealt{Cappi99} fluxes for M82 and NGC 253, scaled to our distances; from \citealt{McDowell03} for Arp 220). \label{fig:BolometricFraction}}
\end{figure}

Our models give the ratio of both synchrotron and IC emission to the bolometric power from star-formation at 2-10 keV, and these are plotted in Figure~\ref{fig:BolometricFraction}.  Both components are always a minority of the standard hard X-ray luminosity, $10^{-4} L_{\rm SF}$, from the FIR-X-ray correlation.  Both the synchrotron and IC fraction increase with $\Sigma_g$.  IC shows less variation in its flux, since it mostly comes from the relatively well-constrained low energy CR electron spectrum, although $L_{2-10}^{\rm IC}$ still varies by a factor of $3 - 35$ for a given $\Sigma_g$ when $h = 100\ \pc$ ($5 - 180$ for $h = 1000\ \pc$).  At the highest $\Sigma_g$, the IC emission is $2\%$ or less of the typical $2 - 10\ \keV$ luminosity for $h = 100\ \pc$ ($6\%$ or less when $h = 1000\ \pc$), falling to $\la 0.1\%$ for $\Sigma_g = 0.01~\gcm2$ (blue shading).

Synchrotron X-ray emission shows a much greater variation with different parameters (see grey shading), ranging from $\la 0.1\%$ (assuming $p = 2.4$ or that $\Sigma_g$ is small) to $24\%$ of the 2 - 10 keV luminosity of typical starbursts.  Models with high $\gamma_{\rm max}^{\rm prim}$ are much more efficient synchrotron radiators at low $\Sigma_g$.  At high $\Sigma_g$, however, the synchrotron fraction starts to converge, as CR proton energy is efficiently converted to pions and $\gamma$-ray photons are efficiently converted to $e^{\pm}$ pairs.  For our fiducial values of $p = 2.2$ and $t_{\rm diff} (3\ \GeV) = 10\ \Myr$, $L_{2-10}^{\rm synch}$ approaches $4\%$ of the total 2-10 keV luminosity in the densest starbursts when $h = 100\ \pc$ (3\% when $h = 1000\ \pc$).  Synchrotron dominates IC for our standard parameters when $\Sigma_g \ga 0.1~\gcm2$ and $h = 100\ \pc$, which are typical of compact starbursts.  Thus, we do not expect synchrotron emission to usually cause large deviations in the X-ray luminosity, although it is present, and at greater levels than IC.

As with the more compact starburst models, the synchrotron fraction of 2-10 keV emission of the puffy starburst models varies over many orders of magnitude with parameters.  For our fiducial $t_{\rm diff} (3\ \GeV) = 10\ \Myr$ from compact starbursts, we actually find that the synchrotron fraction is decreased in puffy starbursts when $\gamma_{\rm max}^{\rm prim}$ is small.  This is because the diffusive escape time is the same, while the density is 10 times lower at a given $\Sigma_g$; the lower pionic yield reduces the amount of secondary $e^{\pm}$ and the pionic $\gamma$-rays that seed pair $e^{\pm}$.  However, for a given diffusion constant, diffusion out of the larger SMGs takes longer.  Setting $t_{\rm diff} (3\ \GeV)$ to 100 Myr restores the synchrotron fraction to near its compact starburst value.  Comparing models with $t_{\rm diff} (3\ \GeV) = 100\ \Myr$ for $h = 1000\ \pc$ and $t_{\rm diff} (3\ \GeV) = 10\ \Myr$ for $h = 100\ \pc$, and $\gamma_{\rm max}^{\rm prim} = 10^6$, there is a $\sim 50\%$ enhancement in the synchrotron emission for $\Sigma_g = 0.03 - 0.3\ \gcm2$, because the $\gamma\gamma$ optical depth is greater along the long sightlines in puffy starbursts.  Given the uncertainties in the parameters and the small contribution of synchrotron at these energies, this is not a big effect.

An interesting effect that occurs in our $h = 1000\ \pc$ models is that the IC fraction actually \emph{increases}.  Unlike the $e^{\pm}$ at multi-TeV energies, the $e^{\pm}$ responsible for 2-10 keV IC emission are relatively low energy ($\sim 100\ \MeV - 1\ \GeV$, emitting synchrotron at $\sim \GHz$ frequencies).  In compact starbursts, these $e^{\pm}$ are cooled significantly by bremsstrahlung and ionization losses, suppressing their IC emission.  In puffy starbursts with the same $\Sigma_g$, however, the gas volume density is lower, so that bremsstrahlung and ionization losses are not as important.  This means that there is more power to go into IC (and possibly synchrotron) losses at these energies, enhancing the X-ray IC (and possibly GHz synchrotron) emission in puffy starbursts relative to compact starbursts \citep[see also][]{Lacki10b}.

In fact, intense starbursts like SMGs, have relatively weak X-ray emission compared to other starbursts, if they lack an AGN, with $L_{X} \approx 10^{-4} L_{\rm SF}$ \citep[][]{Alexander05}.  This suggests that diffuse synchrotron accounts for a few percent of the total hard X-ray emission from SMGs, and not just the diffuse hard X-ray emission.  

\subsection{What Neutrinos and TeV $\gamma$-rays Can Tell Us}
\label{sec:Neutrinos}
There are many uncertainties in our models, leading to a broad range in the synchrotron X-ray luminosities.  These include the hardness of the $e^{\pm}$ spectrum, the strength of $B$ which sets the importance of secondary $e^{\pm}$, and the efficiency of pionic losses at multi-TeV energies.  The synchrotron X-ray emission is enhanced significantly in starbursts where pionic secondary $e^{\pm}$ and pair production $e^{\pm}$ are produced at a high rate, especially if $p \approx 2.0$.  A necessary byproduct of pionic $e^{\pm}$ and $\gamma$-rays is pionic neutrinos.  Starbursts that are bright in neutrinos should therefore be relatively bright in synchrotron X-rays.  Of course, the converse is not necessarily true if the synchrotron X-ray luminosity comes largely from primary $e^{\pm}$.  TeV $\gamma$-rays are also generated by both high energy protons and electrons in the intense environments of starbursts.  While $\sim 10 - 100\ \TeV$ $\gamma$-rays are heavily absorbed by pair-production, there is a window at lower energies below a few TeV (see Figure~\ref{fig:GammaTau}).  Indeed, TeV $\gamma$-rays are the most powerful constraint yet on synchrotron X-ray emission from NGC 253 and M82.

\begin{figure}
\centerline{\includegraphics[width=8cm]{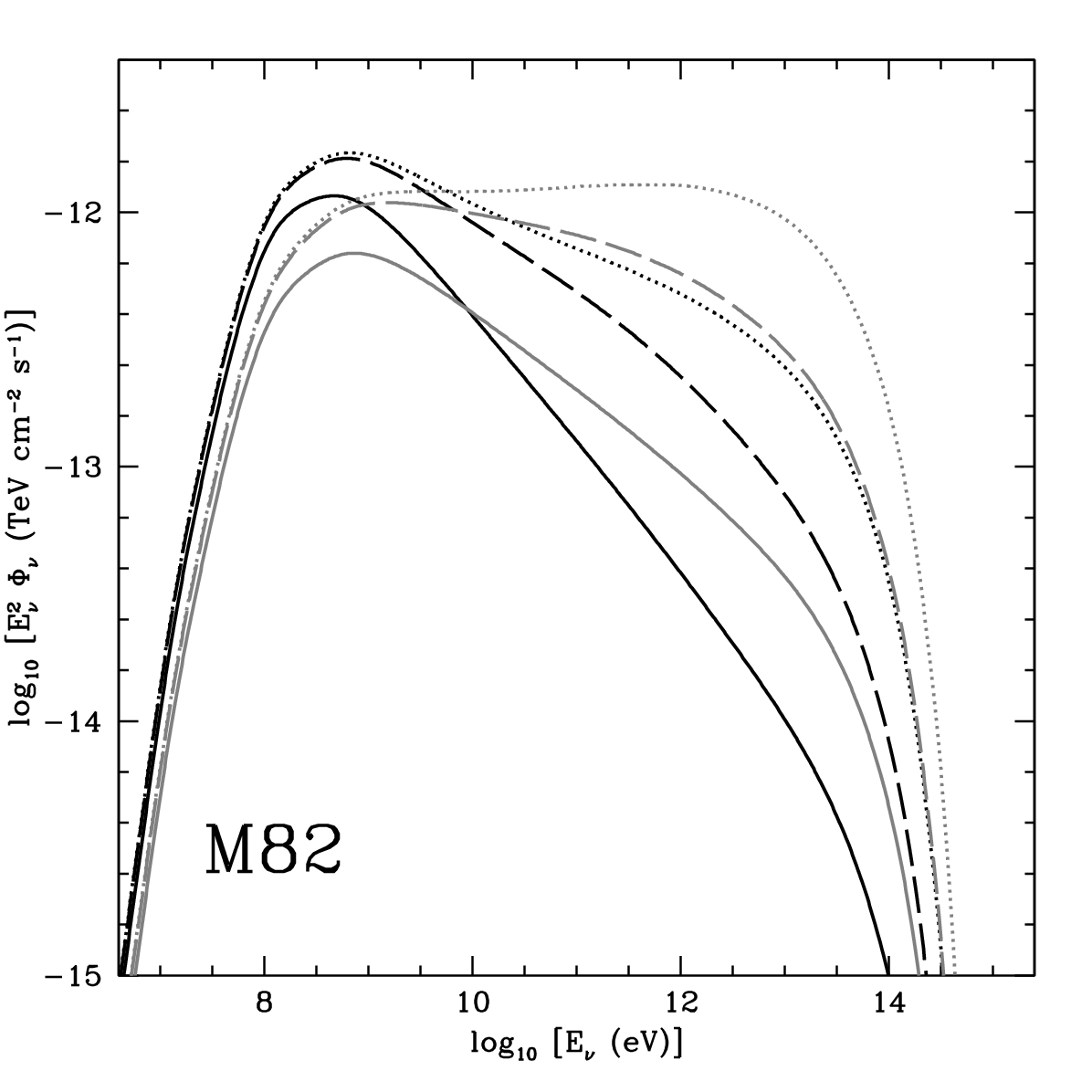}}
\figcaption[figure]{The all-flavor neutrino (plus antineutrino) spectrum of M82 in various models with $\eta = 0.10$. Grey is $p = 2.0$ and black is $p = 2.2$; solid is $t_{\rm diff} (3\ \GeV) = 1\ \Myr$, dashed is $10\ \Myr$, and dotted is $100\ \Myr$.  The neutrino emission directly scales with $\eta$.\label{fig:M82Neutrino}}
\end{figure}

M82 is in the Northern hemisphere and visible to IceCube.  In Figure~\ref{fig:M82Neutrino}, we see that the predicted neutrino spectrum of M82 varies widely between different models.  Models with low $p$ and high $t_{\rm diff}$ naturally predict that M82 is very bright in VHE neutrinos: the more protons at high energies, the more neutrinos are produced.  The magnetic field strength $B$ is also linked to the neutrino flux through the radio and $\gamma$-ray observations.  In models with high $B$, the observed radio flux is mostly secondary $e^{\pm}$, and the accompanying pionic $\gamma$ rays make up almost all of the detected GeV to TeV $\gamma$ rays (as seen in Figure~\ref{fig:M82Spectra}).  The $\gamma$-ray flux therefore translates into a high neutrino flux.  However, in models with low $B$, the observed radio flux is mostly primary $e^{\pm}$, and \emph{leptonic} emission makes up a majority of the GeV (and possibly TeV) $\gamma$-ray flux.  In order not to overproduce the $\gamma$-rays, the CR protons must be less efficiently accelerated, in turn implying lower neutrino flux.

The muon neutrino flux of M82 at 10 TeV is anywhere between $10^{-14}$ and $10^{-12}\ \TeVFUnits$ for $p = 2.0 - 2.2$.   Unfortunately, M82 is at high declination ($\sin \delta \approx 0.93$), where IceCube's sensitivity is relatively weak \citep{Abbasi09}.  Even with a full year of IceCube-80, our most optimistic models predict that M82 is $\sim 5$ times too faint to be observed.  NGC 253 would be even fainter, and since it is located in the Southern Hemisphere sky ($\sin \delta \approx -0.43$), the available existing neutrino detectors are far less sensitive \citep{Abbasi09b,Coyle10,Abbasi11}.  However, the diffuse neutrino background from starbursts may be detectable by IceCube, which may shed some light on the CR spectrum at high energy \citep{Loeb06}.  Stacking searches of nearby, bright starbursts may also lead to a detectable signal \citep{Lacki11}.

\begin{figure}
\centerline{\includegraphics[width=8cm]{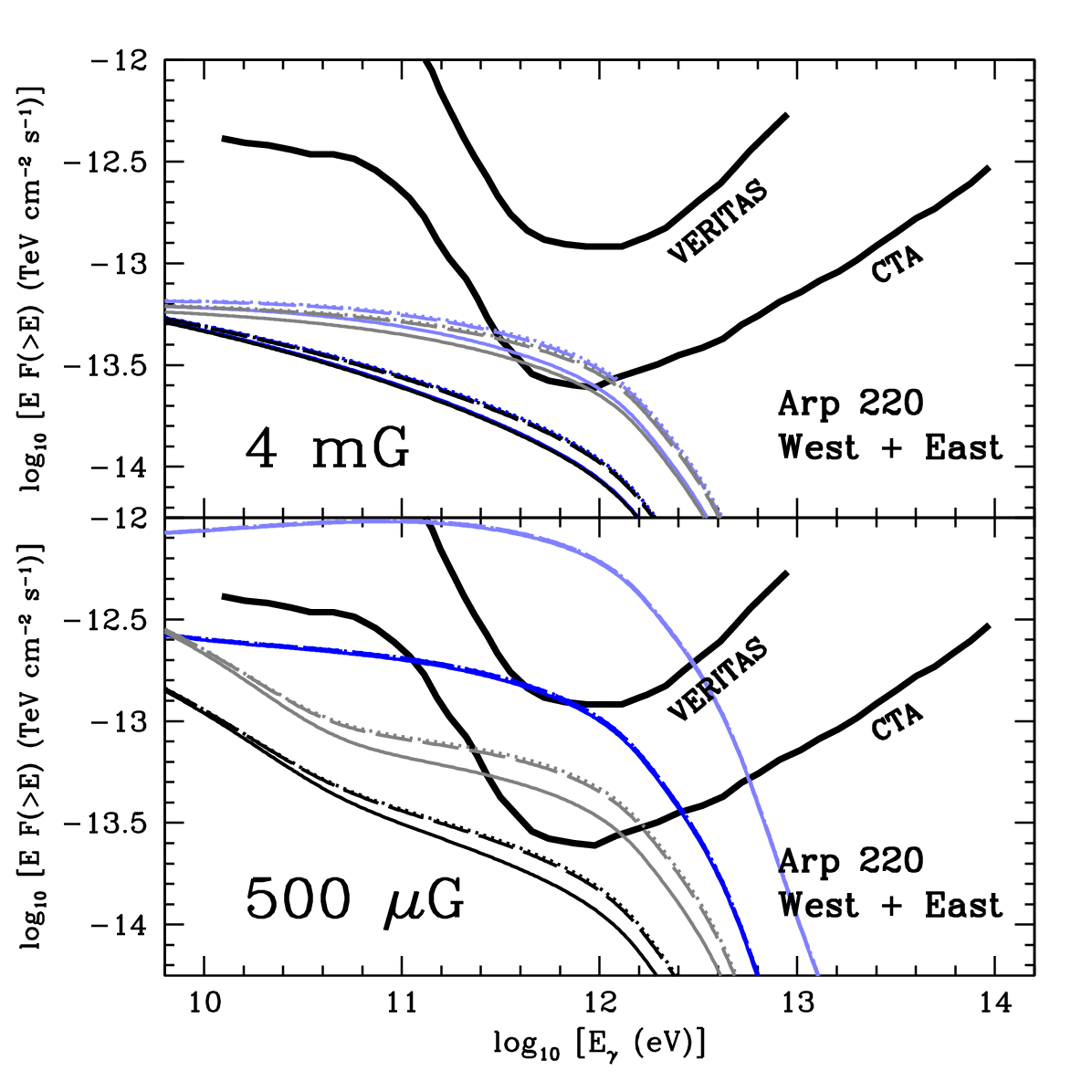}}
\figcaption[figure]{The Earth-observed pair production absorbed VHE $\gamma$-ray spectrum of Arp 220 in various models, for $\gamma_{\rm max}^{\rm prim} = 10^6$ (black for $p = 2.2$ and grey for $p = 2.0$) and $\gamma_{\rm max}^{\rm prim} = 10^9$ (dark blue for $p = 2.2$ and light blue for $p = 2.0$). The line styles are the same as in Fig.~\ref{fig:M82Neutrino}. \label{fig:A220TeVGamma}}
\end{figure}

Arp 220 is much fainter from Earth than M82, far beyond the reach of neutrino detectors like IceCube, but currently has no TeV $\gamma$-ray detection (the current upper limit is set by MAGIC in \citealt{Albert07}).  Thus, high $p$ is allowed by the data.  Just as TeV $\gamma$-rays constrained M82, TeV $\gamma$-rays can constrain the fraction of Arp 220's X-ray emission that comes from synchrotron.  In Figure~\ref{fig:A220TeVGamma}, we show the predicted VHE $\gamma$-ray flux of Arp 220's western and eastern nuclear starbursts in several models compared to the expected $5 \sigma$ sensitivity of VERITAS and Cherenkov Telescope Array (CTA) after 50 hours of integration time \citep{Doro09}.  

VERITAS will not be able to detect Arp 220 at $5\ \sigma$ except in our most optimistic models with high $\gamma_{\rm max}^{\rm prim}$, low $B$, and $p = 2.0$ (Figure~\ref{fig:A220TeVGamma}, lower panel).  However, if $B$ is low, the proposed CTA will be able to detect Arp 220 except in more pessimistic models with $p = 2.2$ and small $\gamma_{\rm max}^{\rm prim}$.  If $B$ is high, even CTA will have trouble detecting Arp 220's nuclear starbursts (Figure~\ref{fig:A220TeVGamma}, upper panel), especially if $p = 2.2$.  We note, though, that our fiducial supernova rate for both of these starbursts ($0.7\ \yr^{-1}$) is about one-third of that expected if the entire TIR luminosity of Arp 220 is due to star-formation (expected supernova rate of $2\ \yr^{-1}$).  Thus, the total TeV luminosity of Arp 220 may be $\sim 3$ times larger, in which case CTA will be able to detect it even in high $B$ cases.  Both the synchrotron X-ray flux and the TeV $\gamma$-ray flux are insensitive to the diffusive escape time: the protons are being efficiently trapped and creating secondaries at TeV energies.  The effects of pair production absorption are clearly visible in Figure~\ref{fig:A220TeVGamma}: the $\gamma$-ray flux should plummet at energies higher than a TeV, an effect first predicted for Arp 220 by \citet{Torres04}.

\section{Conclusion}
\label{sec:Conclusion}
Starbursts are luminous in hard X-rays and accelerate large amounts of CRs.  We have explored whether CR $e^{\pm}$ in starbursts could generate the observed diffuse hard X-ray emission through synchrotron emission.  The strength of this emission depends on the very poorly constrained 10 - 100 TeV CR $e^{\pm}$ spectrum in starburst galaxies.  Using one-zone models to predict this emission, we also reconsider the previously suggested contribution from IC emission of 0.1 - 1 GeV CR $e^{\pm}$, in light of the magnetic field strengths expected in starburst galaxies.  Our conclusions are as follows:

\begin{itemize}
\item We have considered the energetics for synchrotron X-ray emission in section~\ref{sec:Motivation} with simple order of magnitude estimates.  The diffuse hard X-ray emission of M82 could be synchrotron if CR escape is not important at PeV energies and the CR spectrum is hard enough ($p \approx 2.0$).  Much of this emission would come from the previously neglected pair-production $e^{\pm}$, which are efficiently generated from pionic $\gamma$-rays in the intense IR bright environments of starbursts.

\item We have also constructed one-zone models of several starbursts with standard spectra to compare with observations in radio and $\gamma$-rays.  The magnetic field strength in these models is limited at the low end by constraints on leptonic $\gamma$-ray emission and at the high end by constraints on radio emission from secondary $e^{\pm}$.  For the physical conditions we assume (listed in Table~\ref{table:ModelParameters}), these limits are $50\ \muGauss \la B \la 150\ \muGauss$ in NGC 253, $50 \la B \la 200\ \muGauss$ in M82, and $0.5\ \mGauss \la B \la 4\ \mGauss$ for Arp 220's nuclei. In low $B$ models there is a large component of primary electrons, while in high $B$ models the secondary $e^{\pm}$ dominate at GeV energies in M82 and Arp 220's nuclei.

\item The existing $\gamma$-ray data of the modelled starbursts rules out most $p = 2.0$ models, and we find that the synchrotron emission is probably a minority of the hard X-ray emission in the Galactic Center, NGC 253, and M82 with our assumptions.  Nonetheless, there is still enough variation in the parameters for synchrotron to make up from less than one to several tens of percent of the synchrotron emission.  The synchrotron X-ray emission is highest in low $B$ models with large $\gamma_{\rm max}^{\rm prim}$, lower in high $B$ models, and lowest in low $B$ models with small $\gamma_{\rm max}^{\rm prim}$.  In our fiducial models (with high $B$ and large $\gamma_{\rm max}^{\rm prim}$), the fraction of unresolved hard X-ray emission contributed by synchrotron is 6\% in the Galactic Center, 9\% in NGC 253, 2\% in M82, 15\% in Arp 220's western nucleus (Arp 220 X-1), and 34\% in Arp 220's eastern nucleus (Arp 220 X-4).  These results are consistent with previous predictions that X-ray binaries contribute most of the X-ray emission in starburst galaxies at a few keV \citep{Persic02,Persic04}.  We find that the 2 - 10 keV X-ray synchrotron emission is $10^{-7} - 10^{-6} L_{\rm IR}$ in these fiducial models; in models of generic star-forming galaxies it peaks at $\sim 4 \times 10^{-6} L_{\rm IR}$ for $p = 2.2$.\footnote{For comparison, according to the observed FIR-radio correlation, the GHz radio luminosity is $\sim 10^{-6} L_{\rm IR}$ \citep{Yun01}.}

\item Pair production is predicted to contribute significantly to the high energy $e^{\pm}$ population in NGC 253, M82, and Arp 220, with pair $e^{\pm}$ comparable to the pionic $e^{\pm}$.  The huge densities of IR starlight photons efficiently convert $\sim 10 - 100\ \TeV$ $\gamma$-rays into $e^{\pm}$ pairs.  These pairs then can efficiently radiate X-ray synchrotron.  However, the actual density of pair $e^{\pm}$ is directly related to $\gamma$-ray emissivity at very high energies, which in turn depends strongly on the number and escape times of CR protons at these energies.  The Galactic Center region is transparent to VHE $\gamma$-rays, so that pair $e^{\pm}$ are negligible.  

\item We find that Inverse Compton emission in the 2 - 10 keV band is also a minority of the X-ray emission.  In our models, the magnetic field energy density can be greater than some of the values derived with the minimum energy estimate.  This means the IC emission expected by scaling the GHz radio emission is lower than in previous estimates.  In our fiducial models, the fraction of unresolved X-ray emission contributed by IC is 0.1\% in the Galactic Center, 3\% in NGC 253, 1\% in M82, and $\sim 10\%$ in Arp 220's nuclei.  The IC emission at 2 - 10 keV is much harder ($\Gamma_{2 - 10} = 1.2 - 1.7$) than the synchrotron emission at the same energies ($\Gamma_{2 - 10} = 1.8 - 2.5$).

\item Key uncertainties include the maximum energy of primary electrons ($\gamma_{\rm max}^{\rm prim}$), the rate of escape of CR protons at high energies ($t_{\rm diff}$), and the spectral slope of CRs ($p$).  However, some of these quantities might be constrained by future neutrino telescopes, which can determine the pionic luminosities of starbursts at TeV to PeV energies.  Future TeV $\gamma$-ray telescopes like CTA can constrain the electron spectrum at high energies as well, especially in Arp 220 which has relatively weak TeV upper limits.  The effects of hydrogen absorption in dense starbursts also need to be explored, although this applies to IC emission and other sources of the X-ray flux, not just synchrotron.  

\end{itemize}

The large X-ray foregrounds from HMXBs and other sources makes detection of the synchrotron X-rays difficult.  However, if the synchrotron emission can be detected, it will have important implications for our understanding of CRs in starbursts.  Very little is known about the CR population at very high energies in starbursts, such as what the escape rate is or whether there is a ``knee'' in the CR spectrum as there is in the Milky Way.  Combined with upcoming neutrino measurements, detection of synchrotron X-rays can extend our understanding from the directly observed TeV energies.

Even in the Milky Way, we know relatively little about the CR electron population at multi-TeV energies \citep{Kistler09}, and the propagation of electrons at $100\ \GeV - \TeV$ energies is still poorly understood \citep{Chang08,Adriani09,Abdo09,Aharonian09a}.  It is possible that we are \emph{underestimating} the primary population significantly.  We assumed that the primary CR electron injection spectrum is a power law extending to TeV-PeV energies, with one source such as supernova remnants dominating the primary electron spectrum.  An additional hard primary component (from PWNe, for example) on top of a softer component that explains the radio synchrotron emission would increase the synchrotron X-ray emission.  However, a second component such as PWNe would have to be very hard ($p \la 2.0$) to enhance the X-ray flux significantly; otherwise IC emission from the low energy $e^{\pm}$ may exceed the observed $\gamma$-rays.  

If a significant fraction of the hard X-ray emission from starbursts is synchrotron, it may be polarized.  Although the radio synchrotron emission of M82's halo is strongly polarized, the starburst core itself has very little polarization \citep{Reuter94}.  This may be because of Faraday depolarization, in which the back of the starburst is strongly Faraday rotated with respect to the front.  Faraday depolarization, however, will be completely unimportant at X-ray frequencies.  The low polarization could also arise if the magnetic field was turbulent and isotropic on small scales, in which case the X-ray emission will also be unpolarized.  However, even turbulent magnetic fields can be made anisotropic through shear and compression \citep[e.g.,][]{Laing80,Sokoloff98}.  Furthermore, there is observational evidence of anisotropic magnetic fields in M82 from submillimeter and infrared polarization measurements \citep{Greaves00,Jones00}.  \citet{Siebenmorgen01} and \citet{Seiffert07} found that Arp 220 has a low infrared and submillimeter polarization, though their source apertures blend the two nuclear starbursts and the outlying starburst disk together.  In any case, the diffuse hard X-ray emission of M82 only has a flux of $\sim 0.1 - 0.2$ milliCrab \citep{Strickland07,Kirsch05}.  Even with an instrument as sensitive as the now defunct Gravity and Extreme Magnetism Small Explorer (GEMS; \citealt{Swank09}), its polarization could only be detected if it was of order unity.

It is tempting to consider even higher energy synchrotron emission, in the 10 keV - MeV band, to directly probe 30 TeV - PeV $e^{\pm}$.  Presently, there are only relatively weak upper limits (and a claimed detection of NGC 253) of this emission of nearby starbursts and ULIRGs from the OSSE instrument on the Compton Gamma-Ray Observatory \citep{Bhattacharya94,Dermer97}, and a recent detection of M82 by Swift-BAT \citep{Cusumano10}.  With its much greater sensitivity in the 10 - 80 keV range over previous telescopes, \emph{NuSTAR} can improve this situation \citep{Harrison05,Harrison10}.  However, above $\sim 10\ \keV$, we generally expect the very hard IC emission to finally bury the softer synchrotron emission, even if the CR electron spectrum does extend to multi-PeV energies (compare the synchrotron and IC emission in M82 in Figure~\ref{fig:M82Spectra}, right panel).  Furthermore, the angular resolution of \emph{NuSTAR} is 40", corresponding to $750\ \pc$ at the distance of M82, so point sources may contaminate the emission \citep{Harrison05,Harrison10}.  On the other hand, even a measurement of the IC emission itself could be useful in constraining the magnetic field strength.  The synchrotron luminosity may remain within a factor of $\sim 2$ of the IC until MeV energies in strong starbursts like Arp 220, if they have very strong magnetic fields ($B \ga \mGauss$; as seen in Figure~\ref{fig:A220Spectra}). \footnote{Such MeV emission would be unaffected by hydrogen absorption as well, which would be ideal for an extremely dense starburst like Arp 220, although it is beyond the energy range of \emph{NuSTAR}.}  Even if the IC and point source emission was not a significant foreground, ultimately the synchrotron will be buried by bremsstrahlung and finally direct pionic $\gamma$-ray emission above 10 MeV.  Since there is more power in lower energy protons than VHE protons, the synchrotron emission cannot exceed the pionic emission at higher energies.  Thus, synchrotron cannot be used to probe the CR $e^{\pm}$ spectrum above about a PeV, unless an enormous new component of primaries is present.

While we have considered mainly $\gamma$-rays from star-formation, $\gamma$-rays in starburst environments can come from other sources: most notably, an AGN.  In particular, Sgr A$^\star$ in our own Galaxy is known to be a source of VHE $\gamma$-rays \citep{Aharonian04}.  It resides in the Central Cluster, a dense star cluster with a FIR energy density of $\sim 6000\ \eV\ \cm^{-3}$ over a region $\sim 2\ \pc$ in diameter (\citealt{Telesco96}; see also \citealt{Davidson92,Hopkins10}).  Interestingly, the VHE emission appears to have an exponential cutoff at 15 TeV \citep{Aharonian09b}, similar to the Klein-Nishina threshold for pair production on the FIR light of a starburst.  Diffuse hard X-rays are observed from the region near Sgr A$^\star$; these potentially could have a synchrotron contribution from pair $e^{\pm}$.  Similar considerations may apply to other central stellar clusters around VHE-emitting AGNs.

\acknowledgments
We thank Eli Waxman and John Beacom for discussions.  We also thank Felix Aharonian, Markus B\"ottcher, and Dmitry Khangulyan for technical discussions on $\gamma\gamma$ pair production.  This work is supported in part by an Alfred P. Sloan fellowship and NASA grant \#NNX10AD01G.  BCL was supported in part for most of this project by an Elizabeth Clay Howald Presidential Fellowship from the OSU.  BCL would also like to acknowledge a Jansky Fellowship from the NRAO.  NRAO is operated by Associated Universities, Inc., under cooperative agreement with the National Science Foundation.

\appendix
\section{Other Pionic Cross Section and Lifetime Parametrizations}
\label{sec:OtherPion}
The spectrum of pionic secondary $\gamma$-rays and $e^{\pm}$ at TeV energies depends sensitively on the behavior of the pion production process at these energies \citep{Karlsson08}.  The physics of pion production enters two ways in our models: (1) the lifetime of CR protons to pionic processes, which determines the steady-state CR spectrum and (2) the differential cross sections for production of pionic secondaries for protons of each energy.

We run models where we vary the pionic lifetimes or the differential cross sections to consider the effects.  We use the fiducial parameters listed in Table~\ref{table:ModelParameters} for the Galactic Center, NGC 253, M82, and the two nuclei of Arp 220.  We find the thermal fraction and $\xi$ through radio spectrum chi-square fitting (or radio flux normalization for Arp 220) described in \S~\ref{sec:Constraints} for each variation on the pionic physics.  We also compare the purely hadronic flux using the different assumptions of the pionic physics.

\emph{Pionic lifetime variations} -- In most of our models, we calculate the pionic loss lifetime by integrating up the \citet{Kamae06} cross-sections for $\gamma$-ray, neutrino, and $e^{\pm}$ production (equation~\ref{eqn:tPiCS}).  A commonly used pionic loss lifetime comes from \citet{Mannheim94} (MS94) and is given in equation~\ref{eqn:tPion}.  Finally, \citet{Schlickeiser02} recommends a different pionic loss lifetime, which is longer at low energies but substantially shorter at VHE energies:
\begin{equation}
t_{\pi} = 2.2 \times 10^8 \yr\ \gamma^{-0.28} (n / \cm^{-3})^{-1}
\end{equation}
The new $\gamma$ dependence comes from the multiplicity of pions produced per collision.  Formally, this equation is only valid for $\gamma \la 3000$, or below a few TeV.  

The \citet{Schlickeiser02} pionic lifetime predicts substantially less hadronic synchrotron X-ray emission, unless escape dominates the multi-TeV proton lifetimes.  In our Arp 220 models, where pionic losses dominate even at these energies, the hadronic 2 - 10 keV synchrotron flux is only $1/4$ of the predictions using the cross-section derived $t_{\pi}$.   This is because the $\gamma^{-0.28}$ in $t_{\pi}$ becomes very small for 10 - 100 TeV CR protons, reducing the steady-state number of CR protons that are the source of secondary and pair $e^{\pm}$.  The secondary source functions do not correspondingly increase, so the pionic luminosity is underpredicted.  The M82 model also gives only $64\%$ of the hadronic synchrotron flux predicted with the cross-section integrated $t_{\pi}$.  The TeV $\gamma$-ray flux is also reduced in these models by similar amounts.  However, the synchrotron flux in the fiducial models of NGC 253 and the Galactic Center are affected by less than 10\%, because diffusion sets the total number of protons in these models.  Furthermore, the total synchrotron X-ray flux is barely affected in the Galactic Center, NGC 253, and M82, and only reduced to 40\% of nominal in the Arp 220 nuclei models, because primary CR $e^{\pm}$ contribute heavily to the synchrotron emission.  

If we use the MS94 $t_{\pi}$, the hadronic synchrotron X-ray emission is instead slightly enhanced.  In Arp 220 models, the synchrotron X-ray emission is $113\%$ that using our standard pionic lifetime.  The total synchrotron X-ray emission is significantly greater in MS94 models, because higher $\xi$ are picked by our fitting processes: the synchrotron is 1.3 times greater in Arp 220 models using MS94 and 1.2 times greater in M82 models.  There are negligible differences in the TeV $\gamma$-ray fluxes between the MS94 models and eqn.~\ref{eqn:tPiCS} models, but the GeV fluxes are only 80\% as big in the M82 models and 70\% as big in the Arp 220 models.

The \citet{Schlickeiser02} pionic lifetime is only meant to be valid below 2.8 TeV, which is below the energy of the CR protons responsible for hard X-ray emitting $e^{\pm}$.  Furthermore, it is derived by integrating the energy of the pionic secondaries, much like we do to derive eqn.~\ref{eqn:tPiCS}.  Since we used the \citet{Kamae06}~cross sections in eqn.~\ref{eqn:tPiCS}, which are explicitly valid up to several hundred TeV, we believe that eqn.~\ref{eqn:tPiCS} is more likely to be accurate in deriving the X-ray synchrotron emission.  Thus, the pionic lifetime would only contribute less than a factor of $\sim 2$ uncertainty, much less than the uncertainties from $\gamma_{\rm max}^{\rm prim}$, $p$, and $t_{\rm diff}$.  

\emph{Differential cross section variations} -- There are several possible parametrizations of the pionic cross sections.  In addition to the \citet{Kamae06} cross sections, the GALPROP cross sections \citep{Moskalenko98,Strong98,Strong00} based on the work of \citet{Dermer86a} are commonly used (in turn based on \citealt{Stecker70}, \citealt{Badhwar77}, and \citealt{Stephens81}; see also \citealt{Dermer86b}).  

Both the \citet{Kamae06} and GALPROP cross sections are based on lower energy data, and \citet{Kamae06} is only formally valid for CR protons up to 512 TeV.  The pionic cross sections given in \citet{Kelner06} are valid for CR protons at energies at 100 GeV to 100 PeV.  We consider a variation where the \citet{Kamae06} cross sections are used for CR protons at energies below 100 GeV, and the \citet{Kelner06} cross sections are used above 100 GeV.

Using the GALPROP cross sections reduces both the hadronic and total synchrotron X-ray luminosities.  For the Galactic Center, the hadronic (total) GALPROP synchrotron X-ray luminosities are $67\%\ (100\%)$ the \citet{Kamae06} luminosities; for NGC 253, they are $73\%\ (83\%)$ the \citet{Kamae06} luminosities; for M82, the GALPROP synchrotron X-ray luminosities are $79\%\ (75\%)$ the \citet{Kamae06} luminosities; and for Arp 220's nuclei, they are $105\%\ (90\%)$ the \citet{Kamae06} luminosities.  Smaller $\xi$ are preferred using the GALPROP cross sections, reducing the leptonic contribution to the synchrotron X-ray emission.  In the Arp 220 models, using the GALPROP cross sections leads to small increases of the hadronic $\gamma$-ray luminosities, about $\sim 15\%$ at GeV and $\sim 20\%$ at TeV energies.

If we instead use the \citet{Kelner06} cross sections for $E_p > 100\ \GeV$, we find even less of an effect.  For the Galactic Center, the hadronic (total) \citet{Kelner06} synchrotron X-ray luminosities are $84\%\ (100\%)$ the \citet{Kamae06} luminosities; for NGC 253, they are $87\%\ (98\%)$ the \citet{Kamae06} luminosities; for M82, they are $92\%\ (97\%)$ the \citet{Kamae06} luminosities; and for Arp 220's nuclei, they are $114\%\ (110\%)$ the \citet{Kamae06} luminosities.  The TeV $\gamma$-ray luminosity is 27\% brighter in the Arp 220 models using these cross sections instead of our standard cross sections.  

Thus, using other parametrizations of the pionic cross sections does not seem to add even a factor of $\sim 2$ uncertainty. 

\section{Other Maximum Energy Proton Cutoffs}
\label{sec:OtherGammaP}
Throughout this work, we have assumed that primary CR protons are accelerated to a maximum Lorentz factor $\gamma_{\rm max}^p = 10^6$, or a maximum energy of $\sim 938\ \TeV$.  This is loosely based on the observed knee in the CR spectrum in the Galaxy, although given the scarcity of current observational data, the actual high-energy proton cutoff could be at a different energy in starburst regions.  Since the average energy of pionic electrons is about 1/20 of the proton energy \citep[e.g.,][]{Kelner06}, this translates to a pionic $e^{\pm}$ cutoff near 50 TeV.  Thus, models with low $\gamma_{\rm max}^{\rm prim}$ or high B can be affected by the proton energy cutoff.  To consider this effect, we rerun our fiducial models with $\gamma_{\rm max}^p$ of $10^5$, $10^6$, and $10^7$.  We consider both the total and the hadronic fluxes, as in Appendix~\ref{sec:OtherPion}.  As in our fiducial models, we use the \citet{Kamae06} cross sections for proton energies below 500 TeV and the \citet{Kelner06} cross sections for proton energies above 500 TeV.  

In models where $\gamma_{\rm max}^p = 10^5$, the hadronic synchrotron X-ray flux drops by almost an order of magnitude in most of the starbursts we consider.  The hadronic synchrotron X-ray flux is only 9.4\%, 8.5\%, and 11\% of its fiducial value in the Galactic Center, NGC 253, and M82 respectively.  In Arp 220's nuclei, the effect is not so severe since $B$ is $\sim 40$ times higher than in the other starbursts, and synchrotron X-rays probe $e^{\pm}$ energies that are $\sim 6$ times lower (eqn.~\ref{eqn:ECSynch}).  However, the low energy proton cutoff still reduces the hadronic synchrotron X-ray flux to 43\% of its fiducial value.  The 2 - 10 keV hadronic synchrotron X-ray flux is also much softer in these models, because of the cutoff in the secondary $e^{\pm}$ spectrum: the hadronic $\Gamma_{2-10}^{\rm synch}$ is 3.45, 3.29, 3.06, and 2.46 in the Galactic Center, NGC 253, M82, and Arp 220's nuclei.  The total synchrotron X-ray flux is affected less than the hadronic flux, but note that we use $\gamma_{\rm max}^{\rm prim} = 10^9$ in our fiducial models so that primaries dominate in most of the starbursts.  Thus the total synchrotron X-ray flux is 99.7\%, 90\%, 64\%, 58\%, and 49\% of the fiducial values in the Galactic Center, NGC 253, M82, Arp 220 West, and Arp 220 East, respectively.

In models where $\gamma_{\rm max}^p = 10^7$, the hadronic synchrotron X-ray flux in the 2 - 10 keV band is enhanced by $\sim 40\%$.  The hadronic synchrotron X-ray flux is 141\%, 141\%, 136\%, and 112\% of its fiducial value in the Galactic Center, NGC 253, M82, and Arp 220's nuclei respectively.  The hadronic synchrotron X-ray spectrum is also hardened slightly: the hadronic $\Gamma_{2-10}^{\rm synch}$ is 2.32, 2.25, 2.21, and 2.11 in the Galactic Center, NGC 253, M82, and Arp 220's nuclei.  The total synchrotron X-ray flux is affected very little, with total synchrotron fluxes 100.2\%, 103\%, 114\%, 110\%, and 112\% of fiducial in the Galactic Center, NGC 253, M82, Arp 220 West, and Arp 220 East, respectively.

We conclude that if the secondary and pair $e^{\pm}$ dominate at 10 - 100 TeV energies, the maximum proton energy can significantly affect the synchrotron X-ray flux and spectral slope, particularly if the cutoff is much lower than a PeV.

\setlength{\tabcolsep}{0.05in} 

\LongTables
\begin{deluxetable}{ccccccccccccccccccc}
\tabletypesize{\scriptsize}
\tablecaption{Galactic Center X-ray luminosities}
\tablehead{\colhead{$B$} & \colhead{$p$} & \colhead{$t_{\rm diff} (3\ \GeV)$} & \colhead{$\eta$} & \multicolumn{7}{c}{$\gamma_{\rm max}^{\rm prim} = 10^6$} & & \multicolumn{7}{c}{$\gamma_{\rm max}^{\rm prim} = 10^9$} \\ & & & & \colhead{$\xi$} & \colhead {$L_{2-10}^{\rm synch}$} & \colhead{$f_{2-10}^{\rm synch}$\tablenotemark{a}} & \colhead{$\frac{\rm Synch}{\rm IC}$\tablenotemark{b}} & \colhead{$\Gamma_{2-10}^{\rm synch}$} & \colhead{$f_{\rm GeV}$\tablenotemark{c}} & \colhead{$f_{\rm VHE}$\tablenotemark{d}} & & \colhead{$\xi$} & \colhead {$L_{2-10}^{\rm synch}$} & \colhead{$f_{2-10}^{\rm synch}$\tablenotemark{a}} & \colhead{$\frac{\rm Synch}{\rm IC}$\tablenotemark{b}} & \colhead{$\Gamma_{2-10}^{\rm synch}$} & \colhead{$f_{\rm GeV}$\tablenotemark{c}} & \colhead{$f_{\rm VHE}$\tablenotemark{d}} \\ \colhead{$\muGauss$} & & \colhead{(Myr)} & & & \colhead{$\ergps$} & & & & & & & & \colhead{$\ergps$} & & & & }
\startdata
\cutinhead{$\Sigma_g = 0.003\ \gcm2$}
50    & 2.4 & $\infty$ & 0.0088 & \multicolumn{7}{c}{\nodata}                           & & 0.038  & 1.3e35 & 0.018  & 3.3   & 2.22 & 0.65 & 0.67\\
      &     &          & 0.071  & \multicolumn{7}{c}{\nodata}                           & & 0.038  & 1.3e35 & 0.018  & 3.4   & 2.22 & 0.84 & 1.1\\
      &     &          & 0.10   & 0.037  & 1.1e33 & 1.5E-4 & 0.029 & 2.49 & 0.90 & 0.63 & & 0.038  & 1.3e35 & 0.018  & 3.4   & 2.22 & 0.93 & 1.2\\
      &     & 10       & 0.0088 & \multicolumn{7}{c}{\nodata}                           & & 0.038  & 1.3e35 & 0.018  & 3.3   & 2.21 & 0.64 & 0.64\\
      &     &          & 0.10   & \multicolumn{7}{c}{\nodata}                           & & 0.038  & 1.3e35 & 0.018  & 3.3   & 2.21 & 0.92 & 0.99\\
      &     & 1        & 0.0088 & \multicolumn{7}{c}{\nodata}                           & & 0.040  & 1.3e35 & 0.018  & 3.4   & 2.21 & 0.62 & 0.64\\
      &     &          & 0.14   & \multicolumn{7}{c}{\nodata}                           & & 0.040  & 1.3e35 & 0.018  & 3.4   & 2.21 & 0.98 & 0.76\\
100   & 2.0 & $\infty$ & 0.0088 & 0.0049 & 4.6e33 & 6.3E-4 & 0.64  & 2.35 & 0.40 & 0.70 & & \multicolumn{7}{c}{\nodata}\\
      &     &          & 0.0125 & 0.0049 & 6.5e33 & 8.9E-4 & 0.91  & 2.35 & 0.41 & 0.99 & & \multicolumn{7}{c}{\nodata}\\
      &     &          & 0.0250 & 0.0049 & 1.3e34 & 0.0018 & 1.8   & 2.35 & 0.43 & 2.0  & & \multicolumn{7}{c}{\nodata}\\
      &     & 10       & 0.0125 & 0.0049 & 1.9e33 & 2.6E-4 & 0.27  & 2.41 & 0.41 & 0.54 & & \multicolumn{7}{c}{\nodata}\\
      &     &          & 0.0250 & 0.0049 & 3.9e33 & 5.2E-4 & 0.54  & 2.41 & 0.43 & 1.1  & & \multicolumn{7}{c}{\nodata}\\
      &     &          & 0.0354 & 0.0049 & 5.5e33 & 7.4E-4 & 0.76  & 2.41 & 0.45 & 1.5  & & \multicolumn{7}{c}{\nodata}\\
      &     & 1        & 0.0707 & 0.0050 & 1.5e33 & 2.1E-4 & 0.21  & 2.44 & 0.48 & 0.69 & & \multicolumn{7}{c}{\nodata}\\
      &     &          & 0.1000 & 0.0050 & 2.2e33 & 2.9E-4 & 0.30  & 2.44 & 0.53 & 0.98 & & \multicolumn{7}{c}{\nodata}\\
      &     &          & 0.2000 & 0.0049 & 4.3e33 & 5.9E-4 & 0.61  & 2.44 & 0.67 & 2.0  & & \multicolumn{7}{c}{\nodata}\\
      & 2.2 & $\infty$ & 0.0088 & \multicolumn{7}{c}{\nodata}                           & & 0.0066 & 4.3e35 & 0.058  & 48    & 2.12 & 0.24 & 0.60\\
      &     &          & 0.025  & 0.0062 & 2.9e33 & 4.0E-4 & 0.32  & 2.39 & 0.28 & 0.70 & & 0.0066 & 4.3e35 & 0.059  & 48    & 2.12 & 0.29 & 1.1\\
      &     &          & 0.035  & 0.0061 & 4.1e33 & 5.6E-4 & 0.46  & 2.39 & 0.31 & 0.98 & & 0.0066\tablenotemark{e} & 4.3e35 & 0.059  & 48   & 2.12 & 0.33 & 1.3\\
      &     &          & 0.050  & 0.0061\tablenotemark{e} & 5.8e33 & 7.9E-4 & 0.65 & 2.39 & 0.35 & 1.4 & & 0.0065 & 4.3e35 & 0.059  & 48    & 2.12 & 0.37 & 1.7\\
      &     &          & 0.071  & 0.0061 & 8.2e33 & 0.0011 & 0.92  & 2.39 & 0.42 & 2.0  & & \multicolumn{7}{c}{\nodata}\\
      &     & 10       & 0.013  & \multicolumn{7}{c}{\nodata}                           & & 0.0066 & 4.3e35 & 0.058  & 48    & 2.12 & 0.25 & 0.56\\
      &     &          & 0.035  & 0.0062 & 1.3e33 & 1.7E-4 & 0.14  & 2.45 & 0.31 & 0.57 & & 0.0065 & 4.3e35 & 0.058  & 48    & 2.12 & 0.32 & 0.93\\
      &     &          & 0.071  & 0.0061 & 2.5e33 & 3.4E-4 & 0.28  & 2.45 & 0.41 & 1.1  & & \multicolumn{7}{c}{\nodata}\\
      &     &          & 0.10   & 0.0060 & 3.5e33 & 4.8E-4 & 0.40  & 2.45 & 0.50 & 1.6  & & 0.0065 & 4.2e35 & 0.058  & 47    & 2.12 & 0.51 & 2.0\\
      &     & 1        & 0.035  & \multicolumn{7}{c}{\nodata}                           & & 0.0068 & 4.4e35 & 0.060  & 49    & 2.12 & 0.30 & 0.51\\
      &     &          & 0.14   & 0.0062 & 7.1e32 & 1.0E-4 & 0.079 & 2.49 & 0.56 & 0.56 & & 0.0066\tablenotemark{f} & 4.3e35 & 0.059  & 48    & 2.12 & 0.57 & 0.91\\
      &     &          & 0.28   & 0.0059 & 1.4e33 & 1.9E-4 & 0.16  & 2.49 & 0.92 & 1.1  & & 0.0064 & 4.1e35 & 0.056  & 46    & 2.12 & 0.93 & 1.5\\
      & 2.4 & $\infty$ & 0.10   & 0.011  & 1.6e33 & 2.1E-4 & 0.13  & 2.43 & 0.44 & 0.62 & & 0.012  & 4.8e34 & 0.0066 & 4.1   & 2.22 & 0.45 & 0.68\\
      &     &          & 0.14   & 0.011  & 2.2e33 & 3.0E-4 & 0.19  & 2.43 & 0.57 & 0.88 & & 0.011  & 4.8e34 & 0.0066 & 4.1   & 2.22 & 0.58 & 0.93\\
      &     &          & 0.20   & 0.011  & 3.1e33 & 4.3E-4 & 0.27  & 2.43 & 0.76 & 1.2  & & 0.011  & 4.8e34 & 0.0065 & 4.1   & 2.22 & 0.76 & 1.3\\
      &     & 10       & 0.14   & 0.011  & 7.0e32 & 1.0E-4 & 0.059 & 2.50 & 0.57 & 0.55 & & 0.011  & 4.7e34 & 0.0064 & 4.0   & 2.21 & 0.57 & 0.60\\
      &     &          & 0.20   & 0.011  & 9.9e32 & 1.4E-4 & 0.085 & 2.50 & 0.75 & 0.77 & & 0.011  & 4.6e34 & 0.0063 & 3.7   & 2.21 & 0.75 & 0.82\\
\cutinhead{$\Sigma_g = 0.01\ \gcm2$}
100   & 2.0 & 10       & 0.0088 & 0.0052 & 4.5e33 & 6.1E-4 & 0.66  & 2.41 & 0.64 & 1.2  & & \multicolumn{7}{c}{\nodata}\\
      &     &          & 0.013  & 0.0052 & 6.4e33 & 8.7E-4 & 0.94  & 2.41 & 0.66 & 1.8  & & \multicolumn{7}{c}{\nodata}\\
      &     & 1        & 0.018  & 0.0053 & 1.3e33 & 1.7E-4 & 0.19  & 2.44 & 0.66 & 0.57 & & \multicolumn{7}{c}{\nodata}\\
      &     &          & 0.035  & 0.0053 & 2.5e33 & 3.5E-4 & 0.37  & 2.44 & 0.75 & 1.1  & & \multicolumn{7}{c}{\nodata}\\
      &     &          & 0.050  & 0.0052 & 3.6e33 & 4.9E-4 & 0.53  & 2.44 & 0.82 & 1.6  & & \multicolumn{7}{c}{\nodata}\\
      & 2.2 & $\infty$ & 0.0088 & 0.0065 & 3.3e33 & 4.5E-4 & 0.40  & 2.39 & 0.46 & 0.80 & & 0.0069 & 4.5e35 & 0.062  & 55    & 2.13 & 0.48 & 1.2\\
      &     &          & 0.013  & 0.0065 & 4.7e33 & 6.4E-4 & 0.57  & 2.39 & 0.50 & 1.1  & & 0.0069\tablenotemark{e} & 4.5e35 & 0.062  & 55    & 2.13 & 0.51 & 1.5\\
      &     &          & 0.018  & 0.0064 & 6.7e33 & 9.1E-4 & 0.81  & 2.39 & 0.55 & 1.6  & & 0.0069 & 4.5e35 & 0.062  & 55    & 2.13 & 0.56 & 2.0\\
      &     & 10       & 0.0088 & \multicolumn{7}{c}{\nodata}                           & & 0.0070 & 4.5e35 & 0.062  & 54    & 2.13 & 0.48 & 0.85\\
      &     &          & 0.013  & 0.0065 & 1.5e33 & 2.0E-4 & 0.18  & 2.45 & 0.49 & 0.66 & & 0.0070 & 4.5e35 & 0.062  & 54    & 2.13 & 0.51 & 1.0\\
      &     &          & 0.018  & 0.0065 & 2.1e33 & 2.8E-4 & 0.25  & 2.45 & 0.54 & 0.94 & & 0.0069\tablenotemark{e} & 4.5e35 & 0.061  & 54    & 2.13 & 0.56 & 1.3\\
      &     &          & 0.025  & 0.0064\tablenotemark{e} & 2.9e33 & 4.0E-4 & 0.35  & 2.45 & 0.61 & 1.3 & & 0.0069 & 4.5e35 & 0.061  & 54    & 2.13 & 0.63 & 1.7\\
      &     &          & 0.035  & 0.0063 & 4.1e33 & 5.6E-4 & 0.50  & 2.45 & 0.71 & 1.9  & & \multicolumn{7}{c}{\nodata}\\
      &     & 1        & 0.0088 & \multicolumn{7}{c}{\nodata}                           & & 0.0072 & 4.7e35 & 0.063  & 55    & 2.12 & 0.46 & 0.51\\
      &     &          & 0.050  & 0.0066 & 8.4e32 & 1.1E-4 & 0.10  & 2.49 & 0.79 & 0.66 & & 0.0069 & 4.5e35 & 0.061  & 54    & 2.12 & 0.80 & 1.0\\
      &     &          & 0.071  & 0.0064 & 1.2e33 & 1.6E-4 & 0.14  & 2.49 & 0.96 & 0.93 & & 0.0069 & 4.5e35 & 0.061  & 53    & 2.12 & 0.97 & 1.3\\
      & 2.4 & $\infty$ & 0.025  & 0.012  & 1.3e33 & 1.7E-4 & 0.12  & 2.43 & 0.54 & 0.51 & & 0.012  & 5.2e34 & 0.0070 & 4.8   & 2.23 & 0.55 & 0.56\\
      &     &          & 0.050  & 0.012  & 2.5e33 & 3.5E-4 & 0.24  & 2.43 & 0.80 & 1.0  & & 0.012  & 5.1e34 & 0.0070 & 4.8   & 2.23 & 0.80 & 1.1\\
      &     & 10       & 0.035  & \multicolumn{7}{c}{\nodata}                           & & 0.012  & 5.1e34 & 0.0069 & 4.7   & 2.22 & 0.65 & 0.50\\
      &     &          & 0.050  & 0.012  & 8.2e32 & 1.1E-4 & 0.077 & 2.50 & 0.79 & 0.63 & & 0.012  & 5.0e34 & 0.0068 & 4.7   & 2.22 & 0.80 & 0.69\\
      &     &          & 0.071  & 0.012  & 1.2e33 & 1.6E-4 & 0.11  & 2.50 & 1.0  & 0.90 & & \multicolumn{7}{c}{\nodata}\\
\cutinhead{$\Sigma_g = 0.03\ \gcm2$}
      & 2.4 & $\infty$ & 0.0088 & 0.014  & 1.3e33 & 1.7E-4 & 0.13  & 2.43 & 0.99 & 0.51 & & 0.014  & 6.0e34 & 0.0082 & 6.4   & 2.27 & 1.0  & 0.58
\enddata
\label{table:GalCenXRayLuminosities}
\tablenotetext{a}{Fraction of the observed \emph{diffuse} 2 - 10 keV X-ray luminosity (scaled from \citet{Koyama96} as described in \S~\ref{sec:GalCen}) from that synchrotron emission accounts for.}
\tablenotetext{b}{Ratio of the 2 - 10 keV synchrotron and IC luminosities.}
\tablenotetext{c}{Ratio of the predicted GeV $\gamma$-ray emission with the upper limit from EGRET in \citet{Hunter97}.}
\tablenotetext{d}{Ratio of the predicted TeV $\gamma$-ray emission with the observed TeV flux observed from HESS in \citet{Aharonian06} and \citet{Crocker11}.}
\tablenotetext{e}{This model satisfies our constraints (\S~\ref{sec:Constraints}).  While $\eta$ is not the minimum or maximum allowed or the closest to reproducing the TeV luminosity for one of the $\gamma_{\rm max}^{\rm prim}$, this is true for the other $\gamma_{\rm max}^{\rm prim}$ so we list it anyway.}
\tablenotetext{f}{Chosen as our fiducial model.}
\end{deluxetable}

\begin{deluxetable}{ccccccccccccccccccc}
\tabletypesize{\scriptsize}
\tablecaption{NGC 253 X-ray luminosities}
\tablehead{\colhead{$B$} & \colhead{$p$} & \colhead{$t_{\rm diff} (3\ \GeV)$} & \colhead{$\eta$} & \multicolumn{7}{c}{$\gamma_{\rm max}^{\rm prim} = 10^6$} & & \multicolumn{7}{c}{$\gamma_{\rm max}^{\rm prim} = 10^9$} \\ & & & & \colhead{$\xi$} & \colhead {$L_{2-10}^{\rm synch}$} & \colhead{$f_{2-10}^{\rm synch}$\tablenotemark{a}} & \colhead{$\frac{\rm Synch}{\rm IC}$\tablenotemark{b}} & \colhead{$\Gamma_{2-10}^{\rm synch}$} & \colhead{$f_{\rm GeV}$\tablenotemark{c}} & \colhead{$f_{\rm VHE}$\tablenotemark{d}} & & \colhead{$\xi$} & \colhead {$L_{2-10}^{\rm synch}$} & \colhead{$f_{2-10}^{\rm synch}$\tablenotemark{a}} & \colhead{$\frac{\rm Synch}{\rm IC}$\tablenotemark{b}} & \colhead{$\Gamma_{2-10}^{\rm synch}$} & \colhead{$f_{\rm GeV}$\tablenotemark{c}} & \colhead{$f_{\rm VHE}$\tablenotemark{d}} \\ \colhead{$\muGauss$} & & \colhead{(Myr)} & & & \colhead{$\ergps$} & & & & & & & & \colhead{$\ergps$} & & & & }
\startdata
50    & 2.0 & $\infty$ & 0.025  & 0.092  & 2.1e37 & 0.024  & 0.37  & 2.30 & 1.2  & 0.64  & & \multicolumn{7}{c}{\nodata}\\
      &     &          & 0.071  & 0.090  & 5.9e37 & 0.069  & 1.0   & 2.30 & 1.3  & 1.8   & & \multicolumn{7}{c}{\nodata}\\
      &     & 10       & 0.050  & 0.091  & 9.5e36 & 0.011  & 0.17  & 2.35 & 1.2  & 0.68  & & \multicolumn{7}{c}{\nodata}\\
      &     &          & 0.14   & 0.087  & 2.7e37 & 0.032  & 0.48  & 2.35 & 1.3  & 1.9   & & \multicolumn{7}{c}{\nodata}\\
      &     & 1        & 0.20   & 0.092  & 4.8e36 & 0.0057 & 0.084 & 2.37 & 1.3  & 0.66  & & \multicolumn{7}{c}{\nodata}\\
      &     &          & 0.57   & 0.081  & 1.4e37 & 0.016  & 0.24  & 2.37 & 1.6  & 1.8   & & \multicolumn{7}{c}{\nodata}\\
      & 2.2 & $\infty$ & 0.050  & 0.15   & 9.0e36 & 0.011  & 0.10  & 2.33 & 1.0  & 0.53  & & \multicolumn{7}{c}{\nodata}\\
      &     &          & 0.14   & 0.13   & 2.6e37 & 0.030  & 0.32  & 2.33 & 1.1  & 1.5   & & \multicolumn{7}{c}{\nodata}\\
      &     & 10       & 0.0088 & \multicolumn{7}{c}{\nodata}                            & & 0.18  & 3.6e38 & 0.43   & 3.9   & 2.08 & 1.0  & 2.0\\
      &     &          & 0.013  & \multicolumn{7}{c}{\nodata}                            & & 0.18  & 3.6e38 & 0.43   & 3.9   & 2.08 & 1.0  & 2.0\\
      &     &          & 0.10   & 0.14   & 4.2e36 & 0.0050 & 0.051 & 2.38 & 1.1  & 0.63  & & \multicolumn{7}{c}{\nodata}\\
      &     &          & 0.28   & 0.12   & 1.2e37 & 0.014  & 0.15  & 2.38 & 1.4  & 1.8   & & \multicolumn{7}{c}{\nodata}\\
      &     & 1        & 0.10   & \multicolumn{7}{c}{\nodata}                            & & 0.17  & 3.5e38 & 0.41   & 3.8   & 2.03 & 1.1  & 2.0\\
      &     &          & 0.28   & \multicolumn{7}{c}{\nodata}                            & & 0.14  & 2.9e38 & 0.34   & 3.5   & 2.03 & 1.3  & 2.0\\
      &     &          & 0.40   & 0.12   & 2.2e36 & 0.0026 & 0.027 & 2.40 & 1.4  & 0.67  & & \multicolumn{7}{c}{\nodata}\\
      &     &          & 0.57   & 0.11   & 3.1e36 & 0.0036 & 0.038 & 2.40 & 1.7  & 0.94  & & \multicolumn{7}{c}{\nodata}\\
      & 2.4 & $\infty$ & 0.0088 & \multicolumn{7}{c}{\nodata}                            & & 0.43  & 5.1e37 & 0.060  & 0.33  & 2.27 & 0.95 & 0.51\\
      &     &          & 0.050  & \multicolumn{7}{c}{\nodata}                            & & 0.41  & 5.0e37 & 0.059  & 0.34  & 2.27 & 1.0  & 0.61\\
      &     &          & 0.20   & 0.34   & 4.8e36 & 0.0056 & 0.037 & 2.36 & 1.2  & 0.59  & & 0.34\tablenotemark{e} & 4.5e37 & 0.053  & 0.35  & 2.27 & 1.2  & 0.96\\
      &     &          & 0.57   & 0.23   & 1.3e37 & 0.016  & 0.13  & 2.36 & 1.9  & 1.7   & & 0.23  & 4.1e37 & 0.048  & 0.38  & 2.27 & 1.9  & 1.9\\
      &     & 10       & 0.0088 & \multicolumn{7}{c}{\nodata}                            & & 0.44  & 5.1e37 & 0.061  & 0.33  & 2.20 & 0.95 & 0.50\\
      &     &          & 0.050  & \multicolumn{7}{c}{\nodata}                            & & 0.42  & 4.9e37 & 0.058  & 0.33  & 2.20 & 1.0  & 0.56\\
      &     &          & 0.28   & 0.31   & 1.6e36 & 0.0019 & 0.013 & 2.42 & 1.3  & 0.54  & & 0.31\tablenotemark{e} & 3.8e37 & 0.045  & 0.31  & 2.20 & 1.3  & 0.88\\
      &     &          & 0.57   & 0.24   & 3.2e36 & 0.0038 & 0.030 & 2.42 & 1.9  & 1.1   & & 0.24  & 3.1e37 & 0.037  & 0.29  & 2.20 & 1.9  & 1.3\\
      &     & 1        & 0.0088 & \multicolumn{7}{c}{\nodata}                            & & 0.45  & 5.3e37 & 0.062  & 0.33  & 2.12 & 0.95 & 0.50\\
      &     &          & 0.050  & \multicolumn{7}{c}{\nodata}                            & & 0.44  & 5.1e37 & 0.061  & 0.33  & 2.12 & 1.0  & 0.51\\
      &     &          & 0.57   & \multicolumn{7}{c}{\nodata}                            & & 0.27  & 3.2e37 & 0.038  & 0.28  & 2.12 & 1.6  & 0.61\\
100   & 2.0 & 10       & 0.14   & 0.028  & 4.3e37 & 0.050  & 2.2   & 2.33 & 0.53 & 1.8   & & \multicolumn{7}{c}{\nodata}\\
      &     & 1        & 0.14   & \multicolumn{7}{c}{\nodata}                            & & 0.048 & 1.0e39 & 1.2    & 49    & 2.00 & 0.51 & 2.0\\
      &     &          & 0.20   & 0.029  & 7.8e36 & 0.0092 & 0.38  & 2.35 & 0.53 & 0.62  & & \multicolumn{7}{c}{\nodata}\\
      &     &          & 0.57   & 0.017  & 2.2e37 & 0.026  & 1.1   & 2.35 & 0.81 & 1.7   & & \multicolumn{7}{c}{\nodata}\\
      & 2.2 & $\infty$ & 0.14   & 0.035  & 4.0e37 & 0.047  & 1.6   & 2.31 & 0.54 & 1.4   & & 0.038 & 1.4e38 & 0.16   & 5.4   & 2.19 & 0.54 & 1.6\\
      &     &          & 0.20   & 0.030  & 5.7e37 & 0.067  & 2.3   & 2.31 & 0.66 & 2.0   & & \multicolumn{7}{c}{\nodata}\\
      &     & 10       & 0.14   & 0.036  & 9.8e36 & 0.012  & 0.38  & 2.37 & 0.52 & 0.85  & & 0.039 & 1.1e38 & 0.13   & 4.3   & 2.14 & 0.53 & 1.1\\
      &     &          & 0.28   & 0.024  & 2.0e37 & 0.023  & 0.81  & 2.37 & 0.79 & 1.7   & & 0.025 & 8.5e37 & 0.10   & 3.5   & 2.14 & 0.80 & 1.9\\
      &     & 1        & 0.20   & \multicolumn{7}{c}{\nodata}                            & & 0.040 & 1.1e38 & 0.12   & 3.9   & 2.08 & 0.56 & 0.59\\
      &     &          & 0.40   & 0.025  & 3.6e36 & 0.0043 & 0.14  & 2.39 & 0.85 & 0.65  & & 0.027\tablenotemark{e,f} & 7.2e37 & 0.085  & 2.8   & 2.08 & 0.85 & 0.82\\
      &     &          & 0.57   & 0.014  & 5.2e36 & 0.0061 & 0.21  & 2.39 & 1.1  & 0.92  & & 0.015 & 4.3e37 & 0.051  & 1.7   & 2.08 & 1.1  & 1.0\\
      & 2.4 & $\infty$ & 0.20   & 0.068  & 7.6e36 & 0.0090 & 0.23  & 2.34 & 0.62 & 0.56  & & 0.068 & 1.9e37 & 0.022  & 0.56  & 2.29 & 0.62 & 0.61\\
      &     &          & 0.40   & 0.035  & 1.5e37 & 0.018  & 0.51  & 2.34 & 1.0  & 1.1   & & 0.035 & 2.1e37 & 0.025  & 0.70  & 2.29 & 1.0  & 1.2\\
      &     &          & 0.57   & 0.0064 & 2.2e37 & 0.025  & 0.80  & 2.34 & 1.4  & 1.6   & & 0.0064 & 2.3e37 & 0.027 & 0.84  & 2.29 & 1.4  & 1.6\\
      &     & 10       & 0.28   & 0.057  & 2.7e36 & 0.0032 & 0.085 & 2.41 & 0.78 & 0.53  & & 0.058 & 1.2e37 & 0.014  & 0.37  & 2.26 & 0.78 & 0.57\\
      &     &          & 0.40   & 0.039  & 3.9e36 & 0.0046 & 0.13  & 2.41 & 1.0  & 0.75  & & 0.039 & 1.0e37 & 0.012  & 0.33  & 2.26 & 1.0  & 0.78\\
      &     &          & 0.57   & 0.012  & 5.5e36 & 0.0065 & 0.20  & 2.41 & 1.4  & 1.1   & & 0.012 & 7.4e36 & 0.0087 & 0.26  & 2.26 & 1.4  & 1.1\\
150   & 2.0 & 1        & 0.40   & \multicolumn{7}{c}{\nodata}                            & & 0.011 & 2.6e38 & 0.31   & 23    & 2.01 & 0.50 & 1.4\\
      &     &          & 0.57   & 0.0016 & 2.7e37 & 0.032  & 2.5   & 2.33 & 0.63 & 1.7   & & 0.0024 & 8.1e37 & 0.095 & 7.4   & 2.01 & 0.63 & 1.7\\
      & 2.2 & $\infty$ & 0.20   & 0.0078 & 6.8e37 & 0.080  & 5.5   & 2.28 & 0.52 & 1.9   & & 0.0085 & 9.2e37 & 0.11  & 7.4   & 2.20 & 0.52 & 1.9\\
      &     & 10       & 0.20   & 0.0090 & 1.7e37 & 0.020  & 1.3   & 2.35 & 0.50 & 1.2   & & 0.0094 & 4.4e37 & 0.051 & 3.5   & 2.15 & 0.50 & 1.2\\
      &     &          & 0.28   & 0.0014 & 2.4e37 & 0.029  & 2.1   & 2.35 & 0.66 & 1.7   & & 0.0016 & 2.9e37 & 0.034 & 2.4   & 2.15 & 0.66 & 1.7\\
      &     & 1        & 0.40   & 9.7E-4 & 4.6e36 & 0.0054 & 0.35  & 2.38 & 0.72 & 0.64  & & 0.0012 & 8.0e36 & 0.0094 & 0.61 & 2.10 & 0.72 & 0.64\\
      & 2.4 & $\infty$ & 0.20   & 0.021  & 9.3e36 & 0.011  & 0.59  & 2.32 & 0.52 & 0.55  & & 0.020  & 1.3e37 & 0.015 & 0.83  & 2.28 & 0.52 & 0.56\\
      &     &          & 0.28   & 0.0061 & 1.3e37 & 0.016  & 0.92  & 2.32 & 0.69 & 0.78  & & 0.0061 & 1.4e37 & 0.017 & 1.0   & 2.28 & 0.69 & 0.78\\
      &     & 10       & 0.28   & 0.0083 & 3.5e36 & 0.0041 & 0.23  & 2.40 & 0.67 & 0.52  & & 0.0084 & 5.0e36 & 0.0059 & 0.34 & 2.27 & 0.67 & 0.53
\enddata
\tablenotetext{a}{Fraction of the observed \emph{diffuse} 2 - 10 keV X-ray luminosity (from \citealt{Bauer08}) from that synchrotron emission accounts for.}
\tablenotetext{b}{Ratio of the 2 - 10 keV synchrotron and IC luminosities.}
\tablenotetext{c}{Ratio of the predicted 0.3 - 10 GeV $\gamma$-ray emission with the flux detected by \emph{Fermi}-LAT in \citet{Abdo10b}.}
\tablenotetext{d}{Ratio of the predicted $\ge$220 GeV $\gamma$-ray emission with the observed TeV flux observed from HESS in \citet{Acero09}.}
\tablenotetext{e}{This model satisfies our constraints (\S~\ref{sec:Constraints}).  While $\eta$ is not the minimum or maximum allowed or the closest to reproducing the GeV luminosity for $\gamma_{\rm max}^{\rm prim} = 10^9$, this is true for $\gamma_{\rm max}^{\rm prim} = 10^6$ so we list it anyway.}
\tablenotetext{f}{Chosen as our fiducial model.}
\label{table:NGC253XRayLuminosities}
\end{deluxetable}

\begin{deluxetable}{ccccccccccccccccccc}
\tabletypesize{\scriptsize}
\tablecaption{M82 X-ray luminosities}
\tablehead{\colhead{$B$} & \colhead{$p$} & \colhead{$t_{\rm diff} (3\ \GeV)$} & \colhead{$\eta$} & \multicolumn{7}{c}{$\gamma_{\rm max}^{\rm prim} = 10^6$} & & \multicolumn{7}{c}{$\gamma_{\rm max}^{\rm prim} = 10^9$} \\ & & & & \colhead{$\xi$} & \colhead {$L_{2-8}^{\rm synch}$} & \colhead{$f_{2-8}^{\rm synch}$\tablenotemark{a}} & \colhead{$\frac{\rm Synch}{\rm IC}$\tablenotemark{b}} & \colhead{$\Gamma_{2-8}^{\rm synch}$} & \colhead{$f_{\rm GeV}$\tablenotemark{c}} & \colhead{$f_{\rm VHE}$\tablenotemark{d}} & & \colhead{$\xi$} & \colhead {$L_{2-8}^{\rm synch}$} & \colhead{$f_{2-8}^{\rm synch}$\tablenotemark{a}} & \colhead{$\frac{\rm Synch}{\rm IC}$\tablenotemark{b}} & \colhead{$\Gamma_{2-8}^{\rm synch}$} & \colhead{$f_{\rm GeV}$\tablenotemark{c}} & \colhead{$f_{\rm VHE}$\tablenotemark{d}} \\ \colhead{$\muGauss$} & & \colhead{(Myr)} & & & \colhead{$\ergps$} & & & & & & & & \colhead{$\ergps$} & & & & }
\startdata
50    & 2.0 & $\infty$ & 0.013  & 0.067  & 5.4e37 & 0.012  & 0.29  & 2.25 & 1.8  & 0.63 & & \multicolumn{7}{c}{\nodata}\\
      &     &          & 0.035  & 0.066  & 1.5e38 & 0.035  & 0.83  & 2.25 & 1.9  & 1.8  & & \multicolumn{7}{c}{\nodata}\\
      &     &          & 0.035  & 0.066  & 2.5e37 & 0.0056 & 0.14  & 2.31 & 1.8  & 0.61 & & \multicolumn{7}{c}{\nodata}\\
      &     &          & 0.071  & 0.065  & 5.0e37 & 0.011  & 0.27  & 2.31 & 2.0  & 1.2  & & \multicolumn{7}{c}{\nodata}\\
      & 2.2 & $\infty$ & 0.0088 & \multicolumn{7}{c}{\nodata}                           & & 0.12   & 7.1e38 & 0.16   & 2.5  & 2.17 & 1.4  & 1.8\\
      &     &          & 0.025  & \multicolumn{7}{c}{\nodata}                           & & 0.11   & 6.9e38 & 0.16   & 2.5  & 2.17 & 1.5  & 2.0\\
      &     &          & 0.035  & 0.10   & 3.3e37 & 0.0075 & 0.13  & 2.28 & 1.4  & 0.61 & & \multicolumn{7}{c}{\nodata}\\
      &     &          & 0.10   & 0.094  & 9.4e37 & 0.021  & 0.36  & 2.28 & 1.8  & 1.7  & & \multicolumn{7}{c}{\nodata}\\
      &     & 10       & 0.0088 & \multicolumn{7}{c}{\nodata}                           & & 0.12   & 7.1e38 & 0.16   & 2.5  & 2.07 & 1.4  & 1.7\\
      &     &          & 0.071  & \multicolumn{7}{c}{\nodata}                           & & 0.11   & 6.3e38 & 0.14   & 2.4  & 2.07 & 1.7  & 1.9\\
      &     &          & 0.10   & 0.096  & 1.6e37 & 0.0036 & 0.061 & 2.34 & 1.8  & 0.68 & & \multicolumn{7}{c}{\nodata}\\
      &     &          & 0.14   & 0.091  & 2.2e37 & 0.0051 & 0.087 & 2.34 & 2.0  & 0.96 & & \multicolumn{7}{c}{\nodata}\\
      &     & 1        & 0.0088 & \multicolumn{7}{c}{\nodata}                           & & 0.13   & 7.5e38 & 0.17   & 2.6  & 2.01 & 1.5  & 1.7\\
      &     &          & 0.14   & \multicolumn{7}{c}{\nodata}                           & & 0.11   & 6.3e38 & 0.14   & 2.3  & 2.01 & 1.8  & 1.6\\
      & 2.4 & $\infty$ & 0.050  & \multicolumn{7}{c}{\nodata}                           & & 0.28   & 1.0e38 & 0.023  & 0.22 & 2.26 & 1.5  & 0.52\\
      &     &          & 0.14   & 0.22   & 1.7e37 & 0.0040 & 0.047 & 2.31 & 1.9  & 0.56 & & 0.21   & 9.0e37 & 0.020  & 0.24 & 2.26 & 1.9  & 0.81\\
      &     & 10       & 0.14   & \multicolumn{7}{c}{\nodata}                           & & 0.22   & 7.9e37 & 0.018  & 0.20 & 2.19 & 1.9  & 0.50\\
100   & 2.0 & $\infty$ & 0.0088 & \multicolumn{7}{c}{\nodata}                           & & 0.037  & 2.3e39 & 0.52   & 35   & 2.12 & 0.62 & 1.9\\
      &     &          & 0.013  & 0.023  & 8.3e37 & 0.019  & 1.3   & 2.24 & 0.59 & 0.55 & & \multicolumn{7}{c}{\nodata}\\
      &     &          & 0.035  & 0.022  & 2.4e38 & 0.054  & 3.7   & 2.24 & 0.67 & 1.6  & & \multicolumn{7}{c}{\nodata}\\
      &     & 10       & 0.0088 & \multicolumn{7}{c}{\nodata}                           & & 0.037  & 2.2e39 & 0.51   & 35   & 2.03 & 0.61 & 1.7\\
      &     &          & 0.035  & 0.022  & 4.0e37 & 0.0092 & 0.63  & 2.30 & 0.65 & 0.57 & & 0.034  & 2.1e39 & 0.48   & 33   & 2.03 & 0.69 & 2.0\\
      &     &          & 0.10   & 0.018  & 1.1e38 & 0.026  & 1.8   & 2.30 & 0.84 & 1.6  & & \multicolumn{7}{c}{\nodata}\\
      &     & 1        & 0.0088 & \multicolumn{7}{c}{\nodata}                           & & 0.039  & 2.4e39 & 0.54   & 36   & 1.98 & 0.61 & 1.6\\
      &     &          & 0.20   & 0.018  & 2.7e37 & 0.0062 & 0.41  & 2.32 & 0.93 & 0.54 & & 0.028  & 1.7e39 & 0.39   & 26   & 1.98 & 0.96 & 1.7\\
      &     &          & 0.57   & 0.0047 & 7.8e37 & 0.018  & 1.2   & 2.32 & 1.6  & 1.5  & & 0.0071 & 5.1e38 & 0.12   & 7.6  & 1.98 & 1.6  & 1.8\\
      & 2.2 & $\infty$ & 0.025  & \multicolumn{7}{c}{\nodata}                           & & 0.034  & 2.9e38 & 0.067  & 3.5  & 2.20 & 0.52 & 0.65\\
      &     &          & 0.035  & 0.031  & 5.3e37 & 0.012  & 0.63  & 2.27 & 0.57 & 0.56 & & 0.033\tablenotemark{e} & 3.0e38 & 0.068  & 3.6  & 2.20 & 0.58 & 0.80\\
      &     &          & 0.10   & 0.022  & 1.5e38 & 0.034  & 1.9   & 2.27 & 0.94 & 1.6  & & 0.024  & 3.3e38 & 0.074  & 4.1  & 2.20 & 0.94 & 1.8\\
      &     & 10       & 0.050  & \multicolumn{7}{c}{\nodata}                           & & 0.032  & 2.5e38 & 0.057  & 3.0  & 2.13 & 0.64 & 0.55\\
      &     &          & 0.10   & 0.023  & 2.6e37 & 0.0060 & 0.32  & 2.34 & 0.89 & 0.65 & & 0.025  & 2.1e38 & 0.049  & 2.6  & 2.13 & 0.90 & 0.83\\
      &     &          & 0.28   & 7.5E-4 & 7.5e37 & 0.017  & 1.1   & 2.34 & 1.8  & 1.8  & & 9.4E-4 & 8.2e37 & 0.019  & 1.2  & 2.13 & 1.8  & 1.8\\
      &     & 1        & 0.40   & \multicolumn{7}{c}{\nodata}                           & & 0.0064 & 6.1e37 & 0.014  & 0.73 & 2.07 & 1.8  & 0.50\\
      & 2.4 & $\infty$ & 0.14   & 0.042  & 2.9e37 & 0.0065 & 0.28  & 2.31 & 1.1  & 0.52 & & 0.042  & 4.8e37 & 0.011  & 0.47 & 2.28 & 1.1  & 0.55\\
      &     &          & 0.28   & 0.0081 & 5.7e37 & 0.013  & 0.64  & 2.31 & 2.0  & 1.0  & & 0.0081 & 6.1e37 & 0.014  & 0.68 & 2.28 & 2.0  & 1.1\\
150   & 2.0 & 10       & 0.10   & 0.0075 & 1.4e38 & 0.032  & 4.1   & 2.28 & 0.56 & 1.5  & & 0.012  & 9.0e38 & 0.20   & 26   & 2.05 & 0.57 & 1.8\\
      &     & 1        & 0.14   & \multicolumn{7}{c}{\nodata}                           & & 0.014  & 9.2e38 & 0.21   & 25   & 2.00 & 0.55 & 0.70\\
      &     &          & 0.20   & 0.0067 & 3.4e37 & 0.0077 & 0.93  & 2.30 & 0.65 & 0.52 & & 0.010\tablenotemark{e} & 7.0e38 & 0.16 & 19 & 2.00 & 0.65 & 0.77\\
      &     &          & 0.28   & 0.0036 & 4.8e37 & 0.011  & 1.3   & 2.30 & 0.81 & 0.74 & & 0.0056 & 4.1e38 & 0.093  & 11   & 2.00 & 0.81 & 0.87\\
      & 2.2 & $\infty$ & 0.071  & 0.0099 & 1.3e38 & 0.029  & 3.0   & 2.25 & 0.58 & 1.1  & & 0.011  & 2.1e38 & 0.049  & 5.1  & 2.20 & 0.58 & 1.1\\
      &     &          & 0.10   & 0.0061 & 1.8e38 & 0.041  & 4.5   & 2.25 & 0.74 & 1.5  & & 0.0064 & 2.3e38 & 0.053  & 5.9  & 2.20 & 0.74 & 1.5\\
      &     & 10       & 0.10   & 0.0071 & 3.3e37 & 0.0075 & 0.80  & 2.33 & 0.70 & 0.63 & & 0.0077\tablenotemark{f} & 9.7e37 & 0.022  & 2.3  & 2.15 & 0.70 & 0.66\\
      &     &          & 0.14   & 0.0020 & 4.6e37 & 0.011  & 1.2   & 2.33 & 0.91 & 0.89 & & 0.0023 & 6.5e37 & 0.015  & 1.7  & 2.15 & 0.91 & 0.90\\
      & 2.4 & $\infty$ & 0.14   & 0.0060 & 3.5e37 & 0.0079 & 0.74  & 2.29 & 0.97 & 0.51 & & 0.0061 & 3.8e37 & 0.0087 & 0.81 & 2.27 & 0.97 & 0.51\\
200   & 2.0 & 1        & 0.20   & 0.0021 & 3.8e37 & 0.0087 & 1.6   & 2.28 & 0.54 & 0.52 & & 0.0032 & 2.5e38 & 0.057  & 10   & 2.01 & 0.54 & 0.56\\
      & 2.2 & 10       & 0.10   & 7.2E-4 & 3.7e37 & 0.0085 & 1.5   & 2.31 & 0.62 & 0.62 & & 7.0E-4 & 4.4e37 & 0.0099 & 1.7  & 2.16 & 0.62 & 0.62
\enddata
\tablenotetext{a}{Fraction of the observed \emph{diffuse} 2 - 8 keV X-ray luminosity (from \citealt{Strickland07}) from that synchrotron emission accounts for.}
\tablenotetext{b}{Ratio of the 2 - 8 keV synchrotron and IC luminosities.}
\tablenotetext{c}{Ratio of the predicted 0.3 - 10 GeV $\gamma$-ray emission with the flux detected by \emph{Fermi}-LAT in \citet{Abdo10b}.}
\tablenotetext{d}{Ratio of the predicted $\ge$700 GeV $\gamma$-ray emission with the observed TeV flux observed from HESS in \citet{Acciari09}.}
\tablenotetext{e}{This model satisfies our constraints (\S~\ref{sec:Constraints}).  While $\eta$ is not the minimum or maximum allowed or the closest to reproducing the GeV luminosity for $\gamma_{\rm max}^{\rm prim} = 10^9$, this is true for $\gamma_{\rm max}^{\rm prim} = 10^6$ so we list it anyway.}
\tablenotetext{f}{Chosen as our fiducial model.}
\label{table:M82XRayLuminosities}
\end{deluxetable}

\LongTables
\begin{deluxetable}{cccccccccccccccccc}
\tabletypesize{\scriptsize}
\tablecaption{Arp 220 X-ray luminosities}
\tablehead{\colhead{$B$} & \colhead{$p$} & \colhead{$t_{\rm diff} (3\ \GeV)$} & \multicolumn{7}{c}{$\gamma_{\rm max}^{\rm prim} = 10^6$} & & \multicolumn{7}{c}{$\gamma_{\rm max}^{\rm prim} = 10^9$} \\ & & & \colhead{$\xi$} & \colhead {$L_{2-10}^{\rm synch}$} & \colhead{$f_{2-10}^{\rm synch}$\tablenotemark{a}} & \colhead{$\frac{\rm Synch}{\rm IC}$\tablenotemark{b}} & \colhead{$\Gamma_{2-10}^{\rm synch}$} & \colhead{$f_{\rm GeV}$\tablenotemark{c}} & \colhead{$f_{\rm VHE}$\tablenotemark{d}} & & \colhead{$\xi$} & \colhead {$L_{2-10}^{\rm synch}$} & \colhead{$f_{2-10}^{\rm synch}$\tablenotemark{a}} & \colhead{$\frac{\rm Synch}{\rm IC}$\tablenotemark{b}} & \colhead{$\Gamma_{2-10}^{\rm synch}$} & \colhead{$f_{\rm GeV}$\tablenotemark{c}} & \colhead{$f_{\rm VHE}$\tablenotemark{d}} \\ \colhead{$\muGauss$} & & \colhead{(Myr)} & & \colhead{$\ergps$} & & & & & & & & \colhead{$\ergps$} & & & & }
\startdata
\cutinhead{West nucleus}
500    & 2.0 & $\infty$ & 0.81   & 2.6e39 & 0.065  & 0.0416 & 1.90 & 0.60  & 0.016  & & 1.3   & 1.2e41 & 2.9   & 1.9   & 1.85 & 0.67  & 0.23\\
       &     & 10       & 0.81   & 2.4e39 & 0.061  & 0.0388 & 1.90 & 0.60  & 0.016  & & 1.3   & 1.2e41 & 2.9   & 1.9   & 1.85 & 0.67  & 0.23\\
       &     & 1        & 0.81   & 1.5e39 & 0.039  & 0.0246 & 1.93 & 0.60  & 0.012  & & 1.3   & 1.2e41 & 2.9   & 1.9   & 1.84 & 0.66  & 0.23\\
       & 2.2 & $\infty$ & 1.3    & 6.0e38 & 0.015  & 0.0072 & 1.94 & 0.49  & 0.0060 & & 1.3   & 1.7e40 & 0.41  & 0.20  & 1.91 & 0.50  & 0.044\\
       &     & 10       & 1.3    & 5.6e38 & 0.014  & 0.0068 & 1.94 & 0.49  & 0.0058 & & 1.3   & 1.7e40 & 0.41  & 0.20  & 1.91 & 0.50  & 0.044\\
       &     & 1        & 1.3    & 3.6e38 & 0.0091 & 0.0044 & 1.97 & 0.49  & 0.0047 & & 1.3   & 1.6e40 & 0.41  & 0.20  & 1.92 & 0.50  & 0.043\\
       & 2.4 & $\infty$ & 2.8    & 8.5e37 & 0.0021 & 8E-4   & 1.98 & 0.46  & 0.0015 & & 2.8   & 2.3e39 & 0.058 & 0.021 & 1.97 & 0.46  & 0.0085\\
       &     & 10       & 2.8    & 8.0e37 & 0.0020 & 7E-4   & 1.99 & 0.46  & 0.0014 & & 2.8   & 2.3e39 & 0.058 & 0.021 & 1.98 & 0.46  & 0.0085\\
       &     & 1        & 2.8    & 5.3e37 & 0.0013 & 5E-4   & 2.02 & 0.46  & 0.0012 & & 2.8   & 2.3e39 & 0.058 & 0.021 & 2.00 & 0.46  & 0.0082\\
1000   & 2.0 & $\infty$ & 0.25   & 6.5e39 & 0.16   & 0.33   & 1.97 & 0.19  & 0.015  & & 0.38  & 7.3e40 & 1.8   & 3.7   & 1.93 & 0.21  & 0.061\\
       &     & 10       & 0.25   & 6.1e39 & 0.15   & 0.31   & 1.97 & 0.19  & 0.014  & & 0.38  & 7.3e40 & 1.8   & 3.7   & 1.93 & 0.21  & 0.060\\
       &     & 1        & 0.25   & 3.9e39 & 0.098  & 0.20   & 2.00 & 0.19  & 0.011  & & 0.38  & 7.1e40 & 1.8   & 3.6   & 1.93 & 0.21  & 0.057\\
       & 2.2 & $\infty$ & 0.35   & 1.6e39 & 0.039  & 0.063  & 2.01 & 0.16  & 0.0056 & & 0.38  & 1.1e40 & 0.28  & 0.45  & 1.99 & 0.16  & 0.014\\
       &     & 10       & 0.35   & 1.5e39 & 0.037  & 0.060  & 2.01 & 0.16  & 0.0054 & & 0.38  & 1.1e40 & 0.27  & 0.45  & 1.99 & 0.16  & 0.014\\
       &     & 1        & 0.36   & 9.7e38 & 0.024  & 0.040  & 2.04 & 0.16  & 0.0045 & & 0.38  & 1.1e40 & 0.26  & 0.43  & 2.01 & 0.16  & 0.013\\
       & 2.4 & $\infty$ & 0.75   & 2.3e38 & 0.0058 & 0.0075 & 2.06 & 0.15  & 0.0014 & & 0.76  & 1.6e39 & 0.041 & 0.053 & 2.05 & 0.15  & 0.0029\\
       &     & 10       & 0.75   & 2.2e38 & 0.0054 & 0.0071 & 2.07 & 0.15  & 0.0014 & & 0.76  & 1.6e39 & 0.041 & 0.052 & 2.06 & 0.15  & 0.0028\\
       &     & 1        & 0.76   & 1.5e38 & 0.0037 & 0.0048 & 2.10 & 0.15  & 0.0012 & & 0.76  & 1.6e39 & 0.039 & 0.050 & 2.08 & 0.15  & 0.0026\\
2000   & 2.0 & $\infty$ & 0.077  & 1.3e40 & 0.31   & 1.9    & 2.04 & 0.069 & 0.012  & & 0.12  & 4.6e40 & 1.1   & 6.9   & 2.01 & 0.073 & 0.020\\
       &     & 10       & 0.077  & 1.2e40 & 0.29   & 1.8    & 2.04 & 0.069 & 0.012  & & 0.12  & 4.5e40 & 1.1   & 6.8   & 2.01 & 0.073 & 0.020\\
       &     & 1        & 0.077  & 7.9e39 & 0.20   & 1.2    & 2.07 & 0.068 & 0.0096 & & 0.12  & 4.1e40 & 1.0   & 6.2   & 2.02 & 0.048 & 0.0078\\
       & 2.2 & $\infty$ & 0.097  & 3.2e39 & 0.079  & 0.41   & 2.09 & 0.064 & 0.0049 & & 0.10  & 7.6e39 & 0.19  & 1.0   & 2.07 & 0.064 & 0.0062\\
       &     & 10       & 0.097  & 3.0e39 & 0.075  & 0.39   & 2.09 & 0.064 & 0.0048 & & 0.10  & 7.5e39 & 0.19  & 0.98  & 2.08 & 0.065 & 0.0061\\
       &     & 1        & 0.097  & 2.0e39 & 0.051  & 0.26   & 2.12 & 0.063 & 0.0040 & & 0.10  & 6.6e39 & 0.17  & 0.86  & 2.10 & 0.064 & 0.0053\\
       & 2.4 & $\infty$ & 0.20   & 4.9e38 & 0.012  & 0.054  & 2.15 & 0.061 & 0.0013 & & 0.20  & 1.2e39 & 0.030 & 0.13  & 2.14 & 0.061 & 0.0015\\
       &     & 10       & 0.20   & 4.7e38 & 0.012  & 0.052  & 2.15 & 0.035 & 0.0013 & & 0.20  & 1.2e39 & 0.029 & 0.13  & 2.15 & 0.061 & 0.0015\\
       &     & 1        & 0.20   & 3.3e38 & 0.0082 & 0.036  & 2.19 & 0.061 & 0.0011 & & 0.20  & 1.0e39 & 0.026 & 0.11  & 2.17 & 0.061 & 0.0013\\
4000   & 2.0 & $\infty$ & 0.023  & 1.8e40 & 0.46   & 7.8    & 2.07 & 0.030 & 0.010  & & 0.036 & 3.1e40 & 0.79  & 13    & 2.05 & 0.030 & 0.011\\
       &     & 10       & 0.023  & 1.7e40 & 0.44   & 7.3    & 2.07 & 0.030 & 0.0099 & & 0.036 & 3.0e40 & 0.76  & 13    & 2.05 & 0.030 & 0.011\\
       &     & 1        & 0.024  & 1.2e40 & 0.30   & 5.0    & 2.10 & 0.029 & 0.0082 & & 0.036 & 2.5e40 & 0.63  & 11    & 2.07 & 0.030 & 0.0091\\
       & 2.2 & $\infty$ & 0.020  & 4.9e39 & 0.12   & 1.9    & 2.13 & 0.036 & 0.0043 & & 0.021 & 6.2e39 & 0.16  & 2.4   & 2.12 & 0.036 & 0.0044\\
       &     & 10       & 0.020  & 4.6e39 & 0.12   & 1.8    & 2.13 & 0.036 & 0.0042 & & 0.021\tablenotemark{f} & 6.0e39 & 0.15  & 2.3   & 2.12 & 0.036 & 0.0043\\
       &     & 1        & 0.020  & 3.3e39 & 0.082  & 1.3    & 2.16 & 0.035 & 0.0036 & & 0.022 & 4.6e39 & 0.12  & 1.8   & 2.15 & 0.035 & 0.0037\\
       & 2.4 & $\infty$ & 0.039  & 8.0e38 & 0.020  & 0.28   & 2.19 & 0.037 & 0.0012 & & 0.039 & 1.0e39 & 0.026 & 0.36  & 2.19 & 0.037 & 0.0012\\
       &     & 10       & 0.039  & 8.0e38 & 0.020  & 0.28   & 2.19 & 0.037 & 0.0012 & & 0.040 & 9.8e38 & 0.025 & 0.35  & 2.19 & 0.037 & 0.0012\\
       &     & 1        & 0.040  & 5.5e38 & 0.014  & 0.20   & 2.23 & 0.037 & 0.0010 & & 0.040 & 7.7e38 & 0.019 & 0.27  & 2.22 & 0.037 & 0.0010\\
8000   & 2.0 & $\infty$ & 0.0057 & 2.2e40 & 0.56   & 26     & 2.06 & 0.018 & 0.0092 & & 0.0088 & 2.6e40 & 0.65 & 30    & 2.05 & 0.018 & 0.0093\\
       &     & 10       & 0.0057 & 2.1e40 & 0.53   & 24     & 2.06 & 0.018 & 0.0090 & & 0.0088 & 2.5e40 & 0.62 & 28    & 2.05 & 0.018 & 0.0091\\
       &     & 1        & 0.0059 & 1.5e40 & 0.37   & 17     & 2.09 & 0.018 & 0.0075 & & 0.0092 & 1.9e40 & 0.46 & 21    & 2.07 & 0.018 & 0.0076\\
16000  & 2.0 & 1        & 1.8E-4 & 1.6e40 & 0.41   & 57     & 2.06 & 0.015 & 0.0073 & & 2.7E-4 & 1.7e40 & 0.41 & 57    & 2.04 & 0.015 & 0.0073\\
\cutinhead{East nucleus}
500    & 2.0 & $\infty$ & 0.61   & 2.6e39 & 0.17   & 0.055  & 1.90 & 0.46  & 0.016  & & 0.95  & 8.9e40 & 5.9   & 1.9   & 1.85 & 0.51  & 0.18\\
       &     & 10       & 0.61   & 2.4e39 & 0.16   & 0.051  & 1.90 & 0.46  & 0.016  & & 0.95  & 8.9e40 & 5.9   & 1.9   & 1.85 & 0.51  & 0.18\\
       &     & 1        & 0.61   & 1.5e39 & 0.10   & 0.032  & 1.93 & 0.46  & 0.012  & & 0.95  & 8.8e40 & 5.9   & 1.9   & 1.84 & 0.51  & 0.17\\
       & 2.2 & $\infty$ & 0.95   & 6.0e38 & 0.040  & 0.0095 & 1.94 & 0.38  & 0.0060 & & 1.0   & 1.3e40 & 0.84  & 0.20  & 1.91 & 0.39  & 0.035\\
       &     & 10       & 0.95   & 5.6e38 & 0.038  & 0.0089 & 1.94 & 0.38  & 0.0058 & & 1.0   & 1.3e40 & 0.84  & 0.20  & 1.91 & 0.39  & 0.035\\
       &     & 1        & 0.95   & 3.6e38 & 0.024  & 0.0057 & 1.97 & 0.37  & 0.0047 & & 1.0   & 1.2e40 & 0.83  & 0.20  & 1.92 & 0.39  & 0.034\\
       & 2.4 & $\infty$ & 2.1    & 8.5e37 & 0.0057 & 0.0010 & 1.98 & 0.36  & 0.0015 & & 2.1   & 1.8e39 & 0.12  & 0.021 & 1.97 & 0.36  & 0.0068\\
       &     & 10       & 2.1    & 8.0e37 & 0.0054 & 9E-4   & 1.99 & 0.36  & 0.0014 & & 2.1   & 1.8e39 & 0.12  & 0.021 & 1.98 & 0.36  & 0.0067\\
       &     & 1        & 2.1    & 5.3e37 & 0.0035 & 6E-4   & 2.02 & 0.36  & 0.0012 & & 2.1   & 1.8e39 & 0.12  & 0.021 & 2.00 & 0.36  & 0.0065\\
1000   & 2.0 & $\infty$ & 0.18   & 6.5e39 & 0.43   & 0.43   & 1.97 & 0.15  & 0.015  & & 0.29  & 5.6e40 & 3.7   & 3.8   & 1.93 & 0.16  & 0.049\\
       &     & 10       & 0.18   & 6.1e39 & 0.40   & 0.41   & 1.97 & 0.15  & 0.014  & & 0.29  & 5.6e40 & 3.7   & 3.8   & 1.93 & 0.16  & 0.049\\
       &     & 1        & 0.18   & 3.9e39 & 0.26   & 0.26   & 2.00 & 0.15  & 0.011  & & 0.29  & 5.4e40 & 3.6   & 3.6   & 1.93 & 0.16  & 0.046\\
       & 2.2 & $\infty$ & 0.26   & 1.6e39 & 0.10   & 0.084  & 2.01 & 0.13  & 0.0056 & & 0.28  & 8.6e39 & 0.57  & 0.46  & 1.99 & 0.13  & 0.012\\
       &     & 10       & 0.26   & 1.5e39 & 0.098  & 0.079  & 2.01 & 0.13  & 0.0054 & & 0.28  & 8.5e39 & 0.57  & 0.46  & 1.99 & 0.13  & 0.011\\
       &     & 1        & 0.26   & 9.7e38 & 0.065  & 0.052  & 2.04 & 0.13  & 0.0045 & & 0.28  & 8.0e39 & 0.53  & 0.43  & 2.01 & 0.13  & 0.010\\
       & 2.4 & $\infty$ & 0.56   & 2.3e38 & 0.015  & 0.0099 & 2.06 & 0.12  & 0.0014 & & 0.56  & 1.3e39 & 0.085 & 0.054 & 2.05 & 0.12  & 0.0025\\
       &     & 10       & 0.56   & 2.2e38 & 0.015  & 0.0093 & 2.07 & 0.12  & 0.0014 & & 0.56  & 1.3e39 & 0.084 & 0.054 & 2.06 & 0.12  & 0.0025\\
       &     & 1        & 0.56   & 1.5e38 & 0.0099 & 0.0063 & 2.10 & 0.12  & 0.0012 & & 0.56  & 1.2e39 & 0.079 & 0.051 & 2.08 & 0.12  & 0.0023\\
2000   & 2.0 & $\infty$ & 0.054  & 1.3e40 & 0.83   & 2.5    & 2.04 & 0.056 & 0.012  & & 0.084 & 3.6e40 & 2.4   & 7.2   & 2.01 & 0.059 & 0.018\\
       &     & 10       & 0.054  & 1.2e40 & 0.79   & 2.4    & 2.04 & 0.056 & 0.012  & & 0.084 & 3.5e40 & 2.3   & 7.1   & 2.01 & 0.059 & 0.017\\
       &     & 1        & 0.055  & 7.9e39 & 0.52   & 1.6    & 2.07 & 0.055 & 0.0096 & & 0.085 & 3.1e40 & 2.1   & 6.3   & 2.02 & 0.058 & 0.015\\
       & 2.2 & $\infty$ & 0.066  & 3.2e39 & 0.21   & 0.54   & 2.09 & 0.054 & 0.0049 & & 0.070 & 6.3e39 & 0.42  & 1.1   & 2.07 & 0.055 & 0.0058\\
       &     & 10       & 0.066  & 3.0e39 & 0.20   & 0.51   & 2.09 & 0.054 & 0.0048 & & 0.070 & 6.1e39 & 0.41  & 1.0   & 2.08 & 0.055 & 0.0057\\
       &     & 1        & 0.066  & 2.0e39 & 0.14   & 0.35   & 2.12 & 0.054 & 0.0040 & & 0.071 & 5.2e39 & 0.34  & 0.89  & 2.10 & 0.054 & 0.0049\\
       & 2.4 & $\infty$ & 0.14   & 4.9e38 & 0.033  & 0.072  & 2.15 & 0.053 & 0.0013 & & 0.14  & 9.8e38 & 0.065 & 0.14  & 2.14 & 0.053 & 0.0015\\
       &     & 10       & 0.14   & 4.7e38 & 0.031  & 0.069  & 2.15 & 0.053 & 0.0013 & & 0.14  & 9.5e38 & 0.064 & 0.14  & 2.15 & 0.053 & 0.0014\\
       &     & 1        & 0.14   & 3.3e38 & 0.022  & 0.048  & 2.19 & 0.053 & 0.0011 & & 0.14  & 8.2e38 & 0.054 & 0.12  & 2.17 & 0.053 & 0.0013\\
4000   & 2.0 & $\infty$ & 0.014  & 1.8e40 & 1.2    & 10     & 2.07 & 0.026 & 0.010  & & 0.021 & 2.6e40 & 1.7   & 15    & 2.05 & 0.026 & 0.011\\
       &     & 10       & 0.014  & 1.7e40 & 1.2    & 9.7    & 2.07 & 0.026 & 0.0099 & & 0.021 & 2.5e40 & 1.7   & 14    & 2.05 & 0.026 & 0.011\\
       &     & 1        & 0.014  & 1.2e40 & 0.80   & 6.7    & 2.10 & 0.025 & 0.0082 & & 0.022 & 2.0e40 & 1.3   & 11    & 2.07 & 0.026 & 0.0087\\
       & 2.2 & $\infty$ & 0.0077 & 4.9e39 & 0.33   & 2.5    & 2.13 & 0.033 & 0.0043 & & 0.0082 & 5.4e39 & 0.36 & 2.8   & 2.12 & 0.033 & 0.0044\\
       &     & 10       & 0.0078 & 4.6e39 & 0.31   & 2.4    & 2.13 & 0.033 & 0.0042 & & 0.0083\tablenotemark{f} & 5.2e39 & 0.34 & 2.7   & 2.12 & 0.033 & 0.0043\\
       &     & 1        & 0.0082 & 3.3e39 & 0.22   & 1.7    & 2.16 & 0.033 & 0.0036 & & 0.0087 & 3.8e39 & 0.25 & 2.0   & 2.15 & 0.033 & 0.0037\\
       & 2.4 & $\infty$ & 0.017  & 8.0e38 & 0.054  & 0.38   & 2.19 & 0.035 & 0.0012 & & 0.017 & 9.0e38 & 0.060 & 0.43  & 2.19 & 0.035 & 0.0012\\
       &     & 10       & 0.017  & 7.7e38 & 0.051  & 0.37   & 2.20 & 0.035 & 0.0012 & & 0.017 & 8.6e38 & 0.057 & 0.41  & 2.19 & 0.035 & 0.0012\\
       &     & 1        & 0.017  & 5.5e38 & 0.037  & 0.26   & 2.23 & 0.034 & 0.0010 & & 0.018 & 6.5e38 & 0.043 & 0.31  & 2.22 & 0.034 & 0.0010\\
8000   & 2.0 & $\infty$ & 0.00054 & 2.2e40 & 1.5   & 34     & 2.06 & 0.017 & 0.0092 & & 8.4E-4 & 2.3e40 & 1.5  & 35    & 2.05 & 0.017 & 0.0092\\
       &     & 10       & 0.00057 & 2.1e40 & 1.4   & 32     & 2.06 & 0.017 & 0.0090 & & 8.9E-4 & 2.2e40 & 1.4  & 33    & 2.05 & 0.017 & 0.0090\\
       &     & 1        & 0.00083 & 1.5e40 & 0.99  & 23     & 2.09 & 0.017 & 0.0075 & & 0.0013 & 1.5e40 & 1.0  & 23    & 2.07 & 0.017 & 0.0076
\enddata
\label{table:A220XRayLuminosities}
\tablenotetext{a}{Fraction of the observed \emph{diffuse} 2 - 10 keV X-ray luminosity (from \citealt{Clements02}) from that synchrotron emission accounts for.  We assume that Arp 220 X-1 is the west nucleus, and Arp 220 X-4 is the east nucleus.}
\tablenotetext{b}{Ratio of the 2 - 10 keV synchrotron and IC luminosities.}
\tablenotetext{c}{Ratio of the predicted $\ge$100 MeV $\gamma$-ray emission with the \emph{Fermi}-LAT one-year catalog source sensitivity at high Galactic latitude \citep{Abdo10b}.}
\tablenotetext{d}{Ratio of the predicted 0.36 - 1.8 TeV $\gamma$-ray emission with the upper limits on the flux obtained with MAGIC in \citet{Albert07}.}
\tablenotetext{e}{Chosen as our fiducial model.}
\end{deluxetable}


\begin{thebibliography}{}
\bibitem[Abbasi et al.(2009a)]{Abbasi09} Abbasi, R., et al.\ 2009a, \apjl, 701, L47 

\bibitem[Abbasi et al.(2009b)]{Abbasi09b} Abbasi, R., et al.\ 2009b, Physical Review Letters, 103, 221102 

\bibitem[Abbasi et al.(2011)]{Abbasi11} Abbasi, R., et al.\ 2011, \apj, 732, 18 

\bibitem[Abdo et al.(2009)]{Abdo09} Abdo, A.~A., et al.\ 2009, \prl, 102, 181101 

\bibitem[Abdo et al.(2010a)]{Abdo10} Abdo, A.~A., et al.\ 2010a, \apjl, 709, L152 

\bibitem[Abdo et al.(2010b)]{Abdo10b} Abdo, A.~A., et al.\ 2010b, \apjs, 188, 405 

\bibitem[Abramowski et al.(2012)]{Abramowski12} Abramowski, A., Acero, F., Aharonian, F., et al.\ 2012, \apj, 757, 158 

\bibitem[Acero et al.(2009)]{Acero09} Acero, F., et al.\ 2009, Science, 326, 1080 

\bibitem[Acciari et al.(2009)]{Acciari09} Acciari, V.~A., et al.\ 2009, \nat, 462, 770 

\bibitem[Achterberg et al.(2007)]{Achterberg07} Achterberg, A., et al.\ 2007, \prd, 76, 042008 

\bibitem[Adriani et al.(2009)]{Adriani09} Adriani, O., et al.\ 2009, \nat, 458, 607 

\bibitem[Adriani et al.(2011)]{Adriani11} Adriani, O., Barbarino, G.~C., Bazilevskaya, G.~A., et al.\ 2011, Physical Review Letters, 106, 201101 

\bibitem[Aharonian et al.(1983)]{Aharonian83} Agaronyan, F.~A., Atoyan, A.~M., \& Nagapetyan, A.~M.\ 1983, Astrophysics, 19, 187 

\bibitem[Aharonian \& Atoyan(2000)]{Aharonian00} Aharonian, F.~A., \& Atoyan, A.~M.\ 2000, \aap, 362, 937 

\bibitem[Aharonian et al.(1995)]{Aharonian95} Aharonian, F.~A., Atoyan, A.~M., \& Voelk, H.~J.\ 1995, \aap, 294, L41 

\bibitem[Aharonian et al.(2004)]{Aharonian04} Aharonian, F., et al.\ 2004, \aap, 425, L13 

\bibitem[Aharonian(2004)]{Aharonian04b} Aharonian, F.~A.\ 2004, Very high energy cosmic gamma radiation : a crucial window on the extreme 
Universe, River Edge, NJ: World Scientific Publishing, 2004

\bibitem[Aharonian et al.(2006)]{Aharonian06} Aharonian, F., et al.\ 2006, \nat, 439, 695 

\bibitem[Aharonian et al.(2009a)]{Aharonian09a} Aharonian, F., et al.\ 2009a, \aap, 503, 817 

\bibitem[Aharonian et al.(2009b)]{Aharonian09b} Aharonian, F., et al.\ 2009b, \aap, 503, 817 

\bibitem[Albert et al.(2007)]{Albert07} Albert, J., et al.\ 2007, \apj, 658, 245 

\bibitem[Alexander et al.(2005)]{Alexander05} Alexander, D.~M., Bauer, F.~E., Chapman, S.~C., Smail, I., Blain, A.~W., Brandt, W.~N., \& Ivison, R.~J.\ 2005, \apj, 632, 736 

\bibitem[Allen et al.(1997)]{Allen97} Allen, G.~E., et al.\ 1997, \apjl, 487, L97 

\bibitem[Atoyan et al.(1995)]{Atoyan95} Atoyan, A.~M., Aharonian, F.~A., V\"olk, H.~J.\ 1995, \prd, 52, 3265 

\bibitem[Badhwar et al.(1977)]{Badhwar77} Badhwar, G.~D., Golden, R.~L., \& Stephens, S.~A.\ 1977, \prd, 15, 820 

\bibitem[Ballantyne et al.(2007)]{Ballantyne07} Ballantyne, D.~R., Melia, F., Liu, S., \& Crocker, R.~M.\ 2007, \apjl, 657, L13 

\bibitem[Bamba et al.(2010)]{Bamba10} Bamba, A., Anada, T., Dotani, T., Mori, K., Yamazaki, R., Ebisawa, K., \& Vink, J.\ 2010, \apjl, 719, L116 

\bibitem[Bauer et al.(2008)]{Bauer08} Bauer, M., Pietsch, W., Trinchieri, G., Breitschwerdt, D., Ehle, M., Freyberg, M.~J., \& Read, A.~M.\ 2008, \aap, 489, 1029 

\bibitem[Beck(2005)]{Beck05} Beck, R.\ 2005, Cosmic Magnetic Fields, 664, 41 

\bibitem[Beck(2011)]{Beck11} Beck, R.\ 2011, arXiv:1112.1823 

\bibitem[Berezhko \& Ellison(1999)]{Berezhko99} Berezhko, E.~G., \& Ellison, D.~C.\ 1999, \apj, 526, 385 

\bibitem[Bhattacharya et al.(1994)]{Bhattacharya94} Bhattacharya, D., et al.\ 1994, \apj, 437, 173 

\bibitem[Bi et al.(2009)]{Bi09} Bi, X.-J., Chen, T.-L., Wang, Y., \& Yuan, Q.\ 2009, \apj, 695, 883 

\bibitem[Blasi et al.(2005)]{Blasi05} Blasi, P., Gabici, S., \& Vannoni, G.\ 2005, \mnras, 361, 907 

\bibitem[B\"ottcher \& Schlickeiser(1997)]{Boettcher97} B\"ottcher, M., \& Schlickeiser, R.\ 1997, \aap, 325, 866 

\bibitem[Bradford et al.(2003)]{Bradford03} Bradford, C.~M., Nikola, T., Stacey, G.~J., Bolatto, A.~D., Jackson, J.~M., Savage, M.~L., Davidson, J.~A., \& Higdon, S.~J.\ 2003, \apj, 586, 891 

\bibitem[Brogan et al.(2003)]{Brogan03} Brogan, C.~L., Nord, M., Kassim, N., Lazio, J., \& Anantharamaiah, K.\ 2003, Astronomische Nachrichten Supplement, 324, 17 

\bibitem[B{\"u}sching et al.(2007)]{Buesching07} B{\"u}sching, I., de Jager, O.~C., \& Snyman, J.\ 2007, \apj, 656, 841 

\bibitem[Butt(2009)]{Butt09} Butt, Y.\ 2009, \nat, 460, 701 

\bibitem[Calzetti et al.(2000)]{Calzetti00} Calzetti, D., Armus, L., Bohlin, R.~C., Kinney, A.~L., Koornneef, J., \& Storchi-Bergmann, T.\ 2000, \apj, 533, 682

\bibitem[Cappi et al.(1999)]{Cappi99} Cappi, M., et al.\ 1999, \aap, 350, 777 

\bibitem[Chang et al.(2008)]{Chang08} Chang, J., et al.\ 2008, \nat, 456, 362 

\bibitem[Chapman et al.(2004)]{Chapman04} Chapman, S.~C., Smail, I., Windhorst, R., Muxlow, T., \& Ivison, R.~J.\ 2004, \apj, 611, 732 

\bibitem[Chevalier \& Clegg(1985)]{Chevalier85} Chevalier, R.~A., \& Clegg, A.~W.\ 1985, \nat, 317, 44 

\bibitem[Clements et al.(2002)]{Clements02} Clements, D.~L., McDowell, J.~C., Shaked, S., Baker, A.~C., Borne, K., Colina, L., Lamb, S.~A., \& Mundell, C.\ 2002, \apj, 581, 974 

\bibitem[Condon et al.(1991)]{Condon91} Condon, J. J., Huang, Z.-P., Yin, Q. F., \& Thuan, T. X. 1991, \apj~378, 65.

\bibitem[Condon(1992)]{Condon92} Condon, J. J. 1992, \araa~30, 575.

\bibitem[Connell(1998)]{Connell98} Connell, J.~J.\ 1998, \apjl, 501, L59 

\bibitem[Coyle(2010)]{Coyle10} Coyle, P.\ 2010, arXiv:1002.0754 

\bibitem[Crocker et al.(2010a)]{Crocker10} Crocker, R.~M., Jones, D.~I., Melia, F., Ott, J., \& Protheroe, R.~J.\ 2010, \nat, 463, 65 

\bibitem[Crocker et al.(2011a)]{Crocker10b} Crocker, R.~M., Jones, D.~I., Aharonian, F., Law, C.~J., Melia, F., \& Ott, J.\ 2011a, \mnras, 411, L11 

\bibitem[Crocker et al.(2011b)]{Crocker11} Crocker, R.~M., Jones, D.~I., Aharonian, F., Law, C.~J., Melia, F., Oka, T., \& Ott, J.\ 2011b, \mnras, 313 

\bibitem[Cusumano et al.(2010)]{Cusumano10} Cusumano, G., et al.\ 2010, arXiv:1009.0522 

\bibitem[Dahlem et al.(1998)]{Dahlem98} Dahlem, M., Weaver, K.~A., \& Heckman, T.~M.\ 1998, \apjs, 118, 401 

\bibitem[David et al.(1992)]{David92} David, L.~P., Jones, C., \& Forman, W.\ 1992, \apj, 388, 82 

\bibitem[Davidson et al.(1992)]{Davidson92} Davidson, J.~A., Werner, M.~W., Wu, X., Lester, D.~F., Harvey, P.~M., Joy, M., \& Morris, M.\ 1992, \apj, 387, 189 

\bibitem[de Cea del Pozo et al.(2009a)]{deCeaDelPozo09a} de Cea del Pozo, E., Torres, D.~F., \& Rodriguez Marrero, A.~Y.\ 2009a, \apj, 698, 1054 

\bibitem[de Cea del Pozo et al.(2009b)]{deCeaDelPozo09b} de Cea del Pozo, E., Torres, D.~F., Rodriguez, A.~Y., \& Reimer, O.\ 2009b, arXiv:0912.3497 

\bibitem[Dermer(1986a)]{Dermer86a} Dermer, C.~D.\ 1986a, \aap, 157, 223 

\bibitem[Dermer(1986b)]{Dermer86b} Dermer, C.~D.\ 1986b, \apj, 307, 47 

\bibitem[Dermer et al.(1997)]{Dermer97} Dermer, C.~D., Bland-Hawthorn, J., Chiang, J., \& McNaron-Brown, K.\ 1997, \apjl, 484, L121 

\bibitem[di Bernardo et al.(2010)]{DiBernardo10} di Bernardo, G., Evoli, C., Gaggero, D., Grasso, D., \& Maccione, L.\ 2010, Astroparticle Physics, 34, 274 

\bibitem[Domingo-Santamar\'ia \& Torres(2005)]{Domingo05} Domingo-Santamar\'ia, E. \& Torres, D. F. 2005, \aap 444, 403.

\bibitem[Doro(2009)]{Doro09} Doro, M., \ 2009, arXiv:0908.1410

\bibitem[Downes \& Solomon(1998)]{Downes98} Downes, D., \& Solomon, P.~M.\ 1998, \apj, 507, 615 

\bibitem[Downes \& Eckart(2007)]{Downes07} Downes, D., \& Eckart, A.\ 2007, \aap, 468, L57 

\bibitem[Dwek \& Barker(2002)]{Dwek02} Dwek, E., \& Barker, M.~K.\ 2002, \apj, 575, 7 

\bibitem[Ellison et al.(2000)]{Ellison00} Ellison, D.~C., Berezhko, E.~G., \& Baring, M.~G.\ 2000, \apj, 540, 292 

\bibitem[Everett et al.(2008)]{Everett08} Everett, J.~E., Zweibel, E.~G., Benjamin, R.~A., McCammon, D., Rocks, L., \& Gallagher, J.~S., III 2008, \apj, 674, 258 

\bibitem[Fabbiano et al.(1982)]{Fabbiano82} Fabbiano, G., Feigelson, E., \& Zamorani, G.\ 1982, \apj, 256, 397 

\bibitem[Fabbiano(1989)]{Fabbiano89} Fabbiano, G.\ 1989, \araa, 27, 87 

\bibitem[Ferri{\`e}re et al.(2007)]{Ferriere07} Ferri{\`e}re, K., Gillard, W., \& Jean, P.\ 2007, \aap, 467, 611 

\bibitem[Franceschini et al.(2003)]{Franceschini03} Franceschini, A., et al.\ 2003, \mnras, 343, 1181 

\bibitem[Gaisser(1990)]{Gaisser90} Gaisser, T.~K.\ 1990, Cambridge and New York, Cambridge University Press, 1990, p.160.  

\bibitem[Genzel et al.(2008)]{Genzel08} Genzel, R., et al.\ 2008, \apj, 687, 59 

\bibitem[Ginzburg \& Ptuskin(1976)]{Ginzburg76} Ginzburg, V.~L., \& Ptuskin, V.~S.\ 1976, Reviews of Modern Physics, 48, 161 

\bibitem[Goetz et al.(1990)]{Goetz90} Goetz, M., Downes, D., Greve, A., \& McKeith, C.~D.\ 1990, \aap, 240, 52 

\bibitem[Gould \& Schr{\'e}der(1967)]{Gould67} Gould, R.~J., \& Schr{\'e}der, G.~P.\ 1967, Physical Review , 155, 1404 

\bibitem[Greaves et al.(2000)]{Greaves00} Greaves, J.~S., Holland, W.~S., Jenness, T., \& Hawarden, T.~G.\ 2000, \nat, 404, 732 

\bibitem[Greve(2004)]{Greve04} Greve, A.\ 2004, \aap, 416, 67 

\bibitem[Griffiths \& Padovani(1990)]{Griffiths90} Griffiths, R.~E., \& Padovani, P.\ 1990, \apj, 360, 483 

\bibitem[Grimm et al.(2003)]{Grimm03} Grimm, H.-J., Gilfanov, M., \& Sunyaev, R.\ 2003, \mnras, 339, 793 

\bibitem[Hargrave(1974)]{Hargrave74} Hargrave, P.~J.\ 1974, \mnras, 168, 491 

\bibitem[Harrison et al.(1999)]{Harrison99} Harrison, A., Henkel, C., \& Russell, A.\ 1999, \mnras, 303, 157 

\bibitem[Harrison et al.(2005)]{Harrison05} Harrison, F.~A., et al.\ 2005, Experimental Astronomy, 20, 131 

\bibitem[Harrison et al.(2010)]{Harrison10} Harrison, F.~A., et al.\ 2010, \procspie, 7732

\bibitem[Heckman et al.(2000)]{Heckman00} Heckman, T. M., Lehnert, M. D., Strickland, D. K., Armus, L. 2000, \apjs~129, 493.

\bibitem[Heckman(2003)]{Heckman03} Heckman, T. M. 2003, in Rev. Mex. AA Ser. Conf., 17, 47

\bibitem[Heesen et al.(2011)]{Heesen11} Heesen, V., Beck, R., Krause, M., \& Dettmar, R.-J.\ 2011, \aap, 535, A79 

\bibitem[Hooper et al.(2009)]{Hooper09} Hooper, D., Blasi, P., \& Dario Serpico, P.\ 2009, Journal of Cosmology and Astroparticle Physics, 1, 25 

\bibitem[Hopkins et al.(2010)]{Hopkins10} Hopkins, P.~F., Murray, N., Quataert, E., \& Thompson, T.~A.\ 2010, \mnras, 401, L19 

\bibitem[Hunter et al.(1997)]{Hunter97} Hunter, S.~D., et al.\ 1997, \apj, 481, 205 

\bibitem[Inoue(2011)]{Inoue11} Inoue, Y.\ 2011, \apj, 728, 11

\bibitem[Iwasawa et al.(2001)]{Iwasawa01} Iwasawa, K., Matt, G., Guainazzi, M., \& Fabian, A.~C.\ 2001, \mnras, 326, 894 

\bibitem[Iwasawa et al.(2005)]{Iwasawa05} Iwasawa, K., Sanders, D.~B., Evans, A.~S., Trentham, N., Miniutti, G., \& Spoon, H.~W.~W.\ 2005, \mnras, 357, 565 

\bibitem[Iwasawa et al.(2009)]{Iwasawa09} Iwasawa, K., Sanders, D.~B., Evans, A.~S., Mazzarella, J.~M., Armus, L., \& Surace, J.~A.\ 2009, \apjl, 695, L103 

\bibitem[Jones(2000)]{Jones00} Jones, T.~J.\ 2000, \aj, 120, 2920 

\bibitem[Kamae et al.(2006)]{Kamae06} Kamae, T., Karlsson, N., Mizuno, T., Abe, T., \& Koi, T.\ 2006, \apj, 647, 692 

\bibitem[Karlsson(2008)]{Karlsson08} Karlsson, N.\ 2008, American Institute of Physics Conference Series, 1085, 561 

\bibitem[Kelner et al.(2006)]{Kelner06} Kelner, S.~R., Aharonian, F.~A., \& Bugayov, V.~V.\ 2006, \prd, 74, 034018 

\bibitem[Kennicutt(1998)]{Kennicutt98} Kennicutt, R.~C.\ 1998, \apj, 498, 541 

\bibitem[Kirsch et al.(2005)]{Kirsch05} Kirsch, M.~G., et al.\ 2005, \procspie, 5898, 22 

\bibitem[Kistler \& Y\"uksel(2009)]{Kistler09} Kistler, M.~D., \& Y\"uksel, H.\ 2009, arXiv:0912.0264 

\bibitem[Klein et al.(1988)]{Klein88} Klein, U., Wielebinski, R., \& Morsi, H.~W.\ 1988, \aap, 190, 41 

\bibitem[Koyama et al.(1995)]{Koyama95} Koyama, K., Petre, R., Gotthelf, E.~V., Hwang, U., Matsuura, M., Ozaki, M., \& Holt, S.~S.\ 1995, \nat, 378, 255 

\bibitem[Koyama et al.(1996)]{Koyama96} Koyama, K., Maeda, Y., Sonobe, T., Takeshima, T., Tanaka, Y., \& Yamauchi, S.\ 1996, \pasj, 48, 249 

\bibitem[Krivonos et al.(2007)]{Krivonos07} Krivonos, R., Revnivtsev, M., Churazov, E., Sazonov, S., Grebenev, S., \& Sunyaev, R.\ 2007, \aap, 463, 957 

\bibitem[Lacki et al.(2010)]{Lacki10a} Lacki, B.~C., Thompson, T.~A., \& Quataert, E.\ 2010, \apj, 717, 1 

\bibitem[Lacki \& Thompson(2010)]{Lacki10b} Lacki, B.~C., \& Thompson, T.~A.\ 2010, \apj, 717, 196 

\bibitem[Lacki et al.(2011)]{Lacki11} Lacki, B.~C., Thompson, T.~A., Quataert, E., Loeb, A., \& Waxman, E.\ 2011, \apj, 734, 107 

\bibitem[Laing(1980)]{Laing80} Laing, R.~A.\ 1980, \mnras, 193, 439 

\bibitem[Launhardt et al.(2002)]{Launhardt02} Launhardt, R., Zylka, R., \& Mezger, P.~G.\ 2002, \aap, 384, 112 

\bibitem[Law et al.(2009)]{Law09} Law, D.~R., Steidel, C.~C., Erb, D.~K., Larkin, J.~E., Pettini, M., Shapley, A.~E., \& Wright, S.~A.\ 2009, \apj, 697, 2057 

\bibitem[Lehmer et al.(2010)]{Lehmer10} Lehmer, B.~D., Alexander, D.~M., Bauer, F.~E., Brandt, W.~N., Goulding, A.~D., Jenkins, L.~P., Ptak, A., \& Roberts, T.~P.\ 2010, \apj, 724, 559 

\bibitem[Lira et al.(2002)]{Lira02} Lira, P., Ward, M., Zezas, A., Alonso-Herrero, A., \& Ueno, S.\ 2002, \mnras, 330, 259 

\bibitem[Loeb \& Waxman(2006)]{Loeb06} Loeb, A. \& Waxman, E. 2006, Journal of Cosmology and Astroparticle Physics 5,3.

\bibitem[Lonsdale et al.(2006)]{Lonsdale06} Lonsdale, C.~J., Diamond, P.~J., Thrall, H., Smith, H.~E., \& Lonsdale, C.~J.\ 2006, \apj, 647, 185 

\bibitem[Malkov \& O'C Drury(2001)]{Malkov01} Malkov, M.~A., \& O'C Drury, L.\ 2001, Reports on Progress in Physics, 64, 429 

\bibitem[Mannheim \& Schlickeiser(1994)]{Mannheim94} Mannheim, K., \& Schlickeiser, R.\ 1994, \aap, 286, 983 

\bibitem[Mannheim et al.(2010)]{Mannheim10} Mannheim, K., Els{\"a}sser, D., \& Tibolla, O.\ 2010, arXiv:1010.2185 

\bibitem[Mastichiadis et al.(1991)]{Mastichiadis91} Mastichiadis, A., Protheroe, R.~J., \& Stephens, S.~A.\ 1991, Proceedings of the Astronomical Society of Australia, 9, 115 

\bibitem[Mauersberger et al.(1996)]{Mauersberger96} Mauersberger, R., Henkel, C., Wielebinski, R., Wiklind, T., \& Reuter, H.-P.\ 1996, \aap, 305, 421 

\bibitem[McDowell et al.(2003)]{McDowell03} McDowell, J.~C., et al.\ 2003, \apj, 591, 154 

\bibitem[Melia \& Fatuzzo(2011)]{Melia11} Melia, F., \& Fatuzzo, M.\ 2011, \mnras, 410, L23 

\bibitem[Melo et al.(2002)]{Melo02} Melo, V.~P., P{\'e}rez Garc{\'{\i}}a, A.~M., Acosta-Pulido, J.~A., Mu{\~n}oz-Tu{\~n}{\'o}n, C., \& Rodr{\'{\i}}guez Espinosa, J.~M.\ 2002, \apj, 574, 709 

\bibitem[Miyawaki et al.(2009)]{Miyawaki09} Miyawaki, R., Makishima, K., Yamada, S., Gandhi, P., Mizuno, T., Kubota, A., Tsuru, T.~G., \& Matsumoto, H.\ 2009, \pasj, 61, 263 

\bibitem[Moran \& Lehnert(1997)]{Moran97} Moran, E.~C., \& Lehnert, M.~D.\ 1997, \apj, 478, 172 

\bibitem[Moran et al.(1999)]{Moran99} Moran, E.~C., Lehnert, M.~D., \& Helfand, D.~J.\ 1999, \apj, 526, 649 

\bibitem[Moskalenko \& Strong(1998)]{Moskalenko98} Moskalenko, I.~V., \& Strong, A.~W.\ 1998, \apj, 493, 694 

\bibitem[Moskalenko et al.(2006)]{Moskalenko06} Moskalenko, I.~V., Porter, T.~A., \& Strong, A.~W.\ 2006, \apjl, 640, L155 

\bibitem[Murphy(2009)]{Murphy09} Murphy, E.~J.\ 2009, \apj, 706, 482 

\bibitem[Murphy et al.(2012)]{Murphy12} Murphy, E.~J., Porter, T.~A., Moskalenko, I.~V., Helou, G., \& Strong, A.~W.\ 2012, \apj, 750, 126 

\bibitem[Paglione et al.(1996)]{Paglione96} Paglione, T.~A.~D., Marscher, A.~P., Jackson, J.~M., \& Bertsch, D.~L.\ 1996, \apj, 460, 295 

\bibitem[Parker(1966)]{Parker66} Parker, E.~N.\ 1966, \apj, 145, 811 

\bibitem[Peng et al.(1996)]{Peng96} Peng, R., Zhou, S., Whiteoak, J.~B., Lo, K.~Y., \& Sutton, E.~C.\ 1996, \apj, 470, 821 

\bibitem[Perna \& Stella(2004)]{Perna04} Perna, R., \& Stella, L.\ 2004, \apj, 615, 222 

\bibitem[Persic et al.(1998)]{Persic98} Persic, M., et al.\ 1998, \aap, 339, L33 

\bibitem[Persic \& Rephaeli(2002)]{Persic02} Persic, M., \& Rephaeli, Y.\ 2002, \aap, 382, 843 

\bibitem[Persic \& Rephaeli(2003)]{Persic03} Persic, M., \& Rephaeli, Y.\ 2003, \aap, 399, 9 

\bibitem[Persic et al.(2004)]{Persic04} Persic, M., Rephaeli, Y., Braito, V., Cappi, M., Della Ceca, R., Franceschini, A., \& Gruber, D.~E.\ 2004, \aap, 419, 849 

\bibitem[Persic \& Rephaeli(2007)]{Persic07} Persic, M., \& Rephaeli, Y.\ 2007, \aap, 463, 481 

\bibitem[Persic et al.(2008)]{Persic08} Persic, M., Rephaeli, Y., \& Arieli, Y. 2008, \aap~486, 143.

\bibitem[Persic \& Rephaeli(2010)]{Persic10} Persic, M., \& Rephaeli, Y.\ 2010, \mnras, 403, 1569 

\bibitem[Pierce-Price et al.(2000)]{PiercePrice00} Pierce-Price, D., et al.\ 2000, \apjl, 545, L121 

\bibitem[Porter \& Protheroe(1997)]{Porter97} Porter, T.~A., \& Protheroe, R.~J.\ 1997, Journal of Physics G Nuclear Physics, 23, 1765 

\bibitem[Porter et al.(2008)]{Porter08} Porter, T.~A., Moskalenko, I.~V., Strong, A.~W., Orlando, E., \& Bouchet, L.\ 2008, \apj, 682, 400 

\bibitem[Protheroe \& Wolfendale(1980)]{Protheroe80} Protheroe, R.~J., \& Wolfendale, A.~W.\ 1980, \aap, 92, 175 

\bibitem[Ptak et al.(1997)]{Ptak97} Ptak, A., Serlemitsos, P., Yaqoob, T., Mushotzky, R., \& Tsuru, T.\ 1997, \aj, 113, 1286 

\bibitem[Ranalli et al.(2003)]{Ranalli03} Ranalli, P., Comastri, A., \& Setti, G.\ 2003, \aap, 399, 39 

\bibitem[Rengarajan(2005)]{Rengarajan05} Rengarajan, T.~N.\ 2005, Proc. 29th Int. Cosmic Ray Conf. (Pune), 3.

\bibitem[Rephaeli et al.(1991)]{Rephaeli91} Rephaeli, Y., Gruber, D., MacDonald, D., \& Persic, M.\ 1991, \apjl, 380, L59 

\bibitem[Rephaeli et al.(1995)]{Rephaeli95} Rephaeli, Y., Gruber, D., \& Persic, M.\ 1995, \aap, 300, 91 

\bibitem[Rephaeli et al.(2009)]{Rephaeli09} Rephaeli, Y., Arieli, Y., \& Persic, M.\ 2010, \mnras, 401, 473 

\bibitem[Reuter et al.(1994)]{Reuter94} Reuter, H.-P., Klein, U., Lesch, H., Wielebinski, R., \& Kronberg, P.~P.\ 1994, \aap, 282, 724 

\bibitem[Revnivtsev et al.(2006)]{Revnivtsev06} Revnivtsev, M., Sazonov, S., Gilfanov, M., Churazov, E., \& Sunyaev, R.\ 2006, \aap, 452, 169 

\bibitem[Revnivtsev et al.(2009)]{Revnivtsev09} Revnivtsev, M., Sazonov, S., Churazov, E., Forman, W., Vikhlinin, A., \& Sunyaev, R.\ 2009, \nat, 458, 1142 

\bibitem[Reynolds \& Keohane(1999)]{Reynolds99} Reynolds, S.~P., \& Keohane, J.~W.\ 1999, \apj, 525, 368 

\bibitem[Reynolds(2008)]{Reynolds08a} Reynolds, S.~P.\ 2008, \araa, 46, 89 

\bibitem[Reynolds et al.(2008)]{Reynolds08b} Reynolds, S.~P., Borkowski, K.~J., Green, D.~A., Hwang, U., Harrus, I., 
\& Petre, R.\ 2008, \apjl, 680, L41 

\bibitem[Robishaw et al.(2008)]{Robinshaw08} Robishaw, T., Quataert, E., \& Heiles, C.\ 2008, \apj, 680, 981 

\bibitem[Roussel et al.(2003)]{Roussel03} Roussel, H., Helou, G., Beck, R., Condon, J.~J., Bosma, A., Matthews, K., \& Jarrett, T.~H.\ 2003, \apj, 593, 733.

\bibitem[Rovilos et al.(2005)]{Rovilos05} Rovilos, E., Diamond, P.~J., Lonsdale, C.~J., Smith, H.~E., \& Lonsdale, C.~J.\ 2005, \mnras, 359, 827 

\bibitem[Rybicki \& Lightman(1979)]{Rybicki79} Rybicki, G.~B., \& Lightman, A.~P.\ 1979, New York, Wiley-Interscience, 1979

\bibitem[Sakamoto et al.(1999)]{Sakamoto99} Sakamoto, K., Scoville, N.~Z., Yun, M.~S., Crosas, M., Genzel, R., \& Tacconi, L.~J.\ 1999, \apj, 514, 68 

\bibitem[Sakamoto et al.(2008)]{Sakamoto08} Sakamoto, K., et al.\ 2008, \apj, 684, 957 

\bibitem[Sakamoto et al.(2011)]{Sakamoto11} Sakamoto, K., Mao, R.-Q., Matsushita, S., Peck, A.~B., Sawada, T., \& Wiedner, M.~C.\ 2011, \apj, 735, 19

\bibitem[Sanders et al.(2003)]{Sanders03} Sanders, D.~B., Mazzarella, J.~M., Kim, D.-C., Surace, J.~A., 
\& Soifer, B.~T.\ 2003, \aj, 126, 1607 

\bibitem[Schaaf et al.(1989)]{Schaaf89} Schaaf, R., Pietsch, W., Biermann, P.~L., Kronberg, P.~P., \& Schmutzler, T.\ 1989, \apj, 336, 722

\bibitem[Schlickeiser(2002)]{Schlickeiser02} Schlickeiser, R. 2002, \emph{Cosmic Ray Astrophysics}, (New York: Springer).

\bibitem[Schlickeiser \& Ruppel(2010)]{Schlickeiser10} Schlickeiser, R., \& Ruppel, J.\ 2010, New Journal of Physics, 12, 033044 

\bibitem[Seaquist et al.(1985)]{Seaquist85} Seaquist, E.~R., Bell, M.~B., \& Bignell, R.~C.\ 1985, \apj, 294, 546 

\bibitem[Seiffert et al.(2007)]{Seiffert07} Seiffert, M., Borys, C., Scott, D., \& Halpern, M.\ 2007, \mnras, 374, 409 

\bibitem[Shen(1970)]{Shen70} Shen, C.~S.\ 1970, \apjl, 162, L181 

\bibitem[Siebenmorgen \& Efstathiou(2001)]{Siebenmorgen01} Siebenmorgen, R., \& Efstathiou, A.\ 2001, \aap, 376, L35 

\bibitem[Silva et al.(1998)]{Silva98} Silva, L., Granato, G.~L., Bressan, A., \& Danese, L.\ 1998, \apj, 509, 103

\bibitem[Sokoloff et al.(1998)]{Sokoloff98} Sokoloff, D.~D., Bykov, A.~A., Shukurov, A., Berkhuijsen, E.~M., Beck, R., \& Poezd, A.~D.\ 1998, \mnras, 299, 189 

\bibitem[Stawarz et al.(2010)]{Stawarz10} Stawarz, {\L}., Petrosian, V., \& Blandford, R.~D.\ 2010, \apj, 710, 236 

\bibitem[Stecker(1970)]{Stecker70} Stecker, F.~W.\ 1970, \apss, 6, 377 

\bibitem[Steigman \& Strittmatter(1971)]{Steigman71} Steigman, G., \& Strittmatter, P.~A.\ 1971, \aap, 11, 279 

\bibitem[Stephens \& Badhwar(1981)]{Stephens81} Stephens, S.~A., \& Badhwar, G.~D.\ 1981, \apss, 76, 213 

\bibitem[Strickland \& Stevens(2000)]{Strickland00} Strickland, D.~K., \& Stevens, I.~R.\ 2000, \mnras, 314, 511 

\bibitem[Strickland et al.(2000)]{Strickland00-N253} Strickland, D.~K., Heckman, T.~M., Weaver, K.~A., \& Dahlem, M.\ 2000, \aj, 120, 2965 

\bibitem[Strickland et al.(2002)]{Strickland02} Strickland, D.~K., Heckman, T.~M., Weaver, K.~A., Hoopes, C.~G., \& Dahlem, M.\ 2002, \apj, 568, 689 

\bibitem[Strickland \& Heckman(2007)]{Strickland07} Strickland, D.~K., \& Heckman, T.~M.\ 2007, \apj, 658, 258 

\bibitem[Strickland \& Heckman(2009)]{Strickland09} Strickland, D.~K., \& Heckman, T.~M.\ 2009, \apj, 697, 2030 

\bibitem[Strong \& Moskalenko(1998)]{Strong98} Strong, A.~W., \& Moskalenko, I.~V.\ 1998, \apj, 509, 212 

\bibitem[Strong et al.(2000)]{Strong00} Strong, A.~W., Moskalenko, I.~V., \& Reimer, O.\ 2000, \apj, 537, 763 

\bibitem[Strong et al.(2004)]{Strong04} Strong, A.~W., Moskalenko, I.~V., \& Reimer, O.\ 2004, \apj, 613, 956

\bibitem[Strong et al.(2007)]{Strong07} Strong, A.~W., Moskalenko, I.~V., \& Ptuskin, V.~S.\ 2007, Annual Review of Nuclear and Particle Science, 57, 285 

\bibitem[Strong et al.(2010)]{Strong10} Strong, A.~W., Porter, T.~A., Digel, S.~W., J{\'o}hannesson, G., Martin, P., Moskalenko, I.~V., Murphy, E.~J., \& Orlando, E.\ 2010, \apjl, 722, L58 

\bibitem[Swank et al.(2009)]{Swank09} Swank, J., et al. 2009, eds. R. Bellazzini, E. Costa, G. Matt and G. Tagliaferri, in X-ray Polarimetry: A New Window in Astrophysics (Rome)

\bibitem[Tacconi et al.(2006)]{Tacconi06} Tacconi, L.~J., et al.\ 2006, \apj, 640, 228 

\bibitem[Telesco et al.(1996)]{Telesco96} Telesco, C.~M., Davidson, J.~A., \& Werner, M.~W.\ 1996, \apj, 456, 541 

\bibitem[Thompson et al.(2006)]{Thompson06} Thompson, T. A. et al. 2006, \apj~645, 186.

\bibitem[Thompson, Quataert, \& Waxman(2007)]{Thompson07} Thompson, T. A., Quataert, E., Waxman, E. 2007, \apj~654, 219.

\bibitem[Torres et al.(2004a)]{Torres04a} Torres, D.~F., Reimer, O., Domingo-Santamar{\'{\i}}a, E., \& Digel, S.~W.\ 2004, \apjl, 607, L99 

\bibitem[Torres(2004)]{Torres04} Torres, D. F. 2004, \apj~617, 966.

\bibitem[Turner \& Ho(1983)]{Turner83} Turner, J.~L., \& Ho, P.~T.~P.\ 1983, \apjl, 268, L79 

\bibitem[T{\"u}rler et al.(2010)]{Tuerler10} T{\"u}rler, M., Chernyakova, M., Courvoisier, T.~J.-L., Lubi{\'n}ski, P., Neronov, A., Produit, N., \& Walter, R.\ 2010, \aap, 512, A49 

\bibitem[Ulvestad(2000)]{Ulvestad00} Ulvestad, J.~S.\ 2000, \aj, 120, 278 

\bibitem[Ulvestad \& Antonucci(1997)]{Ulvestad97} Ulvestad, J.~S., \& Antonucci, R.~R.~J.\ 1997, \apj, 488, 621

\bibitem[V\"olk et al.(1989)]{Voelk89a} V\"olk, H.~J., Klein, U., \& Wielebinski, R.\ 1989, \aap, 213, L12 

\bibitem[V\"olk(1989)]{Voelk89b} V\"olk, H.~J.\ 1989, \aap, 218, 67 

\bibitem[Warren et al.(2005)]{Warren05} Warren, J.~S., et al.\ 2005, \apj, 634, 376 

\bibitem[Weaver et al.(2002)]{Weaver02} Weaver, K.~A., Heckman, T.~M., Strickland, D.~K., \& Dahlem, M.\ 2002, \apjl, 576, L19 

\bibitem[Webber et al.(2003)]{Webber03} Webber, W.~R., McDonald, F.~B., \& Lukasiak, A.\ 2003, \apj, 599, 582 

\bibitem[Wei{\ss} et al.(2001)]{Weiss01} Wei{\ss}, A., Neininger, N., H{\"u}ttemeister, S., \& Klein, U.\ 2001, \aap, 365, 571 

\bibitem[Westmoquette et al.(2009)]{Westmoquette09} Westmoquette, M.~S., Smith, L.~J., Gallagher, J.~S., Trancho, G., Bastian, N., \& Konstantopoulos, I.~S.\ 2009, \apj, 696, 192 

\bibitem[White et al.(1983)]{White83} White, N.~E., Swank, J.~H., \& Holt, S.~S.\ 1983, \apj, 270, 711 

\bibitem[Williams \& Bower(2010)]{Williams10} Williams, P.~K.~G., \& Bower, G.~C.\ 2010, \apj, 710, 1462 

\bibitem[Wommer et al.(2008)]{Wommer08} Wommer, E., Melia, F., \& Fatuzzo, M.\ 2008, \mnras, 387, 987 

\bibitem[Worrall et al.(1982)]{Worrall82} Worrall, D.~M., Marshall, F.~E., Boldt, E.~A., \& Swank, J.~H.\ 1982, \apj, 255, 111 

\bibitem[Younger et al.(2010)]{Younger10} Younger, J.~D., et al.\ 2010, \mnras, 984 

\bibitem[Y{\"u}ksel et al.(2009)]{Yuksel09} Y{\"u}ksel, H., Kistler, M.~D., \& Stanev, T.\ 2009, Physical Review Letters, 103, 051101 

\bibitem[Yun et al.(2001)]{Yun01} Yun, M. S., Reddy, N. A., \& Condon, J. J. 2001, \apj, 554, 803

\bibitem[Yusef-Zadeh et al.(2009)]{YusefZadeh09} Yusef-Zadeh, F., et al.\ 2009, \apj, 702, 178 

\end{thebibliography}
\end{document}